\documentclass[prb,amsmath,amssymb,floatfix,footinbib,bibnotes,longbibliography, twocolumn]{revtex4-1}
\usepackage{lipsum}
\usepackage[colorlinks=true,citecolor=blue,linkcolor=magenta]{hyperref}
\usepackage{amsmath}
\usepackage{soul}
\usepackage{graphicx}
\usepackage{changepage}
\usepackage{fancyhdr}
\usepackage{amsthm, amssymb}
\usepackage{floatflt}
\usepackage{float}
\usepackage{array}
\usepackage[usenames,dvipsnames]{color}
\usepackage{mathtools}
\usepackage[normalem]{ulem}
\usepackage{flushend}
\usepackage{feynmp}
\usepackage{tikz-feynman}
\tikzfeynmanset{compat=1.1.0}

\def \mr{\mathrm}

\newcommand{\usenomenclature}{}
\ifdefined\usenomenclature
\usepackage[noprefix]{nomencl}
\fi

\newcommand{\beq}{\begin{equation}}
\newcommand{\eeq}{\end{equation}}
\newcommand{\bcen}{\begin{center}}
\newcommand{\ecen}{\end{center}}
\newcommand{\btab}{\begin{tabular}}
\newcommand{\etab}{\end{tabular}}
\newcommand{\bdes}{\begin{description}}
\newcommand{\edes}{\end{description}}

\newcommand{\bea}{\begin{eqnarray}}
\newcommand{\eea}{\end{eqnarray}}

\newcommand{\bary}{\begin{array}}
\newcommand{\eary}{\end{array}}
\newcommand{\benum}{\begin{enumerate}}
\newcommand{\eenum}{\end{enumerate}}
\newcommand{\bitem}{\begin{itemize}}
\newcommand{\eitem}{\end{itemize}}

\makeatletter

\newcommand{\Rmnum}[1]{\expandafter\@slowromancap\romannumeral #1@}
\makeatother

\newcommand{\ci}{\mathfrak{i}}

\newcommand{\tbd}[1]{}

\newcommand{\concept}[1]{}
\newcommand{\figconcept}[1]{}
\newcommand{\eqnconcept}[1]{}
\newcommand{\tabconcept}[1]{}
\newcommand{\appconcept}[1]{}

\newcommand{\G}[2]{G^{#2}_{#1}}

\usepackage{cancel}

\renewcommand{\thesection}{ { \arabic{section}}}
\renewcommand{\theequation}{\arabic{section}.\arabic{equation}}
\renewcommand{\thefigure}{\arabic{section}.\arabic{figure}}
\numberwithin{equation}{section}
\numberwithin{figure}{section}
\newcommand {\apgt} {\ {\raise-.5ex\hbox{$\buildrel>\over\sim$}}\ }
\newcommand {\aplt} {\ {\raise-.5ex\hbox{$\buildrel<\over\sim$}}\ }

\def  \w{\omega}
\def  \Ginv{G^{-1}}
\def  \G{\Gamma}
\def  \q0{\frac{\w_k^2}{\G}+z}

\def \titlename {Information scrambling and butterfly velocity in quantum spin glass chains}
\def \authornames{Venkata Lokesh K. Y$^{1}$, Surajit Bera$^2$ and Sumilan Banerjee$^2$}
\def \affiliations{$^1$Centre of High Energy Physics, Indian Institute of Science, Bangalore 560012, India\\
$^2$Centre for Condensed Matter Theory, Department of Physics, Indian Institute 
	of Science, Bangalore 560012, India}

\begin{document}
	
	\title{\titlename}
	\author{\authornames}
	\affiliation{\affiliations}
    \email{venkatakumar@iisc.ac.in}
	\email{surajit@iisc.ac.in}
	\email{sumilan@iisc.ac.in}
	\date\today

\begin{abstract}
We make lattice generalization of two well-known zero-dimensional models of quantum spin glass, Sachdev-Ye (SY) and spherical quantum $p$-spin glass model, to one dimension for studying crossovers in non-local scrambling dynamics due to glass transition, complex dynamics, and quantum and thermal fluctuations in paramagnetic (PM) and spin glass (SG) phases. In the SY chain of quantum dots, each described by infinite-range random Heisenberg model of $N$ spin-$S$ $SU(M)$ spins, we obtain the quantum Lyapunov exponent $\lambda_\mathrm{L}$ and butterfly velocity $v_B$ as a function of temperature $T$ and the quantum parameter $S$ across the PM-SG phase boundary using a bosonic spinon representation in the large $N,M$ limit. In particular, we extract asymptotic $T$ and $S$ dependence, e.g., power laws, for $\lambda_\mathrm{L}$ and $v_B$ in different regions deep inside the phases and near the replica symmetry breaking SG transition. We find the chaos to be non-maximal almost over the entire phase diagram. Very similar results for chaos indicators are found for the $p$-spin glass chain as a function of temperature and a suitable quantum parameter $\Gamma$, with some important qualitative differences. In particular, $\lambda_\mathrm{L}$ and $v_B$ exhibit a maximum, coinciding with onset of complex glassy relaxation, above the glass transition as a function of $T$ and $\Gamma$ in the PM phase of the $p$-spin glass model. In contrast, the maximum is only observed as a function of $S$, but not with temperature, in the PM phase of SY model. The maximum originates from enhanced chaos due to maximal complexity in the glassy landscape. Thus, the results in the SY model indicate very different evolution of glassy complexity with quantum and thermal fluctuations. 

\end{abstract}

\maketitle 
\section{Introduction}

In a generic quantum many-body system, quantum information encoded in local degrees of freedom in the initial state \emph{scrambles} to spread globally over the entire system due to its time evolution. The dynamics of quantum information, e.g., how rapidly local information scrambles, characterized by a rate or the quantum {\it Lyapunov exponent} $\lambda_\mathrm{L}$, and how fast the information propagates from one spatial region to other, quantified by a speed or the {\it butterfly velocity} $v_B$, can provide crucial insights into the non-trivial dynamics of plethora of quantum many-body systems \cite{Sekino2008,Maldacena2015,KitaevKITP,Sahu2019}. These insights are complementary to that from transport and usual dynamical correlations. As a result, information scrambling has become an important theme of research for understanding holographic duality\cite{Jensen2016}, black hole information scrambling\cite{Sekino2008,Maldacena2015}, non-Fermi liquids %\cite{Kitaev,Sachdev,BanerjeeAltman2016,Debanjan} 
\cite{KitaevKITP,Maldacena2016,Patel2017,BanerjeeAltman2016,Chowdhury2022}
and quantum thermalization 
%\cite{Sahu,Pollman,others,Srednicki,Chaitanya,..}. 
\cite{Sahu2019,Srednicki1994,Murthy2019}. For systems close to a semiclassical limit, information scrambling can often be connected to chaos \cite{Aleiner2016,Stanford2016,Rozenbaum2017,Scaffidi2019,Xu2020}, namely extreme sensitivity of the classical trajectories to initial conditions. 

For systems with many metastable configurations, information scrambling via quantum dynamics may become enhanced when the dynamics samples this multitude of configurations, leading to extreme sensitivity to initial conditions in the classical limit. Glasses \cite{Kob1997,Debenedetti2001,Berthier2011}, which possess extensive number of metastable states \cite{Debenedetti2001,Berthier2011,Buchner1999,Sastry2001,Sciortino2005}, are among the most prominent examples of such systems. At low temperature, the system remains trapped near one of the metastable configurations for a long time and explores the phase space extremely slowly. As a result, glasses exhibit many fascinating dynamical properties ranging from complex relaxation with anomalously large time scales \cite{Binder1986,Kob1997}, aging and memory effects \cite{Vincent2007}, unusual transport properties such as the violation of Stokes-Einstein relations \cite{Stillinger1993}, dynamical heterogeneity \cite{Berthier2011,Karmakar2020} etc. In particular, spin glasses \cite{Edwards1975,Binder1986,MezardBook1987} with quenched randomness undergo ergodicity breaking thermodynamic phase transition at low temperature, e.g., for mean-field or infinite-range models, to replica symmetry broken \cite{Parisi1979,Castellani_2005} spin glass phase. Recent work \cite{Surajit22} on a well-known solvable zero-dimensional or infinite-range model of quantum spin glass, namely the quantum $p$-spin glass model \cite{Nieuwenhuizen1995,Nieuwenhuizen1998,Cugliandolo2000}, have shown that chaos, as characterized by the Lyapunov exponent $\lambda_\mathrm{L}$, becomes stronger over a region above the glass transition where the dynamics exhibits typical signatures~\cite{Kob1997} of complex glassy relaxation. In this region of enhanced chaos, the dynamics samples exponentially large number of saddle points of the classical potential energy landscape and becomes {\it maximally complex}~\cite{Correale2023,Auffinger2011}, before eventually getting trapped within the basin of attraction of one of the glassy minima below the glass transition. In this work, we study scrambling dynamics in two solvable, albeit \emph{spatially extended models}, of quantum spin glasses and show how the measures of information scrambling, $v_B$ and $\lambda_\mathrm{L}$, are influenced by glass transition, complex dynamics, and quantum and thermal fluctuations in various paramagnetic and spin glass phases.

\begin{figure}[h!]
\centering
\includegraphics[width=1.0\linewidth]{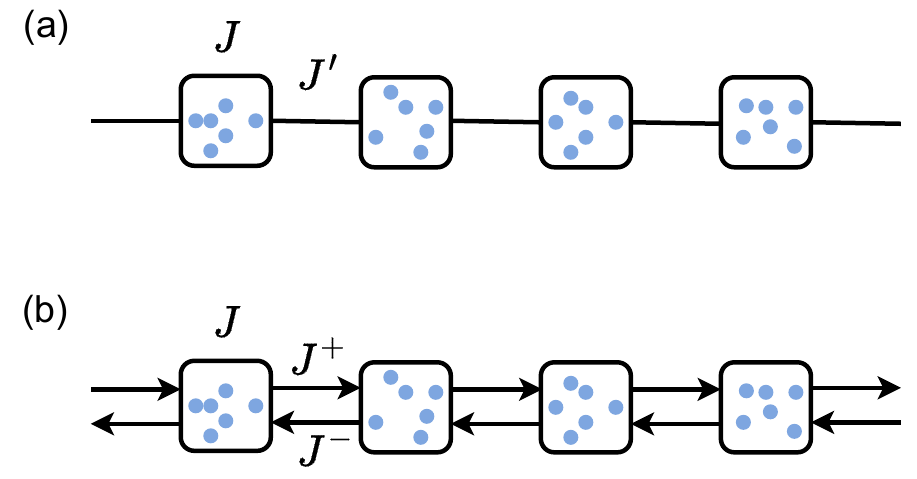}
\caption{An one-dimensional (1D) generalization of (a) SY model, (b) $p$-spin model for $p = 3$. Each lattice site contains a quantum dot described by zero-dimensional SY model in (a) and quantum spherical $p$-spin glass model in (b) with $N \gg 1$ spins (blue circles). In (a), the coupling between nearest neighbor sites are random two-spin interaction with one from each lattice site for the SY model. In the case of $p$-spin model in (b), the random nearest neighbor interaction is between two spins from one site and one from the other and vice versa.} \label{fig:lattice_model}
\end{figure}  

In line with the growing interests~\cite{Surajit22,Anous_2021,Correale2023} in understanding the relation between scrambling and the complex dynamics of spin glass (SG) systems, we consider two well-known large $N$ zero-dimensional ($0+1$ D) quantum spin glass models, namely (a) Sachdev-Ye (SY) model, a random Heisenberg spin glass \cite{Sachdev1993,Georges2000,Georges2001}, and (b) quantum spherical $p$-spin glass model \cite{Nieuwenhuizen1995,Nieuwenhuizen1998,Cugliandolo2000}, and make lattice generalizations of these models to one dimension ($1+1$ D), as shown schematically in Fig.\ref{fig:lattice_model}. These generalizations allow us to meaningfully define a butterfly velocity $v_B$, along with $\lambda_\mathrm{L}$, from an out-of-time-order correlation (OTOC) function \cite{KitaevKITP,Gu2017}. The SY model is parent to the celebrated fermionic Sachdev-Ye-Kitaev (SYK) model \cite{KitaevKITP,Maldacena2016,Sachdev2015}, which has emerged as a paradigm to study quantum many-body chaos, information scrambling \cite{Gu2017,BanerjeeAltman2016,Chowdhury2022}, as well as strongly correlated metals \cite{Song2017}. The SYK model is maximally chaotic, namely it saturates the remarkable upper bound \cite{Maldacena2015,KitaevKITP,Maldacena2016} $2\pi k_\mathrm{B}T/\hbar$ for $\lambda_\mathrm{L}$. There have been numerous lattice and higher dimensional generalizations \cite{Gu2017,Song2017,Davison2017,Jian2017,JianHong2017,Berkooz2017,Zhang2017,Chowdhury2018,Haldar2018} of the zero-dimensional SYK model, with a view of studying transport and information propagation in extended interacting systems. Our lattice generalizations of the spin glass models pave the way to access non-local dynamical correlations and spatio-temporal information scrambling in spatially extended strongly disordered systems that undergo ergodicity-breaking glass transitions.  

% \begin{figure}[h!]
% \begin{center}
%   \begin{subfigure}{\linewidth}
%     \includegraphics[width=1.0\linewidth]{SYchain.pdf}
%     %\caption{Lattice generalization of Sachdev-Ye model}
%   \end{subfigure}
%   \begin{subfigure}{\linewidth}
%     \includegraphics[width=1.0\linewidth]{pspin_chain.pdf}
%     % \caption{Lattice generalization of quantum $p$-spin model}
%   \end{subfigure}  
%   \end{center}
%   \caption{A 1D generalization of (a) SY model, (b) $p$-spin model for $p = 3$. Each site contains a dot model with $N \gg 1$ spins. The coupling between nearest neighbour sites are two spin interaction with one from each site for the SY model. In the case of $p$-spin model, the nearest neighbour interaction is between two spins from one site and one form the other. {SB: Put labels (a) and (b) for subfigures, no subcaptions are needed, remove.}}  
% \end{figure} \label{fig:lattice_model}

Earlier works on quantum chaos in spin glass systems\cite{Surajit22,Anous_2021,Correale2023} have only looked at the Lyapunov exponent, e.g., its dependence on temperature $T$ and quantum fluctuation parameter in the paramagnetic and spin glass phases, in the zero-dimensional quantum $p$-spin glass model, i.e., for a single \emph{$p$-spin glass quantum dot} (see Fig.\ref{fig:lattice_model}). On the other hand, studies on spatially extended large $N$ systems, such as lattice generalizations of the SYK model~\cite{Gu2017,Song2017,Davison2017,Jian2017,JianHong2017,Berkooz2017,Zhang2017,Chowdhury2018,Haldar2018}, have been limited to non (replica) symmetry broken phases, such as Fermi and non-Fermi liquids metals. Our lattice generalizations of the spin glass models retain the same large $N$ saddle points as the zero-dimensional models, and capture replica symmetry broken spin glass (SG) and unbroken paramagnetic (PM) phases, such as a spin liquid for SY quantum dot. Thus, the spatial locality in these extended models permits us to explore information scrambling across replica-symmetry breaking spin glass phase transition. In particular, we obtain the variations of the Lyapunov exponent and butterfly velocity with thermal and quantum fluctuations across the PM-SG phase transition and various crossovers between dynamically different paramagnetic states, e.g., spin liquid and local moment states in the SY model. We also find the asymptotic dependence of $\lambda_\mathrm{L}$ and $v_B$ on temperature $T$ and quantum/classical parameter in different regions deep inside the phases as well as near the replica symmetry breaking SG transitions.

Temperature dependence of chaos indicators, especially that of the butterfly velocity, has been studied in several earlier works in various contexts, for example, for solvable large $N$ models \cite{Gu2017,Guo2019,Chowdhury2017,Sahu2020}, perturbatively in interaction for weakly interacting systems \cite{Patel2017a} and in $1/N$ expansion in strongly interacting systems \cite{Patel2017}, via numerical methods \cite{Sahu2020} as well as in the semiclassical limit of spin models \cite{Bilitewski2018,Ruidas2021,Bilitewski2021}. For a maximally chaotic [$\lambda_\mathrm{L}=2\pi T$ ($\hbar,k_\mathrm{B}=1$)] interacting diffusive metal built from SYK dots~\cite{Gu2017}, $v_B$ has been found to follow a $\sqrt{T}$ dependence. Also, in this case, the quantity, $v_B^2/\lambda_\mathrm{L}$, which has the dimension of the diffusion constant, turns out to be exactly equal to energy diffusion constant, as expected from holographic theories \cite{Blake2016}.  Similar relation, $D=v_B^2/4\lambda_{L}$ has been found to be satisfied for electron diffusion constant $D$ in weakly interacting diffusive metal \cite{Swingle2017,Patel2017a}, where $\lambda_\mathrm{L}\sim T^{d/2},~v_B\sim T^{d/4}$ in $d$ dimension, and a temperature independent $D$ is determined by elastic scattering. A power-law $T$ dependence of $v_B$ is similar to that at a critical point \cite{Wei2019,Ruidas2021}, i.e., $v_B\sim T^{1-1/z}$ for a dynamical exponent $z>1$. In a strongly interacting system with critical Fermi surface of $N$-species fermions coupled with $U(1)$ gauge field, power-law behaviors, $\lambda_\mathrm{L}\sim T$, $v_B\sim T^{1/3}$, with energy diffusion constant $\sim v_B^2/\lambda_\mathrm{L}$, have been obtained~\cite{Patel2017} at leading order in $1/N$-expansion. The Lyapunov exponent and butterfly velocity also have been computed in (2+1 D) $O(N)$ non-linear sigma model, where, in the quantum critical region \cite{Chowdhury2017}, $\lambda_\mathrm{L}\sim T/N$, and $v_B\sim c$, a constant close to the speed of light in the theory. Similar $T$ independent $v_B$ is found in the $O(N)$-symmetry broken ordered phase~\cite{Chowdhury2017} with $\lambda_\mathrm{L}\sim T^3$, whereas $\lambda_\mathrm{L}\sim e^{-E_g/T}$, due to the gap $E_g$ in the disordered phase for $T\ll E_g$, and $v_B$ obeys a $\sqrt{T}$ behavior~\cite{Chowdhury2017,Sahu2020}. The latter is consistent with numerical calculations in 1D quantum Ising model~\cite{Sahu2020}. Power-law temperature dependence of $\lambda_\mathrm{L}$ and $v_B$ is also observed~\cite{Bilitewski2018,Ruidas2021} in the ordered and disordered phases of spin models with semiclassical dynamics.

\begin{figure}[h!]
\centering
\includegraphics[width=1.0\linewidth]{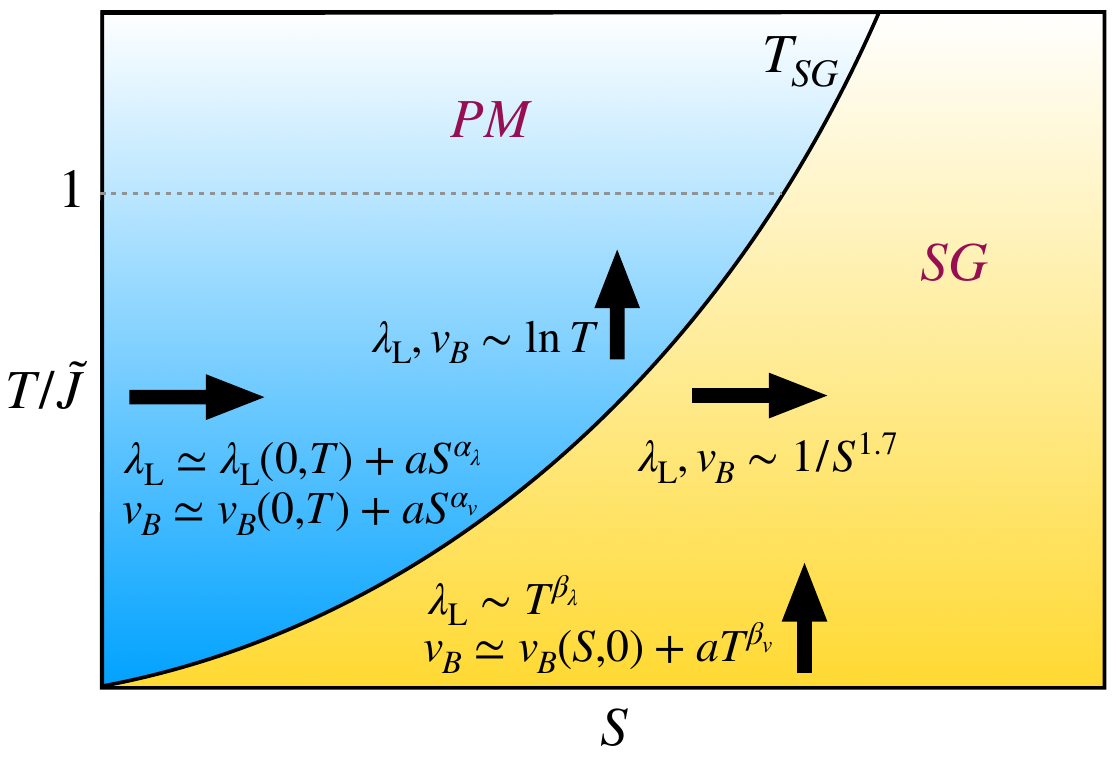}
\caption{{\bf Schematic phase diagram for chaos in SY model.} Dependence of Lyapunov exponent $\lambda_{\mathrm{L}}$ and butterfly velocity $v_B$ on temperature $T$ and quantum parameter $S$ are shown in the paramagnetic (PM) and spin glass (SG) phases. $\lambda_\mathrm{L}$ and $v_B$ increase as a power law for small $S$ in the PM phase, and as a power law in $T$ for low temperature in the SG phase. Both the chaos indicators increase slowly, as $\sim \ln \,{T}$, above the glass transition $T_{SG}$, before saturating to a constant at high temperature. {In the SG phase, both $\lambda_\mathrm{L}$ and $v_B$ decrease as $\sim 1/S^2$}. There is also maximum (not shown) in both $\lambda_\mathrm{L}$ and $v_B$ as a function of $S$ in the PM phase, at intermediate $S$, but there are no maxima as a function of $T$ (see main text). The region bounded by $T_{SG}(S)$ and $T/\tilde{J}=1$ ($\tilde{J}$ coupling constant) line (dotted line) demarcates the quantum critical region within the PM phase. Arrows indicate the direction of the increase of the parameter.} \label{fig:SYchaos_phases}
\end{figure}

\begin{figure}[h!]
\centering
\includegraphics[width=1.0\linewidth]{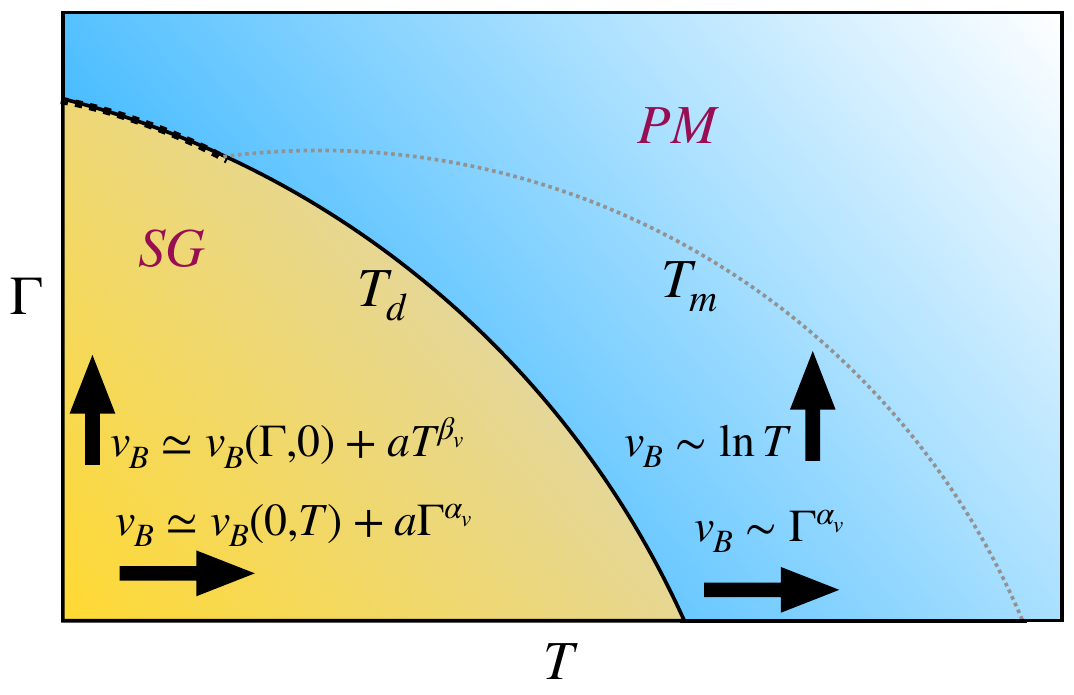}
\caption{{\bf Schematic phase diagram for chaos in $p$-spin glass model.} Dependence of $v_B$ on temperature $T$ and quantum parameter $\Gamma$ are shown in the PM and SG phases. Similar dependence has been seen for $\lambda_\mathrm{L}$ in Ref. \onlinecite{Surajit22}. The continuous PM-SG phase transition $T_d(\Gamma)$ [or $T_{SG}(\Gamma)$] (solid line) terminates at a tricritical point and thereafter the transition is first order (hatched line). There is a maximum in $v_B$ at $T_m(\Gamma)$ (dotted line), both as a function of $T$ and $\Gamma$. $v_B$ increases as power law with $T$ and $\Gamma$ at low temperature and small $\Gamma$, respectively, in the SG phase. The butterfly velocity increases as $\sim \ln \,{T}$ and as a power law in $\Gamma$ above the glass transition in the PM phase, before the maximum. Arrows indicate the direction of the increase of the parameter.} \label{fig:pspinchaos_phases}
\end{figure}  

Our results for the dependence of $\lambda_\mathrm{L}$ and $v_B$ on temperature $T$ and the quantum/classical parameter $S$ in the chain of SY quantum dots are summarized on the schematic phase diagram of Fig.\ref{fig:SYchaos_phases}. In this model, the dots are described by infinite-range random Heisenberg model of spin-$S$ $SU(M)$ spins on $N$ sites, where $N,M\to \infty$. Like in the lattice generalization~\cite{Gu2017} of the SYK model, the SY chain model, in a bosonic spinon representation, is described by the same saddle point of the zero-dimensional quantum dots with the strength of the coupling $J$ renormalized by the inter-dot interaction. As a result, $\lambda_\mathrm{L}$, which is extracted from a local OTOC, has the same $S$ and $T$ dependence of a single dot, or the original SY model~\cite{Sachdev1993}. On the contrary, the butterfly velocity can only be defined in a lattice model like the SY chain of Fig.\ref{fig:lattice_model}(a). To compare with the chaotic properties of the SY chain, we also calculate the butterfly velocity from the non-local OTOC as a function of $T$ and quantum parameter $\Gamma$ (defined later) in a similar lattice model [Fig.\ref{fig:lattice_model}(b)] of quantum spherical $p$-spin spin glass model~\cite{Nieuwenhuizen1995,Nieuwenhuizen1998,Cugliandolo2000}, as summarized in Fig.\ref{fig:pspinchaos_phases} (schematic). Again, $\lambda_\mathrm{L}(\Gamma,T)$ of the $p$-spin glass chain is identical to that of the corresponding zero-dimensional model, and was studied in Ref. \onlinecite{Surajit22}. 

Our results are summarized below.

(1) As shown in Fig.\ref{fig:SYchaos_phases}, we find a power-law increase with $S$ in the SY model, i.e., $\lambda_\mathrm{L}(S,T)\simeq \lambda_\mathrm{L}(0,T)+aS^{\alpha_\lambda}$ and $v_B(S,T)\simeq v_B(0,T)+a S^{\alpha_v}$ for small $S$ in the PM phase with {$\alpha_{\lambda (v)}\approx 0.5$} at a fixed temperature, e.g., in the quantum critical region for $T<\tilde{J}$. In this region, two types of PM behaviors, spin liquid and local moment, compete. In contrast to the power-law or gap-like temperature dependence seen in the disordered phase of many other models~\cite{Chowdhury2017,Sahu2020,Bilitewski2018,Ruidas2021}, a $\ln \,{T}$ variation of both $\lambda_\mathrm{L}$ and $v_B$ are seen in the PM phase of the SY model close to the transition temperature $T_{SG}$. Similar $\ln \,{T}$ dependence is seen for the PM phase of the $p$-spin glass model, where $v_B\sim \Gamma^{\alpha_v}$ with {$\alpha_v\approx 1.5$} [Fig.\ref{fig:pspinchaos_phases}]. 

(2) On the contrary, in the SG phase of SY model, power-law temperature dependence, $\lambda_\mathrm{L}(S,T)\sim T^{\beta_\lambda}$, and a power-law increment, $v_B(S,T)\simeq v_B(S,0)+aT^{\beta_v}$, with $\beta_\lambda\approx 1-1.5$ and {$\beta_v\approx 1.7$}, are observed. { $\lambda_\mathrm{L}(\Gamma,T)$ and $v_B(\Gamma,T)$ behave very similarly with temperature, in the SG phase of $p$-spin glass. In the SG phase of SY model at a fixed $T$, both the Lyapunov exponent and butterfly velocity exhibit similar trends with increasing $S$ or decreasing quantum fluctuations; $\lambda_\mathrm{S}, v_B$ decrease as $\sim 1/S^2$.} These behaviors are similar to the $p$-spin glass model where both $\lambda_\mathrm{L}$~\cite{Surajit22} and $v_B$ decrease with decreasing quantum fluctuation, e.g., $v_B(\Gamma,T)\simeq v_B(0,T)+a\Gamma^{\alpha_v}$ with {$\alpha_v\approx 0.7$}.

(3)  We show that the Lyapunov exponent for the spin liquid solution for small $S$ and low temperature extrapolates to a value much smaller than maximal $2\pi T$ for $S\to 0$, unlike in the SYK-type models \cite{BanerjeeAltman2016}. This indicates large $T$ and $S$ dependent corrections to the expected asymptotic $S\to 0, T\to 0$ spin liquid solution. 

(4) The Lyapunov exponent and butterfly velocity have smooth crossovers across the PM-SG transitions in both SY and $p$-spin glass models.

(5) As mentioned earlier, $\lambda_\mathrm{L}(\Gamma,T)$ exhibits a maximum~\cite{Surajit22} [$T_m(\Gamma)$ in Fig.\ref{fig:pspinchaos_phases}] in the PM phase of the $p$-spin glass model as function of both $\Gamma$ and $T$. The maximum corresponds to enhanced chaos due to the sampling of maximum complexity of the glassy landscape at $T_m$~\cite{Correale2023,Auffinger2011}. The same maximum is found in $v_B$ for $p$-spin glass. However, the maximum in $\lambda_\mathrm{L}$ and $v_B$ is seen in the PM phase of the SY model only as a function of $S$, but not as a function of $T$. This implies a very different evolution of complexity with quantum and thermal fluctuations in the SY model, unlike that in the $p$-spin glass model.

(6) For computing the butterfly velocity, we implement a numerical approach alternative to the more standard semi-analytical \emph{single-mode ansatz}~\cite{Gu:2018jsv,Guo2019}. The results from the numerical method agree very well with those from the single-mode ansatz over the entire phase diagrams of SY and $p$-spin glass chains, thus, independently verifying the applicability of the single-mode ansatz even for the replica symmetry broken spin glass phase. Using the single-mode ansatz, we find that chaos is non-maximal almost over the entire phase diagram for SY and $p$-spin glass models.

We also connect our results for the chaos parameters with spectral characteristics of various regions of the phase diagram of the two models, like spin liquid, local moment, classical and quantum paramagnets, and spin glass. {The dependence of $\lambda_\mathrm{L}$ and $v_B$ on temperature and quantum parameters, like $S$ and $\Gamma$ in our case, can be used to characterize and classify phases of many-body systems as different chaotic fixed points \cite{BanerjeeAltman2016}, in addition to usual characterizations of phases in terms of thermodynamic quantities and transport and other dynamical properties. The temperature dependence of these quantities deep inside a phase as well as close to a phase transition can diagnose the physical process that contributes to the relevant excitations of the system and relaxation mechanisms. In the same vein, the powerlaw temperature dependence of $\lambda_\mathrm{L}$ and $v_B$ on temperature and quantum parameters can detect the processes contributing to chaos and, hence, thermalization. As a result, a large number of recent works \cite{KitaevKITP,Maldacena2016,Patel2017,BanerjeeAltman2016,Bera2023,Anous_2021,Gu2017,Haldar2018,Chowdhury2017,Sahu2020,Patel2017a,Bilitewski2018,Ruidas2021,Bilitewski2021}, have studied temperature dependence of $\lambda_\mathrm{L}$ and $v_B$ in various symmetry broken and unbroken phases and across phase transitions, as we have summarized earlier. The underlying mechanisms giving rise to the power-law temperature dependence in some of the cases, but not for all, are understood. For example, in a Fermi liquid $\lambda_\mathrm{L}\sim T^2$ can be understood from the quasiparticle life time \cite{BanerjeeAltman2016,Kim2020}, whereas the maximal chaos in the SYK model leading to $\lambda_L=2\pi T$ can be connected to the low-energy Schwarzian or \emph{scramblon} mode \cite{KitaevKITP,Maldacena2016,Chowdhury2022}. In our case, the power-law temperature dependence of $\lambda_\mathrm{L}$ and $v_B$ in the spin glass phase at low temperature is due to the gapless nature of the marginal spin glass phase \cite{Bera2023}. However, we do not have any clear understanding yet about the power-law dependence of $\lambda_\mathrm{L}$ and $v_B$ on the quantum parameter $S$ ($\Gamma$). Moreover, as discussed earlier, the exponents for the $T$ dependence of $\lambda_\mathrm{L}$ and $v_B$ in some cases can be interrelated via their relation with the diffusion constant $D\sim v_B^2/\lambda_\mathrm{L}$, that connects chaos with transport. However, in this work, we have not computed the diffusion constant. Thus, we keep the study of the relations between various exponents for future works.} 
% {The asymptotic dependence of $\lambda_\mathrm{L}$ and $v_B$ on temperature $T$ typically provides information about the underlying phases such as disordered or symmetry broken phases~\cite{Chowdhury2017,Bilitewski2018,Ruidas2021} and in some cases the information about the processes contributing to the chaos, such the scramblon mode\cite{Kitaev:2017awl} in case of large $N$ SYK type models. However, we don't know how to interpret various power laws and other asymptotic behavior and if there are any relations between them.} 

The rest of the paper is organized as follows. In Sec.\ref{II}, we discuss the one dimensional lattice generalizations of the zero-dimensional SY and quantum $p$-spin glass model. The large $N$ saddle-point equations for the PM and SG phases for both the models are described in Sec.\ref{sec:PhaseDiagram}, followed by review of the phase diagrams known from earlier works. Sec.\ref{sec:LyapunovExponent} describes the formalism for calculating local OTOC and the Lyapunov exponent using ladder approximation for the PM and SG phases for the SY model. This section reports the results for $\lambda_\mathrm{L}(S,T)$ in the SY model and compares them with those in the $p$-spin glass model. The formalism for computing the butterfly velocity from non-local OTOC using two different methods, a numerical method and the semi-analytical single-mode ansatz~\cite{Gu:2018jsv}, is discussed in Sec.\ref{sec:ButterflyVelocity} for both SY and $p$-spin glass models. The results for $v_B$ as a function of $T$ and quantum parameter for the two models are described in this section. The appendices (\ref{appA}, \ref{appB}) give additional details of the derivations of the saddle-point equations and methods for their numerical self-consistent solutions. \ref{app:SpectralFunction} discusses the results for the spectral properties of the SY model. \ref{app:vB} provides details of the numerical computation of $v_B$.

\section{Lattice generalizations of solvable quantum spin glass models}\label{II}

As mentioned earlier, we consider one-dimensional (1D) chains of zero-dimensional random quantum Heisenberg and spherical $p$-spin glass quantum dots, as shown in Fig.\ref{fig:lattice_model} and discussed below.

\subsection{1D chain of Sachdev-Ye (SY) spin-glass quantum dots}
We make a lattice generalization of the well-known zero-dimensional SY model of random quantum Heisenberg model by arranging $x=1,\dots,L$ quantum dots of SY model, where each dot contains $i=1,\dots,N$ sites. The model is described by the Hamiltonian
\begin{align}\label{eq:SY_1D}
\mathcal{H} = \frac{1}{2\sqrt{MN}}\sum_{x,i\neq j}  J_{ij,x} S_{i\alpha \beta, x} S_{j \beta \alpha, x} \nonumber \\
+ \frac{1}{\sqrt{MN}}\sum_{x,i,j} J^{\prime}_{ij,x} S_{i\alpha \beta, x} S_{j\beta \alpha,x+1}.
\end{align}
Here the first term represents all-to-all infinite-range interactions within a dot, and the second term the nearest-neighbour inter-dot coupling. The couplings $J_{ij,x}$ and $J^{\prime}_{ij,x}$ are drawn from Gaussian distribution, independently at each site with zero mean and variances,
\begin{align}
    {\overline{J_{ij,x}^2} = J^2, \quad \overline{J_{ij,x}^{\prime 2}} = J^{\prime 2}.}
\end{align}

$S_{i\alpha \beta, x}$ denote the spin degrees of freedom, the generators of the SU$(M)$ group for spin $S$, at each $i$ and dot $x$, where $\alpha,\beta=1,\dots,M$. The model is exactly solvable in the large $N$ limit ($N\to\infty$) followed by the large $M$ limit ($M\to\infty$), i.e., thermodynamics, as well as equilibrium and non-equilibrium properties can be exactly obtained in this limit through large $N$ saddle-point equations, as known from previous works \cite{Sachdev1993,Georges2000,Georges2001,Biroli2002} on the zero-dimensional model. The effect of quantum fluctuations in the model can be tuned by varying the spin $S$ and the classical and semiclassical regimes can be probed by taking the large $S$ ($S\to\infty$) limit while appropriately scaling the energy/temperature scales \cite{Georges2001}. The $M=2$ limit of the model is the usual $SU(2)$ Heisenberg model, which has been studied via quantum Monte Carlo (QMC) \cite{Grempel1998} after taking the $N\to\infty$ limit. The large $M$ limit was found to capture the properties of the $SU(2)$ model quite well. 

In Sec.\ref{sec:SY_PhaseDiagram}, we discuss the equilibrium phase diagram obtained from the large $N$ saddle-point equations as a function of temperature $T$ and spin $S$. The large $N$ saddle point is obtained via the standard bosonic representation of the $SU(M)$ spin \cite{Sachdev1993}. Subsequently, our main results on the dynamical phase diagram for chaos, characterized via the Lyapunov exponent $\lambda_\mr{L}$ and butterfly velocity $v_B$, are discussed in Secs. \ref{sec:LyapunovExponent} and \ref{sec:ButterflyVelocity}.

\subsection{1D chain of spherical $p$-spin glass quantum dots}
To compare and contrast the results for $\lambda_\mr{L}$ and $v_B$ in the SY chain, we consider the lattice generalization of another paradigmatic model of quantum glasses, the quantum spherical $p$-spin glass model \cite{Nieuwenhuizen1995,Nieuwenhuizen1998,Cugliandolo2000,Cugliandolo2001}. The 1D chain of the $p$-spin glass quantum dots ($x=1,\dots,L$) is described by the Hamiltonian 
\begin{align} \label{eq:pSpin_1d}
&\mathcal{H} = \sum_{x,i}\frac{\pi_{i,x}^2}{2\mathcal{M}} + \frac{1}{3!}\sum_{x,ijk}J_{ijk,x} s_{i,x} s_{j,x} s_{k,x} \nonumber \\
&+ \frac{1}{2!}\sum_{x,ijk}s_{i,x}(J^+_{ijk,x} s_{j,x+1} s_{k,x+1}+J^-_{ijk,x} s_{j,x-1} s_{k,x-1}),
% &+ \frac{1}{2!}\sum_{x=0}^{L-1}\sum_{ijk} J^{\prime}_{1 ijk,x}s_{i,x+1} s_{j,x} s_{k,x}
\end{align}
for $p=3$. The intra- and inter-dot all-to-all couplings among sites $i=1,\dots,N$ in the dots, $J_{ijk,x}$ and $J_{ijk,x}^\pm$, are Gaussian random numbers with zero mean and variances, 
\begin{align}
   { \overline{J_{ijk,x}^2} = 3J^2/(2N^2), \quad \overline{J^{\pm 2}_{ijk,x}} = 3J'^2/(2N^2).}
\end{align}
The commutation relation of the spin $s_{i,x}$ with the momentum $\pi_{j,x}$, $[s_{i,x}, \pi_{j,x'} ]= \ci \hbar \delta_{ij}\delta_{xx'}$ induces the quantum dynamics. The spherical constraint $\sum_i s^2_{i,x} = N$ makes the model nontrivial. Essentially, when the quantum dots are uncoupled ($J'=0$), the Hamiltonian describes quantum particles
with mass $\mathcal{M}$ moving on the surface of an $N$-dimensional
hypersphere. The inter-dot interaction couples particles in neighboring dots. The advantage of this model lies in the fact that one can take a classical limit of the model by continuously tuning a quantum fluctuation parameter $\Gamma=\hbar^2/\mathcal{M}J$ to zero \cite{Surajit22,Anous_2021}. The classical model for $\hbar\to 0$ limit (fixed $\mathcal{M}$) for a single dot gives rise to dynamics identical to the mode coupling theory (MCT) dynamics in the super-cooled liquid regime of structural glasses \cite{GotzeBook, Kob1997,Reichman2005,Berthier2011,Surajit22}. The limit $\mathcal{M}\to \infty$ while keeping $\hbar$ fixed leads to a different classical limit of infinitely heavy mass \cite{Surajit22,Anous_2021}. In this work, we vary the dimensionless quantum parameter $\Gamma$.

In Sec.\ref{sec:pSpin_PhaseDiagram}, we discuss the equilibrium phase diagram of the model as a function of $T$ and $\Gamma$. The results for $\lambda_\mathrm{L}$ and $v_B$ are discussed in Secs. \ref{sec:LyapunovExponent} and \ref{sec:ButterflyVelocity} and compared with those from the lattice SY model.

% At each site, couplings are randomly drawn from Gaussian distribution $P(J_{ijk,x}) = \mathrm{exp}\Big( -\frac{2 N^2}{3!}\frac{J_{ijk,x}^2}{2 J^2}\Big)$ and similarly nearest neighbour couplings are drawn from Gaussian distribution $P(J_{1ijk,x}) = \mathrm{exp}\Big(-\frac{2 N^2}{2!}\frac{J_{1ijk,x}^2}{2 J_1^2}\Big)$ and $P(J_{1ijk,x}^\prime ) = \mathrm{exp}\Big(- \frac{2 N^2}{2!}\frac{J_{1ijk,x}^{\prime 2}}{2 J_1^2}\Big)$.

% This is a generalization of the SU$(2)$ Heisenberg model\cite{Grempel1998}, for which it's hard to obtain the equilibrium dynamics without resorting to quantum Monte Carlo methods.
% However the large $N$ limit, followed by large $M$ limit enables us to solve the problem under mean-field approximation, which captures the physics of the $M=2$ mean-field model\cite{Subir2001, Camjayi2003}.

% In the large $N$ limit followed by large $M$ limit, this model retains the same saddle point equations as the SY dot model, as shown in \ref{appD}, with only the coupling constant being scaled as $ J^2 + J^{\prime 2} \rightarrow J^2$.

% In this section, we introduce the zero-dimensional models of the two glassy systems of interest, discuss the saddle point equations and the resulting phase diagrams. This will set up the notation for the rest of the paper.
%in a random Heisenberg model \cite{Subir2001, Camjayi2003,Biroli2002},  also known as Sachdev-Ye model(SY) model and $p$-spin[] glass model. 

\section{Equilibrium phase diagrams of the spin glass chains} \label{sec:PhaseDiagram}
\begin{figure}[h!]
\centering
\includegraphics[width=1.2\linewidth]{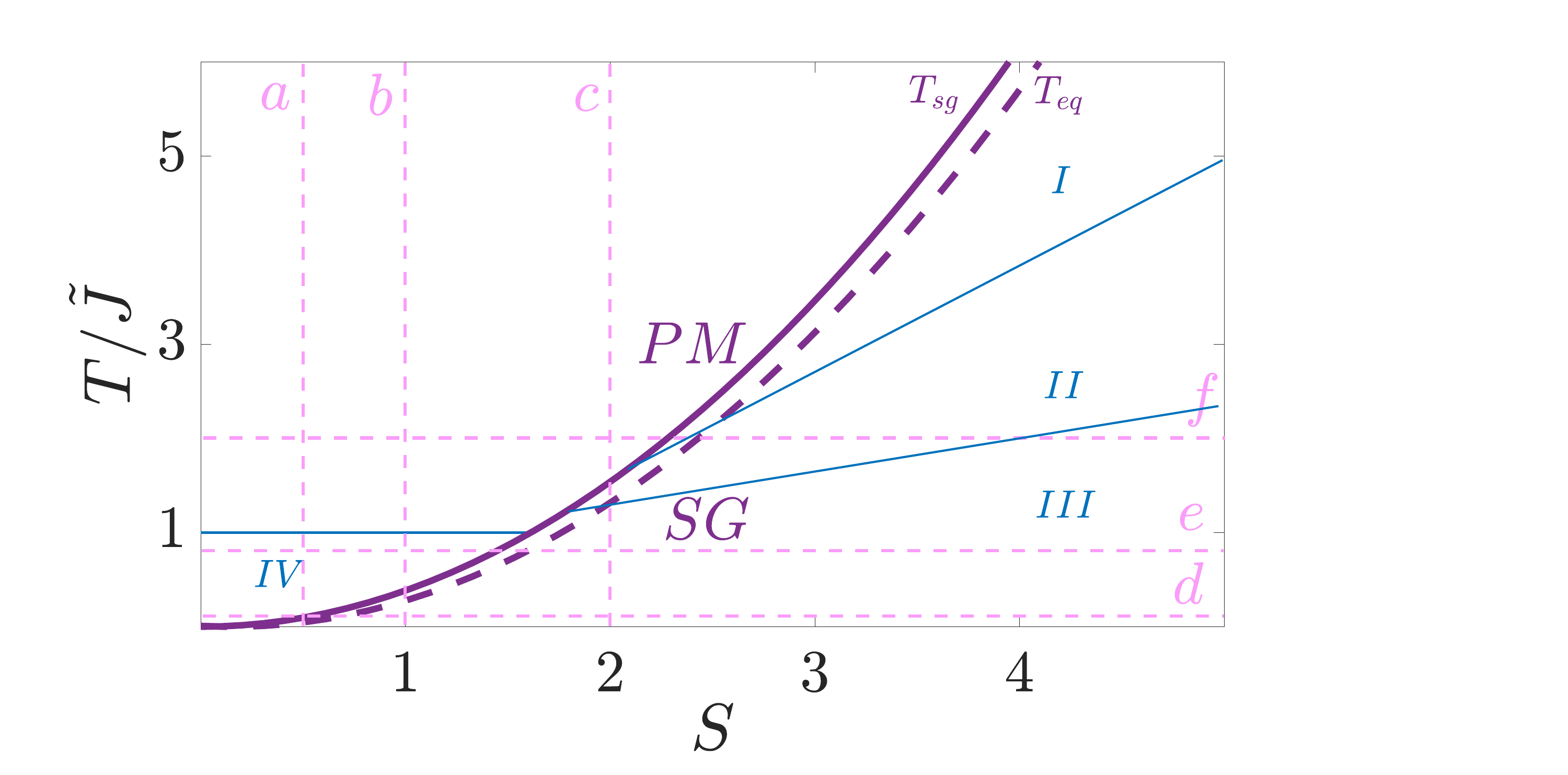}
\caption{Mean-field phase diagram of the bosonic spinon model \cite{Georges2001} showing the paramagnetic and spin glass phases. Spin glass phase below $T_{SG}$ (solid line) is determined by the \emph{marginality} criterion (see main text). $T_{eq}$ (thick dashed line) is determined by the equilibrium criterion. The regimes within spin glass phase separated by blue lines are identified via a large $S$ expansion of saddle point equations \cite{Georges2001} $I$, $II$ and $III$ are classical, semi-classical and quantum spin glass regions respectively. $IV$ is the quantum critical regime in PM phase. We perform the calculations of Lyapunov exponent and butterfly velocity across the SG-PM transition for several $S$ as a function of $T$ (along vertical dashed lines $a$, $b$, and $c$) and for several $T$ as a function of $S$ (along horizontal dashed lines $d$, $e$ and $f$).}\label{fig:phasediagramSY}
\end{figure}

We first discuss the equilibrium phase diagram of the quantum spin glass chains [Eq.\eqref{eq:SY_1D} and Eq.\eqref{eq:pSpin_1d}] as a function of $T$ and quantum fluctuation parameters, $S$ or $\Gamma$ [Figs.\ref{fig:phasediagramSY}, \ref{fig:pSpin_PhaseDiagram}]. The equilibrium phase diagrams are obtained by solving large $N$ saddle-point equations. As we discuss below, due to the particular choice of inter-dot couplings in Eqs.\eqref{eq:SY_1D},\eqref{eq:pSpin_1d}, the large $N$ saddle-point equations of the lattice models are identical to the zero-dimensional saddle-point equations \cite{Sachdev1993,Cugliandolo2001} of the single dot with renormalized intra-dot coupling. 

\subsection{Sachdev-Ye chain} \label{sec:SY_PhaseDiagram}
To capture both (replica) symmetry broken ordered SG and disordered paramagnetic (PM) phases we use bosonic spinons to represent spin-$S$ SU$(M)$ generators as $S_{i\alpha\beta, x}=b_{i\alpha, x}^\dagger b_{i\beta, x}-S\delta_{\alpha\beta}$, where $b_{i \alpha, x}$, $b_{i \alpha, x}^\dagger$ ($\alpha=1,\dots,M$) are bosonic annihilation and creation operators with the constraint $\sum_\alpha b_{i \alpha, x}^\dagger b_{i \alpha, x}=SM$ ($S>0$) on the number of bosons at each $i,x$ to fix the spin of the representation to $S$. The saddle-point equations, their solutions and resulting phase diagram for the bosonic representation have been discussed in detail for the zero-dimensional model in the pioneering works \cite{Sachdev1993,Georges2000, Georges2001}. We briefly discuss the derivation of saddle point equations for the SY chain model [Eq.\eqref{eq:SY_1D}] in \ref{appA}. 
% {\textcolor{magenta}{and for the zero dimenstional model in the rest of this section}. 

The disorder averaged Green's function $G_x^{ab}(\tau, \tau^\prime)=G^{ab}(\tau, \tau^\prime)= -\frac{1}{M} \sum_{\alpha} \overline{\langle \mathcal{T}_\tau b_{i \alpha, x}^a(\tau) \bar{b}_{i \alpha, x}^{b}(\tau^\prime)\rangle}$ is independent of $i,x$ and determines the equilibrium dynamics of the model. The Green's function is obtained from the saddle point equations in the limit $N\rightarrow \infty$ followed by $M \rightarrow \infty$ as
\begin{subequations} \label{eq:SY_SaddleEq}
    \begin{align}
(\Ginv)^{ab}(\ci\w_k) &= \left(\ci \w_k + \lambda \right)\delta_{ab} - \Sigma^{ab}(\ci\w_k)\\
\Sigma^{ab}(\tau) &= \tilde{J}^2[G^{ab}(\tau)]^2 G^{ba}(-\tau)
\end{align}
\end{subequations}
where $\w_k = 2\pi k T$ is bosonic Matsubara frequency with $k$ an integer ($k_B=1$), $T$ the temperature in units of $\tilde{J}$. We have introduced the replicas $a = 1,\dots,n$ to carry out the disorder average.  $G(\ci\w_k) = \int_0^\beta d\tau e^{\ci \w_k \tau} G(\tau)$ and $\beta$ is the inverse temperature. Lagrange multiplier, $\lambda$ is introduced to impose the constraint, $G^{aa}(\tau = 0^-) = -S$. The above saddle-point equations are same as the zero-dimensional ones \cite{Sachdev1993,Georges2000} with renormalized coupling,
\begin{align}
   { \tilde{J}^2=J^2+2J'^2 \label{Eq:Jtilde}}
\end{align}
 due to the nearest-neighbor inter-dot coupling in Eq.\eqref{eq:SY_1D}. In the limit $n\rightarrow 0$, the replica structure of Green's function determines the PM and SG phases as we discuss below.

\subsubsection{Paramagnetic phase}

In the PM phase one employs a replica diagonal and symmetric ansatz, $G^{ab}(\tau) = G(\tau)\delta_{ab}$. Thus, the saddle-point equations reduce to,
\begin{subequations}\label{Eq:SaddlePM}
\begin{align}
[G(\ci \omega_k )]^{-1} &= (\ci \omega_k + \lambda) - \Sigma(\ci \omega_k)\\ 
\Sigma(\tau) &= \tilde{J}^2 [G(\tau)]^2 G(-\tau) \label{Eq:Sigma_tau}
\end{align}
\end{subequations}
Analytic continuation from Matsubara to real frequency $\ci \w_k\rightarrow \w + \ci 0^+$ gives the saddle point equation for retarded Green's function
\begin{align}\label{Eq:SaddlePM_real}
[G_R(\w)]^{-1} = \w + \lambda - \Sigma_R(\w) 
\end{align}
where $\Sigma_R(\w) = \Sigma(\ci\w_k\to \omega+\ci 0^+)$ and $\Sigma(\ci \w_k) = \int_0^\beta d \tau e^{\ci \w_k \tau} \Sigma(\tau)$. $\lambda$ is determined by the constraint on $ G(\tau=0^-)$. The procedure to numerically solve for the retarded Green's function is discussed in \ref{appA}.

\subsubsection{Spin glass phase} \label{sec:SG_SY}

Following earlier works \cite{Georges2001}, we describe the spin glass phase via the following one-step replica symmetry breaking (1RSB) ansatz for the bosonic Green's function that appears in Eq.\eqref{eq:SY_SaddleEq}.
\begin{align}\label{Eq:1rsb}
G^{ab}(\tau)&=\tilde{G}(\tau)\delta_{ab}-g\epsilon_{ab}
\end{align}
or, $G^{ab}(\ci\omega_k) = \tilde{G}(\ci\omega_k) \delta_{ab} - \beta g \epsilon_{ab}\delta_{\omega_k,0}$, and 
\begin{equation}
\Sigma^{ab}(\tau) =  \tilde{\Sigma}(\tau)\delta_{ab} -\tilde{J}^2 g^3\epsilon_{ab},
\end{equation}
where $\epsilon_{ab}=1$ in the diagonal block of size $m\times m$, and 0 otherwise, for the $n\times n$ matrices. Here $\tilde{G}(\tau)$ is the regular part of the Green's function such that $\tilde{G}(\tau\to\infty)\to 0$ at $T=0$. 
%In the limit $n\rightarrow 0$ this matrix becomes a function $g(u)$ of a continuous variable $u$ with $0 \leq u \leq 1$. 
{Working in the $n\rightarrow 0$ limit}, the Edward-Anderson (EA) spin-glass order parameter is obtained from $q_\mathrm{EA}=\lim_{\tau\to\infty}(1/M^2)\sum_{\alpha\beta}\langle S_{i\alpha\beta,  x}(\tau)S_{i\beta\alpha, x}(0)\rangle$$=\lim_{\tau\to\infty}G^{aa}(\tau)G^{aa}(-\tau)=g^2$ at zero temperature.
% The Green's function with SG order can be written as \cite{Georges},
% \begin{equation}
% G^{ab}(\tau) = \begin{cases} 
% 			 \tilde{G}(\tau) - g_1 & (a=b), \\
% 			-g_{ab} & (a\neq b)
% 		  \end{cases}
% \end{equation}
% where $g_1$ is a constant and $g_{ab}$ a constant matrix. In the limit, $n \rightarrow 0$, the matrix becomes a function $g(u)$ of continuous variable $u$ with $0\leq u \leq 1$. In this limit, the widely used order parameter in spin glass literature, $q_{ab} = g^2_{ab}$ becomes $q(u) = g^2(u)$. The Edwards-Anderson parameter is $q_{EA} = q(1) = g^2(1)$. In the long time limit, $\mathrm{lim}_{\tau\rightarrow \infty}G^{aa}(\tau)G^{aa}(-\tau) = q_{EA}$ at $T=0$, and thus $g_1 = g(1)$. For one-step replica symmetry (1RSB) breaking \cite{Georges}, 
% \begin{equation}\label{Eq:1rsb} 
% G^{ab}(\tau) =   \tilde{G}(\tau)\delta_{ab} - g\epsilon_{ab}, 
% \end{equation}
% or, $G^{ab}(i\omega_k) = \tilde{G}(i\omega_k) \delta_{ab} - \beta g \epsilon_{ab}$,
% and
% Thus, from Eq.\eqref{eq:SY_SaddleEq}
% \begin{equation}
% \Sigma^{ab}(\tau) =  \tilde{\Sigma}(\tau)\delta_{ab} -J^2 g^3\epsilon_{ab}
% \end{equation}
% where $\epsilon_{ab}=1$ in a diagonal block of size $m$ and $0$ otherwise, $\tilde{G}$ and $\tilde{\Sigma}$  are the regular parts of the green's function and self energy. 
Using $[G^{-1}]^{ab}(\ci \omega_k)  =  A(\ci\omega_k)\delta_{ab} + B(\ci \omega_k)\epsilon_{ab} $ and the standard replica matrix inversion { in $n\rightarrow 0$ limit}, we get
\begin{subequations}
\begin{align}
A(\ci \omega_k)&= \frac{1}{\tilde{G}(\ci \omega_k)}, \\
B(\ci \omega_k)&= \frac{\beta g}{[\tilde{G}(\ci \omega_k) -m\beta g] \tilde{G}(\ci \omega_k)} \label{eq:Biwk}.
\end{align}
\end{subequations}
The saddle point equation for $\tilde{G}$ is therefore,
\begin{subequations}
\begin{align}
[\tilde{G}(\ci \omega_k)]^{-1} = \ci\omega_k -\frac{\tilde{J} g }{\Theta} -[\tilde{\Sigma}(\ci \omega_k) -\tilde{\Sigma}(\ci \omega_k=0)] \label{Eq:SaddleSG_imag}
\end{align}
where 
\begin{align}\label{Eq:SG_Sigmatau}
 \tilde{\Sigma}(\tau) = & \tilde{J}^2\Big[\tilde{G}^2(\tau) \tilde{G}(-\tau) -2g \tilde{G}(\tau) \tilde{G}(-\tau)-g \tilde{G}^2(\tau) \nonumber \\
 &+2 g^2 \tilde{G}(\tau) + g^2 \tilde{G}(-\tau)\Big]
\end{align}
\end{subequations}
In the above, we have eliminated $\lambda $ using the equation for $\ci\omega_k =0$ and the parameterization $\tilde{G}(\ci \omega_k=0) = -\Theta/\tilde{J} g$. 
% \begin{subequations}
%      \begin{align}
%  [\tilde{G}(i \omega_k)]^{-1} &= i\omega_k + \lambda -\tilde{\Sigma}(i \omega_k), \\
%   \tilde{\Sigma}(\tau) &= J^2[\tilde{G}(\tau) -g]^2[\tilde{G}(-\tau) -g] + J^2 g^3
%  \end{align}
% \end{subequations}
The saddle point equation for $g$ is obtained from Eq.\eqref{eq:Biwk} as, 
% \begin{align}
% -\frac{\beta g}{(G(i\omega_k =0)-m \beta g)G(i \omega_k =0)} =  J^2\int_0^\beta d\tau( -g)^2(-g) 
% %= -J^2 g^3 \beta \implies \frac{1}{J^2 g^2 G(i \omega_k =0)} = G(i\omega_k =0) - x g \beta 
% \end{align}
% which on simplifying gives
\begin{equation}\label{Eq:breakpoint}
 m\beta = \frac{1}{\tilde{J}g^2}\Big(\frac{1}{\Theta}- \Theta \Big)
\end{equation}

The equations Eq.\ref{Eq:SaddleSG_imag}, Eq.\ref{Eq:SG_Sigmatau} and Eq.\ref{Eq:breakpoint} along with the constraint, $\tilde{G}(\tau= 0^{-}) = g - S$ provide a closed set of self-consistent saddle-point equations for the SG phase. 
The characteristic of 1RSB solution is that the saddle point equations form a one-parameter family, parameterized here by $\Theta$ or equivalently by the breaking point $m$ where { $m \in (0,1]$ in the $n\to 0$ limit. To obtain this parameter we impose the usual marginality criterion \cite{Georges2001}, where  we study the fluctuations of the free energy.} 
{
\begin{align}
    \mathbb{F}[G^{ab}, \lambda] =& \frac{1}{\lambda}\sum_{k}\mathrm{Tr}\ln \,[\ci \w_k +\lambda -\Sigma^{ab}(\ci \w_k)] \\ \nonumber
&+    \frac{3\tilde{J}^2}{4}\sum_{ab}\int_0^\beta d \tau [G^{ab}(\tau) G^{ab}(-\tau)]^2 -\lambda S 
\end{align}
}
{in the replica space around the one-step solution and impose the condition that lowest eigenvalue of the fluctuation matrix in the replica space must vanish. The variations in free energy due to small variation $g^{ab}$ in the saddle point Green's function $G^{ab}$ for $a\neq b$, at second order, takes a generic form,}

{ 
\begin{align}
   \delta \mathbb{F} = \sum_{a> b, c> d} M_{ab, cd}\delta g^{ab} \delta g^{cd} 
\end{align}
}
{ Although, in principle one has to consider the variation in diagonal component, $\tilde{G}(\tau)$ and allow for the fluctuations due to the coupling between $\tilde{G}(\tau)$ and $g^{ab}$ in the spin glass phase, it can be shown that these variations leave the eigenvalue $e_1$  intact\cite{Georges2001}. The diagonalization of the $n(n-1)/2 \times n(n-1)/2$  matrix $M$ leads to the following three eigenvalues in the $n\to 0$ limit,}

\begin{subequations}
{
\begin{align}
          e_1 =& 3\beta \tilde{J}^2 g^2(1 - 3 \Theta^2) \\
    e_2 =& \frac{3 \beta \tilde{J}^2 g^2}{\Theta^2}[\Theta ^2 - 3 + 3\beta \tilde{J} g^2\Theta (1+\Theta)] \\
    e_3 =& 6\beta \tilde{J}^2 g^2(3 \beta \tilde{J}g^2\Theta -1 ) 
\end{align}}
\end{subequations}

 {
Setting $e_1=0$ gives,
\begin{align}
   \Theta = \Theta_R = \frac{1}{\sqrt{3}}
\end{align}
}
Coincidentally, the marginality criterion leads to a gapless $\chi^{\prime\prime}(\w) = (1/\pi) \mathrm{Im}\,  \chi(\w)\propto \omega$ with a delta function peak at $\omega=0$, where $\chi^{aa}(\tau) = G^{aa}(\tau)G^{aa}(-\tau)$$=(1/M^2)\sum_{\alpha\beta}\langle S_{i\alpha\beta, x}(\tau)S_{i\beta\alpha, x}(0)\rangle$ is the local spin susceptibility~\cite{Georges2001}. This is unlike the equilibrium criterion, which minimizes free energy with respect to the breaking point $m$ that leads to a gapped spectrum \cite{Georges2001}. In  \ref{app:SpectralFunction} we calculate $\chi^{\prime \prime}(\w)$ and compare with the results from Ref. \onlinecite{Camjayi2003}. Analytic continuation of Eq.\eqref{Eq:SaddleSG_imag} gives the equation for retarded Green's function $G_R^{ab}(\omega)=G_R(\omega)\delta_{ab}$ with
\begin{align} 
[G_R(\omega)]^{-1} = \omega -\frac{\tilde{J} g }{\Theta_R} -[\Sigma_R(\omega) -\Sigma_R(\omega=0)]
\end{align}\label{Eq:SaddleSG_real}
where $\Sigma_R(\omega) = \tilde{\Sigma}(\ci\w_k\to\omega+\ci 0^+)$ and $\tilde{\Sigma}(\ci\w_k) = \int_0^\beta d \tau e^{\ci \w_k \tau }\tilde{\Sigma}(\tau)$. $\lambda$ is fixed by the constraint $\tilde{G}(\tau = 0^-) =  g -S$, $g$ is calculated from the relation Eq.\ref{Eq:breakpoint}. The procedure for numerically calculating the retarded Green's function $G_R(\omega)$ is discussed in \ref{appA}.
For a fixed $S$, the temperature $T_{SG}$, below which the spins freeze into a SG order is identified by $m(T_{SG})=1$, where the SG order parameter $q_\mathrm{EA}$ jumps discontinuously to zero in the PM phase \cite{Georges2001}. A similar procedure is followed to obtain $S_{SG}$ for a given temperature $T$. Unlike earlier works \cite{Georges2001,Camjayi2003} , in this work, we mostly use the real-frequency saddle-point solutions to find the Lagrange multiplier $\lambda$, EA order parameter $q_\mathrm{EA}$ in the SG phase, the PM-SG phase boundary, as well as the retarded Green's function $G_R(\omega)$, and chaos diagnostics $\lambda_\mathrm{L}$ and $v_\mathrm{B}$. The numerical methods employed for obtaining the real-frequency saddle-point solutions are discussed in \ref{sec:RealFreqNumerics}. The imaginary-time saddle-point equations are only used for deriving the real-frequency saddle-point equations via analytical continuation, and occasionally, to check the consistency of the thermodynamic quantities like $\lambda,~q_{EA}$ and $T_{SG}(S)$ computed from the real-frequency solution. For the sake of completeness, we discuss the numerical method used for the imaginary-time solution in \ref{sec:ImagTimeNumerics}. The same procedure is followed for the $p$-spin glass model.

\subsubsection{Phase transition and crossovers} \label{sec:SYcrossover}
The equilibrium phase diagram of the SY chain model is identical to the zero-dimensional case, which has been studied in detail in the $N,M\to\infty$ limit in both bosonic and fermionic representation of the $SU(M)$ spins for general $S$ \cite{Sachdev1993,Georges2000,Georges2001}. The phase diagram has been also studied for finite $M$ in the $N\to \infty$ limit, using $1/M$ expansion \cite{Christos2022} and for small $S$, e.g., $S=1/2$, through QMC \cite{Grempel1998} in the $N\to\infty$ limit, as well as via exact diagonalization for finite $N$ \cite{Arrachea2002}. In Fig.\ref{fig:phasediagramSY}, we reproduce the schematic large $N,M$ phase diagram obtained via bosonic representation in Ref. \onlinecite{Georges2001}. This is consistent with QMC result \cite{Grempel1998} for $S=1/2$ for $N\to\infty$. The system undergoes a PM to SG phase transition at a temperature $T_{SG}\sim (2/3\sqrt{3})\tilde{J}S^2$, where the transition temperature, $T_{SG}^{eq}$ obtained from the equilibrium criterion is slightly smaller than the temperature calculated via the marginal stability criterion. We refer to this latter temperature as $T_{SG}$. This corresponds to a dynamical transition, similar to that in the $p$-spin glass model \cite{Cugliandolo2001,Nieuwenhuizen1998}, where the spin relaxation time diverges and glassy aging behavior commences in the non-equilibrium dynamics \cite{Biroli2002}, approaching from the PM side. 

The point $S=0,~T=0$ corresponds to a quantum critical point with a spin liquid ground state, akin to SYK non-Fermi liquid \cite{KitaevKITP,Maldacena2016}. The spin liquid state exhibits so-called `marginal spectrum' \cite{Sachdev1993,Georges2000,Georges2001}, i.e., $\chi''(\omega)\sim \mathrm{sgn}(\omega)$ at $T=0$, or $\chi''(\omega,T)\sim \tanh(\omega/2T)$ at finite temperature, and diverging spin susceptibility $\chi(T)\sim \ln \,(T/\tilde{J})$, whereas the spectral density of the bosons $\rho(\omega)=-(1/\pi)\mathrm{Im}\, G_R(\omega)\sim -\mathrm{sgn}(\omega)/\sqrt{|\omega|}$ has a characteristic power-law divergence at $T=0$. At any finite $S$ the ground state is a spin glass and the spin liquid state becomes metastable up to a small maximum $S=S_{max}\simeq 0.052$ \cite{Georges2001} at $T=0$. However, as shown in Fig.\ref{fig:phasediagramSY}, a quantum critical regime with marginal spectrum persists at finite temperature, $T_{SG}<T\aplt \tilde{J}$, up to $S\aplt 2$. Apart from the spin liquid, another paramagnetic solution to the Eqs.\eqref{Eq:SaddlePM}, the so-called `local moment' solution \cite{Georges2001} coexists in the phase diagram. This leads to a high-temperature ($T\gg \tilde{J}$) Curie susceptibility $\chi(T)\simeq S(S+1)/T$ which gradually decreases to $\chi(T)\simeq S^2/T$ as $T\to 0$, i.e., below $T_{SG}$, due to partial screening of the spins due to spin-spin interaction. Of course, for $T<T_{SG}$, the local moment solution is metastable. 
% The local moment behavior is expected to be observed~\cite{Subir2001} above the quantum critical regime $T>\tilde{J}$ for $S\lesssim 1.5$, and for $T>T_{SG}$ for larger $S$ (see Fig.\ref{fig:phasediagramSY}). 
In appendix \ref{app:SpectralFunction}, we discuss the numerically obtained bosonic spectral function $\rho(\omega)$ across the PM-SG transition as a function of $S$ for fixed $T$, and as a function of $T$ for fixed $S$, in the regime of interest for our chaos calculations { (\ref{sec:ImagTimeNumerics} and \ref{sec:ButterflyVelocity})}.

Based on the large $S$ expansions in Ref. \onlinecite{Georges2001}, the SG phase in Fig.\ref{fig:phasediagramSY} can be divided into three regimes -- $I$.~a classical regime for $\tilde{J}S\lesssim T\lesssim \tilde{J}S^2$ where all non-zero Matsubara frequency components of the Green's function in eq.\eqref{eq:SY_SaddleEq} can be neglected and spins become commuting vectors of length $S$, $II$.~a semiclassical regime for $\tilde{J}\sqrt{S}\lesssim T\lesssim \tilde{J}S$, where $\rho(\omega)\sim \omega$ for $\omega\to 0$ is obtained by $1/S$ expansion of the saddle-point equations in the SG phase (Sec.\ref{sec:SG_SY}) after scaling $\omega$ and $T$ by $\tilde{J}S$, and $III$.~a quantum regime for $0\leq T\lesssim \tilde{J}\sqrt{S}$, deduced from the $1/S$ expansion of the internal energy and linear in $T$ specific heat. In this work, we do not study the chaos across these various quantum-classical crossovers in detail. As shown in Fig.\ref{fig:phasediagramSY}, we focus on changes in the chaotic properties, like $\lambda_\mathrm{L}$ and $v_B$, from different regions of the PM phase to the SG phase across the SG transition. 

% To get further insights into the spin-glass phase we perform a large $S$ expansion of saddle point equations. Various large limits which differ in the temperature and frequency are scaled with $S$ can be considered. Below we mention these expansions which are discussed in detail in \cite{Subir2001}. 

% At a temperature where $T$ is of order $JS^2$, we get a classical limit which allows us to ignore all nonzero Matsubara frequency. This results in a classical problem that can be solved to obtain the spin glass temperature $T_{SG} = \frac{2}{3\sqrt{3}} JS^2$. At lower temperature where $T$ and $\w$ are of order $JS$ one can expand $\tilde{G}(\w, T)$ in orders of $1/JS$. At the leading order saddle point equation becomes a quartic equation which describes the semiclassical regime of spin glass as shown in Fig.\ref{fig:phasediagram}. A low temperature expansion of interner energy $U(T)$ establishes the existence of a yet another regime below the temperature where $T$ is of order $J\sqrt{S}$. This regime is identified as the Quantum spinglass. From the nature of these regimes, it's understood that it's not possible to take a classical limit by working at a fixed temperature. One has to scale the temperature with $S$ to define a meaningful classical limit. Hence, in this work we don't make such attempt.

Here, we only study the chaos in the SY chain model [Eq.\eqref{eq:SY_1D}] using bosonic representation. The model can also be studied in the large $N,M$ limit via fermionic representation \cite{Sachdev1993,Georges2000,Christos2022}. In this case, the ground state is a spin liquid for all $0\leq S\leq 1$ accessible in the fermionic representation. Thus system is maximally chaotic with $\lambda_\mathrm{L}=2\pi T$ ($\hbar=1$) for $T\to 0$. However, for finite $M$ the SG order can be recovered below $T_{SG}\sim \tilde{J}\exp(-\sqrt{M\pi})$ \cite{Christos2022,Georges2001}. In the SG ground state at $T=0$, one obtains the marginal spectrum $\chi''(\omega)\sim \mathrm{sgn}(\omega)$ above a characteristic frequency $\omega^*=\tilde{J}q_\mathrm{EA}$, i.e., over a range $\omega^*<|\omega|\ll \tilde{J}$, and the marginal SG behavior for $\chi''(\omega)$ for smaller frequencies $0<|\omega|<\omega^*$. The EA order parameter in the fermionic case is $q_\mathrm{EA}\sim 1/M$.

% 1. Fermionic spinons: Spin operator $S$ is represented by Abrikosov fermions $f$ as $S_{\alpha \beta} = f^\dagger_{\alpha} f_{\beta} - q \delta_{\alpha \beta}$, along with the constraint $\sum_\alpha f^\dagger_{\alpha} f_{\alpha} = q M$ $(0\leq q \leq 1)$.

% 2. Bosonic spinons: Spin operator $S$ is represented by bosons $b$ as $S_{\alpha \beta} = b^\dagger_{\alpha} b_{\beta} - S \delta_{\alpha \beta}$, along with the constraint $\sum_\alpha b^\dagger_{\alpha} b_{\alpha} = S M$.

% The ground state of fermionic representation is a spin-liquid\cite{Subir93}, governed by the same large $N$ equation as the SYK model\cite{Gu:2019jub, Maria2021}. While the Bosonic representation also has spin-liquid state, it transitions into a SG order for temperature below $T_{SG}$ in the $M\rightarrow$ limit \cite{Subir93, Subir2001}. One has to work at large $N$ but a general $M$ to reveal the SG order within fermionic version\cite{Christos22}. This is in line with the large $N$ limit numerical studies on $M=2$ Heisenberg model Eq.\ref{Heisenberg} with spin $S = 1/2$ which also indicate a ground state with SG order\cite{Grempel1998, Arrachea2002}. 

% The equivalence of the fermion and boson spinon therory has been shown by qualitatively matching the spin spectral density, $\chi^{\prime \prime}(\w)$ between the two\cite{Christos22}. 

\subsection{$p$-spin glass chain} \label{sec:pSpin_PhaseDiagram}
 As in the case of SY model, the equilibrium dynamics of the $p$-spin glass chain [Eq.\eqref{eq:pSpin_1d}] is described by the disorder averaged Green's function $Q_{ab}(\tau, \tau^\prime) = (1/N)\sum_i\overline{\langle \mathcal{ T}_\tau s_{i,x}^a(\tau ) s_{i,x}^b(\tau^\prime )\rangle }$. This model is solvable in the large $N$ limit and is described by the Dyson equation, 
\begin{subequations} \label{eq:SaddlePoint}
\begin{align}
(Q^{-1})^{ab}(\ci \omega_k) &= \left(  \w_k^2/\Gamma + z \right)\delta_{ab} - \Sigma^{ab}(\ci \omega_k) \\
\Sigma^{ab}(\tau) &= \frac{3 \tilde{J}^2}{2}[Q_{ab}(\tau)]^2,
\end{align}
\end{subequations}
where the $\tilde{J}^2$ takes the same definition as in Eq.\eqref{Eq:Jtilde}. The derivation of the above saddle-point equations, which are identical to that of the zero-dimensional $p$-spin glass model \cite{Cugliandolo2001}, is discussed in \ref{appB}. The spherical constraint $Q_{aa}(\tau = 0) = 1$ is imposed by the Lagrange multiplier $z$. Similar to the case of SY model, in the limit $n \rightarrow 0$, we have the replica diagonal PM and 1RSB SG phases. In the PM phase, the saddle point equations, after analytical continuation, simplify to
% \begin{subequations}
%     \begin{align}
%      &[Q(\ci\w_k)]^{-1} = \mathcal{M} \w_k^2 + z - \Sigma_R(\ci\w_k),\nonumber \\
%     &\Sigma_R(\tau) = \frac{3 J^2}{2}[Q(\tau)]^2
%  \end{align} 
% \end{subequations}
% where $\w_k = 2\pi k T$. After analytic continuation this becomes, 
{
\begin{subequations}
    \begin{align}
     &[Q_R(\w)]^{-1} = -\w^2/\Gamma + z - \Sigma_R(\w), \\
    &\Sigma_R(\w) = \Sigma_R(\ci\w_k\to \omega+\ci 0^+)
 \end{align}
\end{subequations}
}

In the SG phase, we consider the 1RSB solution, which is exact \cite{Cugliandolo2001}. Using this ansatz, we can write the Dyson equation as, 
\begin{align}
    Q^{ab}(\tau) = (q_d(\tau)-q_\mathrm{EA})\delta_{ab} + q_\mathrm{EA} \epsilon_{ab}
\end{align}
 which after Fourier transformation becomes,
 \begin{align}
    Q^{ab}(\ci\w_k) = (q_{d}(\ci\w_k)-\tilde{q}_{EA})\delta_{ab} + \tilde{q}_{EA}\epsilon_{ab},
\end{align}
where $\tilde{q}_\mathrm{EA} = \beta q_\mathrm{EA} \delta_{\w_k, 0}$. Further, it is convenient to write $q_{d}(\tau) = \tilde{Q}(\tau) + q_\mathrm{EA}$, i.e., in terms of a regular part, such that $\tilde{Q}(\tau\to \infty)\to 0$ at $T=0$, and EA order parameter $q_\mathrm{EA}$. This can be inverted in the replica space as,
\begin{align}
    [Q^{-1}]^{ab}(\ci\w_k) = A(\ci\w_k)\delta_{ab} + B(\ci\w_k)\epsilon_{ab}
\end{align}
with
\begin{subequations}
    \begin{align}
    &A(\ci\w_k)  = \frac{1}{q_d(\ci\w_k)-\tilde{q}_\mathrm{EA}}, \\
    &B(\ci\w_k) = \frac{-\tilde{q}_\mathrm{EA}}{q_d(\ci\w_k)^2+(m-2)\tilde{q}_{EA} q_d(\ci\w_k)-(m-1)\tilde{q}_\mathrm{EA}^2}
\end{align}
\end{subequations}
where $q_\mathrm{EA}$ is Edward-Anderson order parameter and $\epsilon_{ab}=1$ for the diagonal blocks and zero otherwise.

\begin{figure}[h!]
\centering
\includegraphics[width=1.0\linewidth]{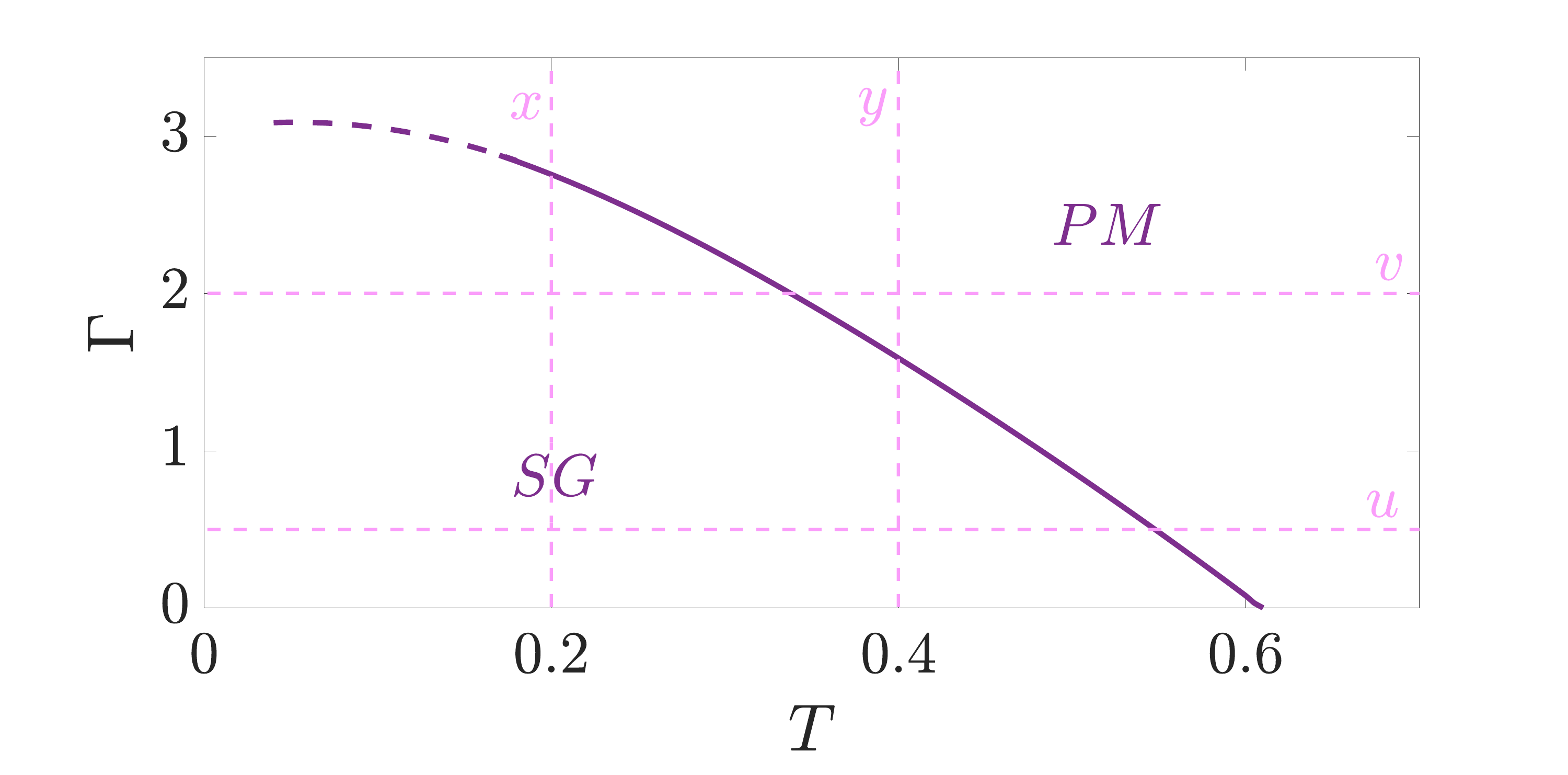}
\caption{Mean-field phase diagram of quantum spherical $p$-spin  glass model~\cite{Cugliandolo2001} as a function of temperature $T$ and dimensionless quantum fluctuation parameter $\Gamma$ (see main text). The solid line denotes second order phase transition between SG and PM phases. The second order line terminates at a tricritical point and the transition becomes first order for low temperature and larger $\Gamma$. Spin glass phase is determined by the marginality criterion. We calculate the Lyapunov exponent and butterfly velocity across the SG-PM transition for two values of $T$ as a function of $\Gamma$ (along vertical dashed lines $x$ and $y$) and for two values of $\Gamma$ as a function of $T$ (along horizontal dashed lines $u$ and $v$). }\label{fig:pSpin_PhaseDiagram}
\end{figure}

This leads to the saddle point equations, 
\begin{align}
    \w_k^2/\Gamma + z = \frac{1}{\tilde{Q}(\ci\w_k)} + \tilde{\Sigma}(\ci\w_k),
\end{align}
and 
\begin{subequations}
    \begin{align}
    &\frac{3(\beta m)^2}{2}q_\mathrm{EA}^3 = \frac{x_p^2}{1+x_p}, \\
     &y={\beta q_\mathrm{EA}}/{q_d(0)} \quad \text{and} \quad x_p={my}/{(1-y)}.
\end{align}
\end{subequations}
where $\tilde{\Sigma}(\ci\w_k) = \frac{3\tilde{J}^2 }{2}\int_{0}^{\beta} d\tau e^{\ci \w_k\tau} (q_d^2(\tau)-q_\mathrm{EA}^2)$.
The marginality criterion of setting the eigenvalue of the transverse fluctuation matrix around the saddle point to zero leads to $x_p = p-2$ \cite{Cugliandolo2001}, which determines the breaking point $m$ in the SG phase. The saddle-point equations along with the constraint $\tilde{Q}(\tau = 0) = 1 - q_\mathrm{EA}$ , can be analytically continued to real frequency and solved numerically. The details are discussed in the \ref{appB}. We show the phase diagram determined by the solutions of the saddle point equation in {Fig.\ref{fig:pSpin_PhaseDiagram}}. 
% {Similar to the SY model, we use the imaginary-time saddle-point solutions to obtain thermodynamic quantities and phase diagram, whereas the real-frequency solutions are used to compute retarded Green's function, $\lambda_\mathrm{L}$ and $v_B$.}

%By tuning the parameter $\hbar$ or equivalently $1/M$ we can tune the model between its quantum and classical limits. 
%

%
%while (ii) the SG phase has one-step replica symmetry breaking $G^{ab}(\tau) =\tilde{G}(\tau) \delta_{ab} -\epsilon^{ab} g $ where $\epsilon^{ab} =1$ with in the diagonal block of size $m$ else $\epsilon^{ab} =0$. The Edward-Anderson parameter, $g$ vanishes in PM phase but has a finite value in SG phase. The breaking point, defined as the value $m$  takes in the limit $n\rightarrow 0$ is obtained by the replicon(marginal stability) criterion.  This value also happens to give a gapless spectrum in the susceptibility, 
% The spin-glass model, exactly solvable in large $N$ limit, shares many common features with other models of quantum spin glass \cite{Goldschmidt1990,Dobrosavljevic1990,Nieuwenhuizen1998,Obuchi2007}. 

%{\it OTOC and Lyapunov exponent ---} 
\section{Out-of-time-ordered correlator (OTOC) and Lyapunov exponent}\label{sec:LyapunovExponent}
To characterize the many-body quantum chaos in the SY and $p$-pin glass chain models [Eqs.\eqref{eq:SY_1D},\eqref{eq:pSpin_1d}], we first extract Lyapunov exponent from \emph{on-site} OTOC. The calculation of Lyapunov exponent in the chain models is identical, with the effective coupling $\tilde{J}=\sqrt{J^2+2J'^2}$, to that in the corresponding zero-dimensional models, as has been done already for the $p$-spin glass in Ref. \onlinecite{Surajit22}. Thus, we focus on the effective zero-dimensional SY model, where the dot indices $x$ in Eq.\eqref{eq:SY_1D} drop out, and compare the results for $\lambda_\mathrm{L}(T)$ across the SG transition with that in the zero-dimensional $p$-spin glass model \cite{Surajit22}.

\begin{figure}[h!]
	\centering
	\includegraphics[width=0.9\linewidth]{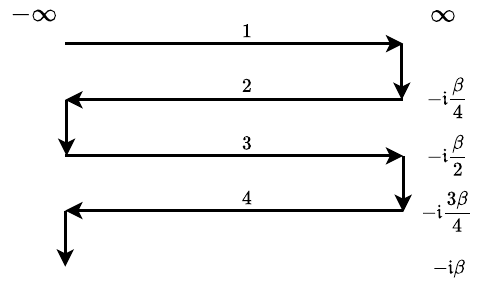}
	\caption{Schwinger-Keldysh contour with four real-time branches separated by $\beta/4$ in imaginary time, used for calculating OTOC. } \label{fig:contour} 
\end{figure}

To diagnose the quantum chaos in the SY model, we consider two  disorder-averaged four-point `regularized' OTOCs \cite{KitaevKITP,Maldacena2016,Aleiner2016} for the bosonic operators on a Schwinger Keldysh (SK) contour Fig.\ref{fig:contour}, with two forward and two backward branches, equally spaced in imaginary time with separation $\beta/4$, namely
\begin{subequations}\label{eq:OTOC_SY}
\begin{align}
&F_1(t_1, t_2) \nonumber \\
&= \frac{1}{N^2 M^2}\sum_{ij,\alpha\beta} \overline{\mathrm{Tr}[y b^{\dagger}_{i\alpha }(t_1)yb^{\dagger }_{j\beta}(0)y b_{i \alpha}(t_2) y b_{j \beta}(0)]}\\
&F_2(t_1, t_2) \nonumber \\
&= \frac{1}{N^2M^2}\sum_{ij,\alpha\beta} \overline{\mathrm{Tr}[y b_{i\alpha}(t_1)yb^{\dagger }_{j\beta}(0)y b_{i \alpha}^{\dagger}(t_2)y b_{j\beta}(0)]},
\end{align}
\end{subequations}
where $y^4 = \mathrm{exp}(-\beta \mathcal{H})/\mathrm{Tr}[\mathrm{exp}(-\beta \mathcal{H})]$. In this work, we do not consider the OTOC of the physical spin operators $S_{i\alpha\beta x}$. Calculations of such an OTOC would involve 8-point bosonic correlation functions. We assume the maximum Lyapunov exponent in the large $N,M$ SY model to be the same for OTOCs for bosonic and spin operators.

% These two correlation functions can be calculated using a path integral on
% Schwinger-Keldysh(SK) contour with four real-time branches that are separated by $\beta/4$ in the imaginary time. 
The calculation of the OTOCs in Eq.\eqref{eq:OTOC_SY} on the SK contour (Fig.\ref{fig:contour}) can be done in the replica diagonal PM phase following a procedure very similar to SYK and related models \cite{KitaevKITP,Maldacena2015,BanerjeeAltman2016}. However below the dynamical transition temperature $T_{SG}$, in the SG phase, calculating any dynamic correlation function is more involved since the system always stays out of equilibrium \cite{Cugliandolo1998,Biroli2002}. Alternatively, as discussed in Ref. \onlinecite{Surajit22} for $p$-spin glass, the OTOCs can be calculated in the marginal SG phase by replicated generating function~\cite{Surajit22} $\mathcal{Z}^n$ on the four-branch contour [Fig.\ref{fig:contour}], i.e., 
\begin{align} \label{eq:ReplicaKeldysh}
\mathcal{Z}^n=& \frac{1}{Z^n} \int \mathcal{D}(\bar{b}_{i \alpha}^a,b_{i\alpha}^a) \mathcal{D}\lambda_{i}^a  \mathrm{exp}(\ci S)
\end{align}
where $a=1,\dots,n$ are the replica indices and the action
\begin{align}
\ci S =& \Bigg[\ci \int_\mathcal{C} d z \sum_{i\alpha a} \left\{  \bar{b}_{i \alpha}^a \ci \partial_z b_{i \alpha}^a -  \lambda_{i}^a\left(\bar{b}_{i \alpha}^a(z) b_{i \alpha}^a(z) -S M\right) \right\} \nonumber \\
&- \frac{1}{\sqrt{M N}} \frac{1}{2!} \ci\int_{\mathcal{C}} dz \sum_{ij,\alpha\beta,a} J_{ij} \bar{b}_{i\alpha}^{a}(z) b_{i \beta}^a(z) \bar{b}^a_{j\beta}(z) b_{j \alpha}^a(z)\Bigg], 
\end{align}
where $z$ is the complex time variable along the SK contour $\mathcal{C}$ [Fig.\ref{fig:contour}]. Here $Z$ is the equilibrium partition function which appears due to the initial thermal density matrix \cite{Surajit22}. The initial density matrix corresponds to the replica symmetric saddle point above $T_{SG}$, and 1RSB marginal SG saddle point below $T_{SG}$, as discussed in Sec.\ref{sec:SY_PhaseDiagram}.  After performing the disorder averaging, we can take the $n\rightarrow 0$ limit. The resulting time-translation invariant dynamical correlations and responses
are identical to those obtained from the 1RSB saddle point solutions in the marginal SG phase. Using this replicated generating function, the OTOC can now be obtained as,
\begin{subequations}
    \begin{align}
&F_1^{aa bb}(t_1, t_2) =\nonumber\\
& \frac{1}{N^2 M^2}\sum_{ij,\alpha \beta} \langle y b^{\dagger a}_{i\alpha 4}(t_1)yb^{\dagger a }_{j\beta 3}(0)y b_{i \alpha 2}^{b }(t_2) y b_{j \beta 1}^{b }(0)\rangle\\
&F_2^{aa bb}(t_1, t_2) =\nonumber\\
& \frac{1}{N^2M^2}\sum_{ij,\alpha \beta} \langle y b_{i\alpha 4}^{a}(t_1)yb^{\dagger a}_{j\beta 3}(0)y b_{i \alpha 2}^{\dagger b}(t_2)y b^{b }_{j\beta 1}(0)\rangle
\end{align}
\end{subequations}

where the superscripts $1,2,3,4$ denote the branch on the SK contour and the average $\langle \dots \rangle$ is with respect to replicated generating function [Eq.\eqref{eq:ReplicaKeldysh}].

\begin{figure}[h!]
	\centering
	\includegraphics[width=1.0\linewidth]{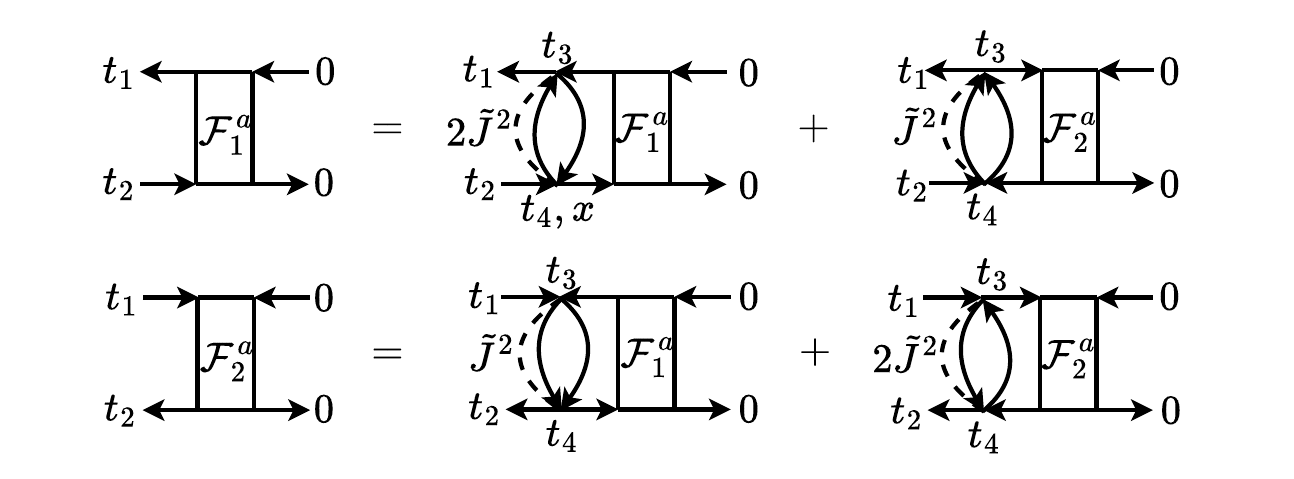}
	\caption{ Diagrammatic representation of the kernel equation in Eq.\eqref{kerneleqpm} for $\mathcal{O}(1/(NM))$ term, ($\mathcal{F}_{1,2}^a$) in the OTOC $F_{1,2}^a(t_1, t_2)$. The solid horizontal lines denote dressed retarded and advanced propagators $G_R$, $G_A$, and the vertical rung denotes the Wightmann correlations $G_{lr}^+$ and $G_{lr}^-$. The dashed line represents disorder averaging. We suppress replica indices of the vertices in the ladder diagram to avoid cluttering.}\label{fig:LadderDiagram}
\end{figure}

The diagrams that contribute to the OTOC can be arranged in powers of $1/(NM)$. $\mathcal{O}(1)$ diagrams are disconnected and do not contribute to the growth of OTOC. The contribution, $\mathcal{F}_\mu^{aabb}$ at $\mathcal{O}(1/(NM))$ form the ladder diagrams (Fig.\ref{fig:LadderDiagram}) that grow exponentially, i.e., for $F_\mu^{aabb}(t_1, t_2)= F_\mu^{0 aabb}(t_1, t_2) +(1/(NM)) \mathcal{F}_\mu^{aabb}(t_1, t_2) + \mathcal{O}(1/(N^2M^2))$ ($\mu = 1,2$), $\mathcal{F}_\mu\sim e^{\lambda_\mathrm{L}t}$. The ladder diagrams can be written in the form of a Bethe-Salpeter-like equation, similar to SYK-type models \cite{Gu:2019jub,Maldacena2016,BanerjeeAltman2016},
\begin{subequations}
\begin{align}
\mathcal{F}_1^{aabb}&(t_1, t_2) = \sum_{c}\int_{t_3,t_4} \left[ K_{11}^{aacc}(t_1, t_2, t_3, t_4)\mathcal{F}_1^{ccbb}(t_3, t_4)\right. \nonumber\\
 &\left. + K_{12}^{aacc}(t_1, t_2, t_3, t_4)\mathcal{F}_2^{ccbb}(t_3, t_4)\right]  \\
\mathcal{F}_2^{aabb}&(t_1, t_2) = \sum_{c}\int_{t_3,t_4} \left[K_{21}^{aacc}(t_1, t_2, t_3, t_4)\mathcal{F}_1^{ccbb}(t_3, t_4) \right. \nonumber \\
&\left. +  K_{22}^{aacc}(t_1, t_2, t_3, t_4)\mathcal{F}_2^{ccbb}(t_3, t_4) \right].
\end{align}
\end{subequations}
Here $\int_{t_3,t_4}=\int dt_3dt_4$. In the chaotic growth regime, $\lambda_\mathrm{L}^{-1}\lesssim t<  \lambda_\mathrm{L}^{-1} \ln \, N$, the propagators along the horizontal lines, from $t_1$ to $t_3$ and from $t_2$ to $t_4$ in Fig.\ref{fig:LadderDiagram} can be approximated by retarded and advanced propagators \cite{KitaevKITP,Maldacena2016,Gu:2018jsv}. Since the retarded and advanced propagators are replica-diagonal in both PM and SG phases, the kernel equations become,
\begin{subequations} \label{eq:kerneleq}
\begin{align}
\mathcal{F}_1^{a}&(t_1, t_2) = \int dt_3 dt_4 K_{11}^{a}(t_1, t_2, t_3, t_4)\mathcal{F}_1^{a}(t_3, t_4) \nonumber \\
&+ \int dt_3 dt_4 K_{12}^{a}(t_1, t_2, t_3, t_4)\mathcal{F}_2^{a}(t_3, t_4)  \\
\mathcal{F}_2^{a}&(t_1, t_2) = \int dt_3 dt_4 K_{21}^{a}(t_1, t_2, t_3, t_4)\mathcal{F}_1^{a}(t_3, t_4) \nonumber \\
&+ \int dt_3 dt_4 K_{22}^{a}(t_1, t_2, t_3, t_4)\mathcal{F}_2^{a}(t_3, t_4), \label{kerneleqpm}
\end{align}
\end{subequations}
where $\mathcal{F}_\mu^{a}(t_1, t_2) \equiv \mathcal{F}_\mu^{aaaa}(t_1, t_2)$ and the kernels
are given by

\begin{align} \label{eq:KernelSY_Local}
 K_{11}^a(t_1, t_2, t_3, t_4) =& 2 \tilde{J}^2 G_{A}(t_{31})  G_R(t_{24}) G_{lr}^+(t_{43})G_{lr}^-(t_{34}) \nonumber \\
K^a_{12}(t_1, t_2, t_3, t_4) =& \tilde{J}^2 G_{A}(t_{31})G_R(t_{24}) G_{lr}^+(t_{43}) G_{lr}^+(t_{43})  \nonumber \\
K^a_{21}(t_1, t_2, t_3, t_4) =& \tilde{J}^2 G_R(t_{13})G_{A}(t_{42})G_{lr}^-(t_{34})G_{lr}^-(t_{34}) \nonumber  \\
K^a_{22}(t_1, t_2, t_3, t_4) =&   2\tilde{J}^2 G_R(t_{13})G_{A}(t_{42})G_{lr}^-(t_{34})G_{lr}^+(t_{43}).
\end{align}

Here all Green's functions are replica diagonal and $G_{lr}^\pm(t)$ refers to Wightmann correlations functions (\ref{app:SpectralFunction}). 

Using the exponential ansatz in the growth regime, $\mathcal{F}_\mu^{a}(t_1, t_2) = e^{\lambda_\mathrm{L}(t_1+t_2)/2}f_\mu^a(t_1-t_2)$ and taking the Fourier transform of the kernel equations [Eq.\eqref{eq:kerneleq}], leads to  eigenvalue equations in the frequency space with eigenvalue, $\lambda_e=1$. For the PM phase, these eigenvalue equations are
\begin{subequations} \label{eq:kernelpm}
\begin{align}
\tilde{J}^2G_A\left(-\omega -\ci \frac{\lambda_\mathrm{L}}{2}\right) G_R\left(-\omega + \ci \frac{\lambda_\mathrm{L}}{2}\right) &
\left[ 2 \int d \omega^\prime g_{1}(\omega - \omega^\prime) \right. \nonumber \\
\left. f_1^a(\omega^\prime) + \int d \omega^\prime g_{2}(-(\omega - \omega^\prime)) f_2^a(\omega^\prime)\right]& = f_1^a(\omega) \\
\tilde{J}^2G_R\left(\omega + \ci \frac{\lambda_\mathrm{L}}{2}\right) G_A\left(\omega - \ci\frac{\lambda_\mathrm{L}}{2}\right)&\left[\int d \omega^\prime g_{2}(\omega - \omega^\prime)\right.\nonumber \\
\left. f_1^a(\omega^\prime)+ 2 \int d \omega^\prime g_{1}(\omega -\omega^\prime) f_2^a(\omega^\prime)\right]& = f_2^a(\omega)
\end{align}
% where 
% \begin{align}
% g_{1}(\omega)& = \frac{1}{4\pi^2} \int d\omega_1 G_{lr}^+(\omega_1)G_{lr}^-(\omega+ \omega_1) \nonumber\\
% &= \frac{1}{4} \int d\omega_1 \frac{ \rho(\omega_1) \rho(\omega + \omega_1)}{ \mathrm{sinh}\frac{\beta\omega_1}{2} \mathrm{sinh}\frac{\beta(\omega+ \omega_1)}{2}} \label{Eq. g1}\\
% g_{2}(\omega) &=\frac{1}{4\pi^2} \int d\omega_1 G_{lr}^+(\omega_1)G_{lr}^+(\omega- \omega_1) \nonumber \\
% &= \frac{1}{4} \int d\omega_1 \frac{ \rho(\omega_1) \rho(\omega - \omega_1)}{ \mathrm{sinh}\frac{\beta\omega_1}{2} \mathrm{sinh}\frac{\beta(\omega- \omega_1)}{2}} \label{Eq. g1}
% \end{align}
\end{subequations}
where
%\begin{widetext}
% \begin{align}
% &J^2G_A(-\omega -i \lambda_\mathrm{L}/2) G_R(-\omega + i \lambda_\mathrm{L}/2) \bigg[ 2 \int d \omega^\prime g_{1}(\omega - \omega^\prime)f_1^a(\omega^\prime) 
%  + \int d \omega^\prime g_{2}(-(\omega - \omega^\prime)) f_2^a(\omega^\prime)\bigg] = f_1^a(\omega) \nonumber\\
% &J^2G_R(\omega + i \lambda_\mathrm{L}/2) G_A(\omega - i \lambda_\mathrm{L}/2) \bigg[\int d \omega^\prime g_{2}(\omega - \omega^\prime)f_1^a(\omega^\prime)
%  + 2 \int d \omega^\prime g_{1}(\omega -\omega^\prime) f_2^a(\omega^\prime)\bigg] = f_2^a(\omega)\label{kernelpm}
% \end{align}
% where 
\begin{subequations}
\begin{align}\label{Eq. g1}
g_{1}(\omega) &= \frac{1}{4\pi^2} \int d\omega_1 G_{lr}^+(\omega_1)G_{lr}^-(\omega+ \omega_1) \nonumber \\
&= \frac{1}{4} \int d\omega_1 \frac{ \rho(\omega_1) \rho(\omega + \omega_1)}{ \mathrm{sinh}\frac{\beta\omega_1}{2} \mathrm{sinh}\frac{\beta(\omega+ \omega_1)}{2}} 
\end{align}
\begin{align}\label{Eq. g2}
g_{2}(\omega) &=\frac{1}{4\pi^2} \int d\omega_1 G_{lr}^+(\omega_1)G_{lr}^+(\omega- \omega_1) \nonumber \\
&= \frac{1}{4} \int d\omega_1 \frac{ \rho(\omega_1) \rho(\omega - \omega_1)}{ \mathrm{sinh}\frac{\beta\omega_1}{2} \mathrm{sinh}\frac{\beta(\omega- \omega_1)}{2}}.
\end{align}
\end{subequations}
%\end{widetext}
Here $\rho(\omega)=-(1/\pi)\mathrm{Im}\, G_R(\omega)$ is the bosonic spectral function (\ref{appA}).
We discretize the kernel equations [Eqs.\eqref{eq:kernelpm}] in frequency and diagonalize the resulting matrix kernel for different trial values of $\lambda_\mathrm{L}$. Eventually, the Lyapunov exponent $\lambda_\mathrm{L}$ is obtained from the one for which the kernel has at least one eigenvalue $\lambda_e=1$.

Similarly, for the 1RSB SG phase, the kernel equations are
\begin{widetext}
\begin{subequations} \label{eq:kernelsg}
\begin{align} 
&\tilde{J}^2 G_A\left(-\omega -\ci \frac{\lambda_\mathrm{L}}{2}\right) G_R\left(-\omega + \ci \frac{\lambda_\mathrm{L}}{2}\right) \bigg[2 \Big(g^2 f_1^a(\omega) 
+ g \int \frac{d\omega^\prime}{2\pi} \tilde{g}_{lr}(\omega-\omega^\prime)f_1^a(\omega^\prime) + g \int \frac{d\omega^\prime}{2\pi} \tilde{g}_{lr}(\omega^\prime-\omega)f_1^a(\omega^\prime)\nonumber \\
&+ \int d\omega^\prime \tilde{g}_1(\omega-\omega^\prime)f_1^a(\omega^\prime)\Big) 
  +g^2 f_2^a(\omega) +2g \int \frac{d\omega^\prime}{2\pi} \tilde{g}_{lr}(-\omega +\omega^\prime)f_2^a(\omega^\prime) 
  + \int d\omega^\prime \tilde{g}_2(-\omega +\omega^\prime)f_2^a(\omega^\prime) \bigg] = f_1^a(\omega) \\
&\tilde{J}^2 G_A\left(\omega -\ci \frac{\lambda_\mathrm{L}}{2}\right) G_R\left(\omega + \ci \frac{\lambda_\mathrm{L}}{2}\right) \bigg[2 \Big(g^2 f_2^a(\omega) + g \int \frac{d\omega^\prime}{2\pi} \tilde{g}_{lr}(\omega-\omega^\prime)f_2^a(\omega^\prime) + g \int \frac{d\omega^\prime}{2\pi} \tilde{g}_{lr}(\omega^\prime-\omega)f_2^a(\omega^\prime) \nonumber \\
&+ \int d\omega^\prime \tilde{g}_1(\omega-\omega^\prime)f_2^a(\omega^\prime)\Big) 
  +g^2 f_1^a(\omega) +2g \int \frac{d\omega^\prime}{2\pi} \tilde{g}_{lr}(\omega -\omega^\prime)f_1^a(\omega^\prime) + \int d\omega^\prime \tilde{g}_2(\omega -\omega^\prime)f_1^a(\omega^\prime) \bigg] = f_2^a(\omega)
 \end{align}
 \end{subequations}
\end{widetext}

% By discretizing frequency, we diagonalize the matrix kernel for a given value of $\lambda_\mathrm{L}$. The value of $\lambda_\mathrm{L}$ is eventually obtained by varying it such that the eigenvalue $\lambda_e=1$. 
% In the PM phase we get Eq.\ref{kernelpm} and in SG phase we get Eq.\ref{kernelsg},
% where 
In the above equations, $\tilde{g}_1(\w)$ and $\tilde{g}_2(\w)$ are the same as in Eqs.\eqref{Eq. g1},\eqref{Eq. g2} in terms of the SG spectral function $\rho(\omega)$.
% with $\rho(\w)$ replaced by $\tilde{\rho}(\w)$, and $\tilde{g}_{lr}(\w) = \pi \tilde{\rho}(\w)/\sinh(\beta \w/2)$, where the spectral function, $\tilde{\rho}(\omega)=-(1/\pi)\mathrm{Im}\, \tilde{G}_R(\omega)$ is obtained from the regular part of the Green's function (\ref{appA}). 
We obtain $\lambda_\mathrm{L}$ as a function of $S$ and $T$ by separately solving Eq.\eqref{eq:kernelpm} for the PM phase, and Eq.\eqref{eq:kernelsg}] for the SG phase, with the retarded, advanced and Wightmann functions obtained from the large $N$ saddle point equations (\ref{appA}).

The crucial difference between the PM and SG phases is encoded in the ladder kernerl [Eq.\eqref{eq:KernelSY_Local}] through the Wightmann correlators. In the SG phase, 
\begin{align}
G_{lr}^+(\w) =G_{lr}^-(\w) = 2\pi g\delta(\w)\epsilon_{ab} +
\pi\rho(\omega)/ (\mathrm{sinh}(\beta \omega/2))\delta_{ab}\label{eq:WightmannSG}
\end{align}
 In contrast, the delta function term is absent in the PM phase, where $g=\sqrt{q_\mathrm{EA}}=0$.

%\begin{subequations}
%\begin{align}
%&J^2G_A(-\omega -i \lambda_\mathrm{L}/2) G_R(-\omega + i \lambda_\mathrm{L}/2) \bigg[2 \int d \omega^\prime g_{a}(\omega - \omega^\prime)f_1(\omega^\prime)  + \int d \omega^\prime g_{b}(-(\omega - \omega^\prime)) f_2(\omega^\prime)\bigg] = f_1(\omega) \\
%&J^2G_R(\omega + i \lambda_\mathrm{L}/2) G_A(\omega - i \lambda_\mathrm{L}/2) \bigg[ \int d \omega^\prime g_{b}(\omega - \omega^\prime)f_1(\omega^\prime)  + 2 \int d \omega^\prime g_{a}(\omega -\omega^\prime) f_2(\omega^\prime)\bigg] = f_2(\omega)
%\end{align}
%\end{subequations}
% 
%where 
%\begin{align}
%G_R(-\omega + i \lambda_\mathrm{L}/2) = \int d\omega^\prime \frac{\rho(\omega^\prime)}{-\omega + i \lambda_\mathrm{L}/2 - \omega^\prime + i0^+} \quad G_A(-\omega - i \lambda_\mathrm{L}/2) = \int d\omega^\prime \frac{\rho(\omega^\prime)}{-\omega - i \lambda_\mathrm{L}/2 - \omega^\prime - i0^+}
%\end{align} 

\begin{figure}[h!]
	\centering
    \includegraphics[width=\linewidth]{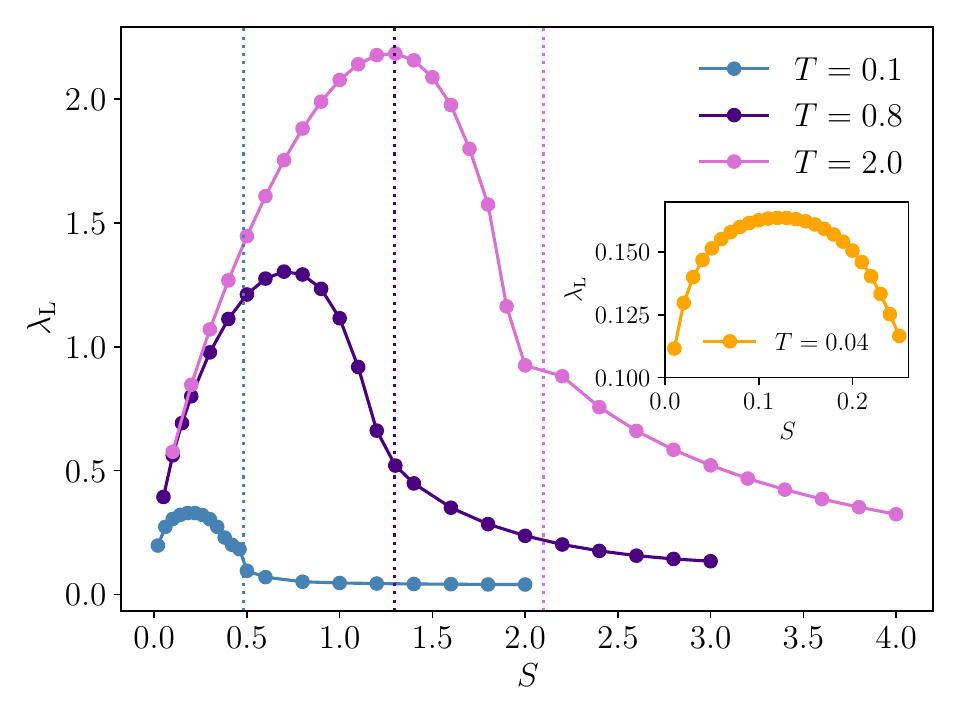}
	\caption{{\bf Variation of Lyapunov exponent with $S$ in the SY model.} $\lambda_\mathrm{L}(S)$ for $T=0.1$, $T=0.8$ and $T=2.0$ (in units of $\tilde{J}$), across the PM-SG phase transition marked by the vertical lines at $S_{SG}\simeq 0.48,~1.29,~2.1$, respectively. $\lambda_\mathrm{L}$ has a maximum in the PM phase and changes non-monotonically. $\lambda_\mathrm{L}\sim 1/S^{1.7}$ in the SG phase. The inset shows $\lambda_{\mathrm{L}}(S)$ for $T = 0.04$ in the PM phase }\label{fig:SY_lambdaL_S}
\end{figure} 

We compare the results for $\lambda_\mathrm{L}(S,T)$ obtained using the above formalism for the {effective} zero-dimensional SY model with the $\lambda_\mathrm{L}(\Gamma,T)$ for the effective zero-dimensional $p$-spin glass model. The OTOC formalism for the latter is very similar to that of SY model, but somewhat simpler. To extract $\lambda_\mathrm{L}(\Gamma,T)$, we compute the following regularized OTOC~\cite{Surajit22} for the $p$-spin glass model,
\begin{align}
     F(t_1, t_2) =& \frac{1}{N^2} \sum_{i,j} \overline{\mathrm{Tr} \Big[y s_{i}(t_1)ys_{j}(0)y s_{i}(t_2)ys_{j}(0)\Big]},
 \end{align}
We skip the details of the formalism here since those have been already described in Ref. \onlinecite{Surajit22}. We briefly discuss a few additional details for the non-local OTOC for $p$-spin glass chain model [Eq.\eqref{eq:pSpin_1d}] in Sec.\ref{sec:ButterflyVelocity} along with those for the SY chain [Eq.\eqref{eq:SY_1D}].

\subsection{Lyapunov exponent across spin glass transition in the SY model}
We obtain $\lambda_\mathrm{L}(S,T)$ in the PM and SG phases by separately solving the corresponding large $N$ {kernel equations (Sec.\ref{sec:LyapunovExponent})}  for the two phases. Specifically, as shown in Fig.\ref{fig:phasediagramSY}, we take cuts along several $S$ and $T$, namely, from small $S$ (low-$T$) quantum critical region $a$.~$S=0.5$ ($d$.~$T=0.1$), from intermediate-$S$ (intermediate-$T$) crossover region between quantum critical and local moment behavior $b$.~$S=1$ ($e$.~$T=0.8$), and from large $S$ (high-$T$) local moment region $c$.~$S=2$ ($f$.~$T=2$), as a function of $T$ ($S$). We set $\tilde{J}=1$ as the unit of temperature (energy). The results for $\lambda_\mathrm{L}(S,T)$ are qualitatively similar for all the cuts, as we discuss below.

\textit{$S$ dependence of Lyapunov exponent.}---We first show the Lyapunov exponent as a function of $S$ in Fig.\ref{fig:SY_lambdaL_S} from PM to spin glass phase for four temperatures, $T=0.04,0.1,0.8,2.0$.  In all the cases, the Lyapunov exponent $\lambda_\mathrm{L}$ has a non-monotonic dependence on $S$ within the PM phase; $\lambda_\mathrm{L}$ initially increases with $S$ till a value $S_m(T)$ where $\lambda_\mathrm{L}$ has a broad peak, and then it decreases with further increase of $S$ approaching the critical value of $S$, $S_{SG}(T)$ for the spin glass transition, which is obtained from the breakpoint criterion $m(T,S_{SG})=1$. The non-monotonic $\lambda_\mathrm{L}(S)$ in the PM phase can be heuristically understood from the behavior of the spectral function. As discussed in \ref{app:SpectralFunction}, in our numerical solution of the real-frequency saddle-point equation [Eq.\eqref{Eq:SaddlePM_real}], e.g., at $T=0.1,0.8$, for small $S\simeq 0.05-0.2$, the spectral function has either a spin liquid or a local moment like form [see Figs.\ref{fig:rho_T0.1}(a), \ref{fig:rho_T0.8}(a)], with a peak at $\omega\propto T$, but still sufficiently larger than $T$, and very little spectral weight around $\omega=0$.
% in the quantum critical PM region at low temperature (e.g., $T=0.1$, Fig.\ref{fig:rho_T0.1}), for small values of $S$, the spectral function $\rho(\w)$ for $\omega>0$ has a peak away from $\omega=0$ with little spectral weight near zero energy. 
The peak moves closer to zero energy with increasing $S$ and the spectral density increases around $\omega=0$. As a result, $\lambda_\mathrm{L}$ is small at small $S$ and increases with $S$ till a maximum value in the PM phase. However, beyond the maximum, closer to the SG transition, there is a loss of spectral weight at intermediate energies or \emph{pseudogap}~\cite{Camjayi2003} like feature, in between a narrower low-energy peak and a broad peak at high energies. This presumably leads to decreases of $\lambda_\mathrm{L}$ approaching the SG transition. Thus, $\lambda_\mathrm{L}$ exhibits a non-monotonic dependence on $S$ in the PM region. 
% At higher temperatures ($T=0.8,~2.0$) the spectral function is consistent with a local moment-like behavior~\cite{Subir2001}, where a peak at $\w_0 \sim T ln[(S+1)/S] > T$, far away from zero energy appears, and, as a result, $\rho(\omega)$ has very little spectral weight at low frequencies ($\omega\sim 0$). This gapped or pseudogapped nature leads to reduced $\lambda_\mathrm{L}$ or weak chaos. As $S$ increases, $\rho(\w)$ again gains weight near zero energy [Fig.\ref{fig:rho_T0.8}], and thus $\lambda_\mathrm{L}$ increases reaching a broad peak.  Beyond that, approaching the phase transition, a loss of spectral weight appears in the spectrum at intermediate frequencies, as in the case of $T=0.1$ discussed above, leading to decrease of $\lambda_\mathrm{L}$ with further increase of $S$ and a broad peak in $\lambda_\mathrm{L}(S)$ overall. 

\begin{figure}[h!]
	\centering
     \includegraphics[width=\linewidth]{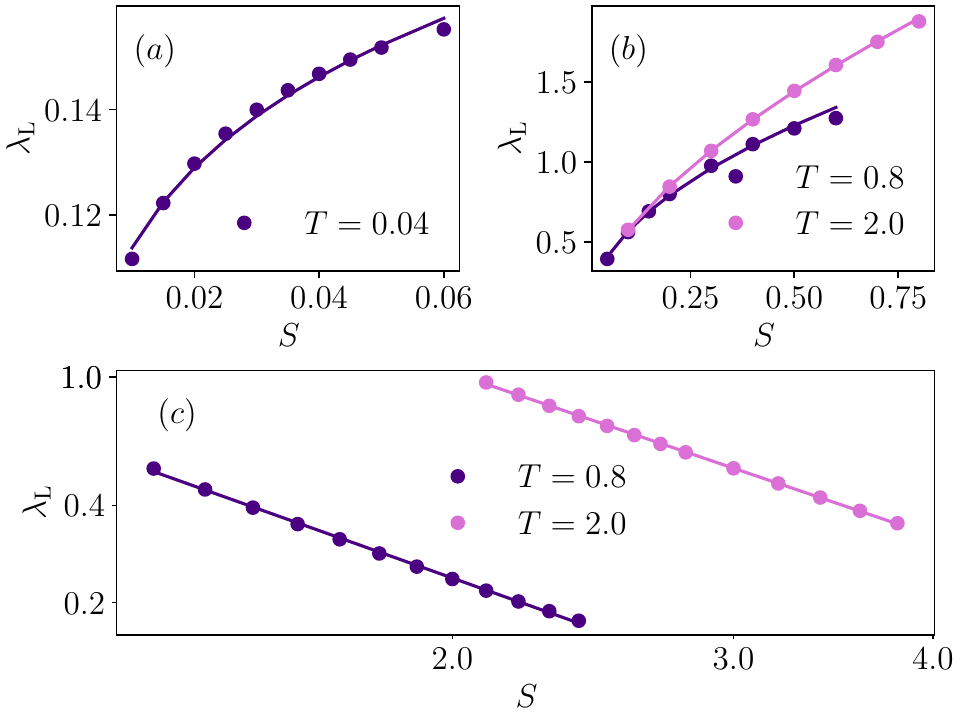}
	\caption{{\bf $S$ dependence of $\lambda_\mathrm{L}$ in PM and SG phases of SY model.} $\lambda_\mathrm{L}$ is shown as a function of $S$. (a) In the PM phase, before reaching the maximum [Fig.\ref{fig:SY_lambdaL_S}], $\lambda_\mathrm{L}$ varies as, $\lambda_\mathrm{L}(S,T)\simeq \lambda_\mathrm{L}(0,T)+a S^{\alpha_\lambda}$, shown as line fit to the data (circles) for $T = 0.04$ with $\alpha_\lambda\approx 0.2 $, and (b) for $T=0.8,2.0$ with $\alpha_\lambda\approx 0.5, 0.6$, respectively.
 % ($\alpha_\lambda, \lambda_\mathrm{L}(0,T)$) are obtained as ($0.48, 0$) for $T=0.8$ and ($0.6, 0.043$) for $T=2.0$. 
 (c) $\lambda_\mathrm{L}$ decays with $S$ as a power law in the SG phase, shown as straight line fit to the data (circles) on log-log scale. $\lambda_\mathrm{L}\sim 1/S^{1.7}$ for $T=0.8,2.0$.}\label{fig:SY_lambdaLfit_S}
\end{figure}

In the PM phase, especially in the quantum critical region $IV$ in Fig.\ref{fig:phasediagramSY}, for $S\to 0$, we expect dominantly spin liquid behavior and $\lambda_\mathrm{L}(S\to0,T)\to \lambda_\mathrm{L}^\mathrm{SYK}(T)$, i.e., $\lambda_\mathrm{L}$ should asymptotically approach the $T$-dependent Lyapunov exponent~\cite{BanerjeeAltman2016}, $\lambda_\mathrm{L}^\mathrm{SYK}(T)$, of the SYK spin liquid~\cite{Georges2001} at $S=0$. Based on this expectation we fit our data with $\lambda_\mathrm{L}(S,T)\simeq \lambda_\mathrm{L}(0,T)+aS^{\alpha_\lambda}$ in the PM phase for small values of $S$, before reaching the peak [Fig.\ref{fig:SY_lambdaL_S}]. As shown in Fig.\ref{fig:SY_lambdaLfit_S}(a,b) for $T=0.04,0.8,1.0$, we find reasonable fit with $\alpha_\lambda\simeq 0.2-0.6$. However, due to this $S$ dependence, there is a rapid decrease of $\lambda_\mathrm{L}(S)$ for $S\to 0$, as seen in Fig.\ref{fig:SY_lambdaLfit_S}(a). As a result, the extrapolated $\lambda_\mathrm{L}(0,T)$ for $S\to 0$ from the fit consistently turns out to be very low, and much smaller than $\lambda_\mathrm{L}^{SYK}(T)$ \cite{BanerjeeAltman2016}, even at low temperatures $T\simeq 0.04-0.1$. Nevertheless, our real frequency saddle-point solutions, e.g., the computed bosonic spectral function, for $S\leq 0.05\approx S_{max}$ at low $T$ are consistent with the conformal spin liquid solutions~\cite{Sachdev1993,Georges2001,Sachdev2015} for $|\omega|,T\ll J$ as well as the numerical spin liquid solutions obtained in Ref.  \onlinecite{Camjayi2003}, as discussed in \ref{app:SpectralFunction}. Thus, our results indicate strong $S$ (and $T$) dependent corrections to the conformal limit of the spectral function, and, consequently, $\lambda_\mathrm{L}\ll 2\pi T$, in the spin liquid region, even for small $S\lesssim 0.05$. 
% This deviation from the expected spin liquid behavior can be traced to the local moment like spectral function [Figs.\ref{fig:rho_T0.1}(a), \ref{fig:rho_T0.8}(a)] that we obtain from our numerical real-frequency saddle-point solutions for $S\lesssim 0.2$, as mentioned above. 
% % This could be due to numerical error associated with the solution of saddle-point equations [Eq.\eqref{Eq:SaddlePM}] with the constraint $G(\tau=0^-)=-S$ for small values of $S\lesssim 0.1$, where number of bosons per index $\alpha=1,\cdots,M$ becomes very small. 
% We note that special care~\cite{Sachdev1993,Tikhanovskaya2021} needs to be taken to obtain the real-frequency spin liquid solution at low temperature due to the $1/\sqrt{|\omega|}$ divergence [\ref{app:SpectralFunction}]. Such methods~\cite{Sachdev1993,Tikhanovskaya2021} were not implemented in our numerical calculations, as we do not probe the small $S\lesssim 0.2$ region and the relative stability of the spin liquid vs. local moment solutions in more detail in this work. 
We also note that our numerical saddle-point solutions at small $S$ smoothly  to the solutions at higher $S$, as evident from the smooth evolution of $\lambda_\mathrm{L}(S)$ in Figs.\ref{fig:SY_lambdaL_S}, \ref{fig:SY_lambdaLfit_S}(a) for all $T$.
% with a sharp peak around $\w \sim 0$ separated from the spectral weight at higher frequencies. \textcolor{magenta}{The appearance of peak is attributed to the formation of local moments of spins\cite{Subir2001}.}{SB: In Georges et al PRB paper, such local moment peak shown for the local-moment solution only at low temperature. It is not clear whether the peak seen here at $T\sim 1$ is related to that}. As a result, $\lambda_\mathrm{L}$ decreases as the pseudogap deepens with increasing $S$ towards the phase transition, leading to a broad maximum in an intermediate range of $S$, as seen in Fig.\ref{fig:SY_lambdaL_T}. 

% We obtain $\lambda_\mathrm{L}(S,T)$ in the SG phase by solving the large $N$ saddle-point equations (Sec.\ref{sec:pSpin_PhaseDiagram}), as shown in Fig.\ref{fig:SY_lambdaL_T}. 

{The Lyapunov exponent appears to have weak signature of singularity in the form discontinuity/change of slope at $S_{SG}$ for $T=0.1,~2.0$. The Lyapunov exponent, like other usual observable, e.g., the free-energy or relaxation time, can have genuine singularity across a phase transition. However, in contrast, we find $\lambda_\mathrm{L}(S)$ to be continuous for $T=0.8$ across the SG transition. Moreover, we cannot rule out numerical artifact due to problem in
obtaining a converged solution of the saddle-point equations very close to the transition, as well as, due to difficulty in matching the PM and 1RSB SG solutions approaching from two different sides of the SG transition. As a result, we refrain from making any strong statement about the singularity of $\lambda_\mathrm{L}$ across the SG transition.}
% could be a numerical artifact due to problem in obtaining a converged solution of the saddle-point equations very close to the transition, as well as, due to difficulty in matching the PM and 1RSB SG solutions approaching from two different sides of the SG transition. In contrast, we find $\lambda_\mathrm{L}(S)$ to be continuous for $T=0.8$ across the SG transition. {Due to this difference in the behavior of $\lambda_{\mathrm{L}}(S)$ at $S_{SG}$ for different temperatures, we are unable to make a strong statement about the continuity of the same.} 
In the SG phase, $\lambda_\mathrm{L}(S)$ decreases monotonically as $\sim 1/S^{1.7}$, as shown in Fig.\ref{fig:SY_lambdaLfit_S}(b). The reduction of $\lambda_\mathrm{L}$ with increasing $S$ or its increase with decreasing $S$ (increasing quantum fluctuations) can be understood qualitatively by drawing analogy with the similar dependence of $\lambda_\mathrm{L}$ on $\Gamma$ in the $p$-spin glass model. In the latter case, the kernel equations for $\lambda_\mathrm{L}$ can be mapped to a solvable one-dimensional Schr\"{o}dinger equation \cite{Surajit22}. As a result, it can be shown by that $\lambda_\mathrm{L}\sim T^2/q_\mathrm{EA}^{3/2}$. Thus $\lambda_\mathrm{L}$ increases with increasing quantum fluctuations in the SG phase due to the reduction of EA order parameter. Similar relation between $\lambda_\mathrm{L}$ and $q_\mathrm{EA}$ can be expected for
the SY model, however, the kernel equations [Eq.\eqref{eq:kernelsg}] are much more complicated in this case, compared to the $p$-spin glass model \cite{Surajit22}, and we could not map them into a solvable quantum mechanical problem.
\begin{figure}[h!]
	\centering
    \includegraphics[width=0.95\linewidth]{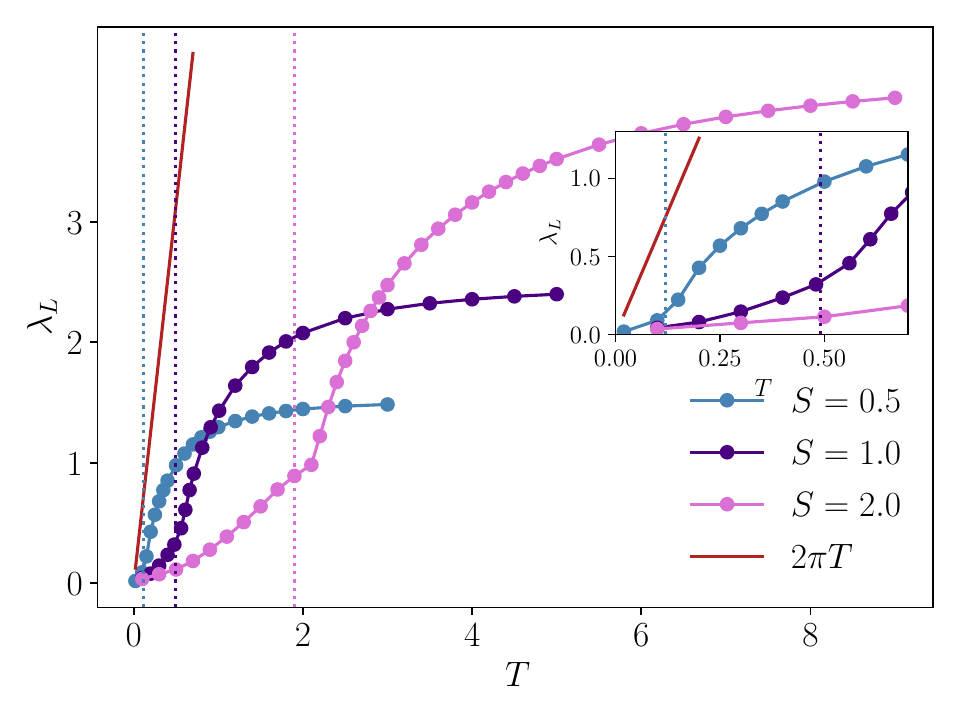}
	\caption{{\bf Temperature dependence of Lyapunov exponent in SY model.} $\lambda_\mathrm{L}$ as function of $T$ for $S=0.5$, $S=1.0$ and $S=2.0$ across the SG to PM phase transitions, marked by the vertical lines at $T_{SG}\simeq 0.12,~0.49,~1.9$, respectively.  The inset shows zoomed in view near the transitions for smaller $S$ values.
 % As discussed in the main text, for large $S$, $\lambda_\mathrm{L}$ has a power law growth, $T^\alpha$ within the SG phase while for small $S$ it grows exponentially with $T$, $\sim \exp{E_g T}$. In the PM phase, $\lambda_\mathrm{L}(T)$ initially increases as $a\ln \,{T}+c$ close to the transition, but eventually tends to saturate to a constant value at high temperature.
 }\label{fig:SY_lambdaL_T}
\end{figure}

\begin{figure}[h!]
	\centering
	\includegraphics[width=0.9\linewidth]{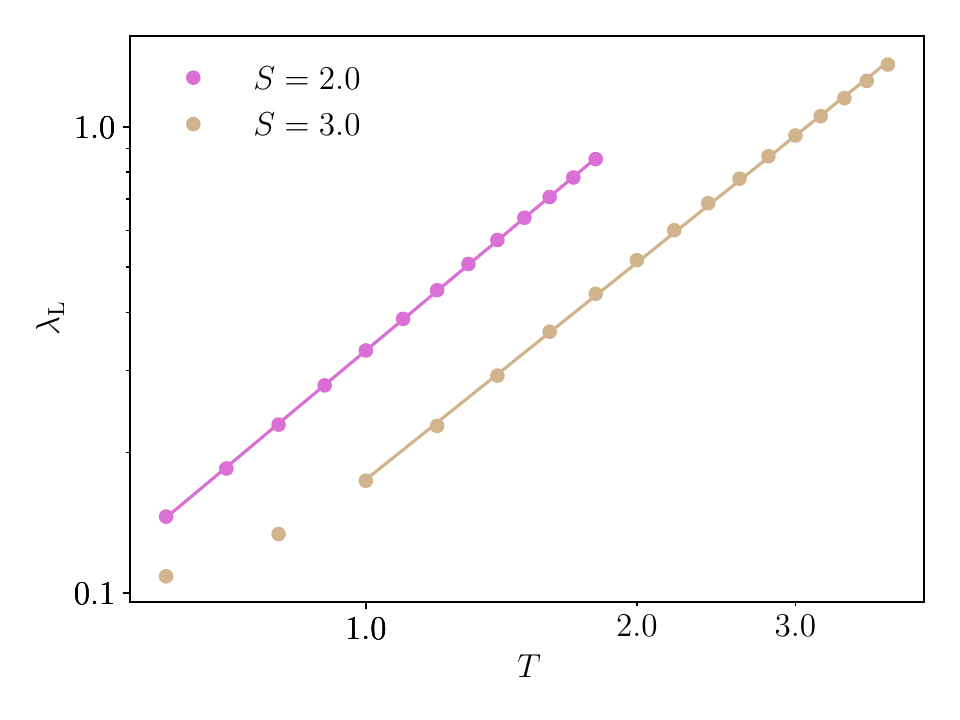}
	\caption{ {\bf Power-law temperature dependence of $\lambda_\mathrm{L}$ in the SG phase for large $S$ in SY model}. The straight lines show the power-law fit, $\lambda_\mathrm{L}\sim T^{\beta_\lambda}$ to the data (circles), on log-log plot with $\beta_\lambda=1.6,1.5$ for $S=2.0 ,3.0$, respectively.}\label{fig:lambdaLfit_T_power}
\end{figure}

\begin{figure}[h!]
	\centering
	\includegraphics[width=\linewidth]{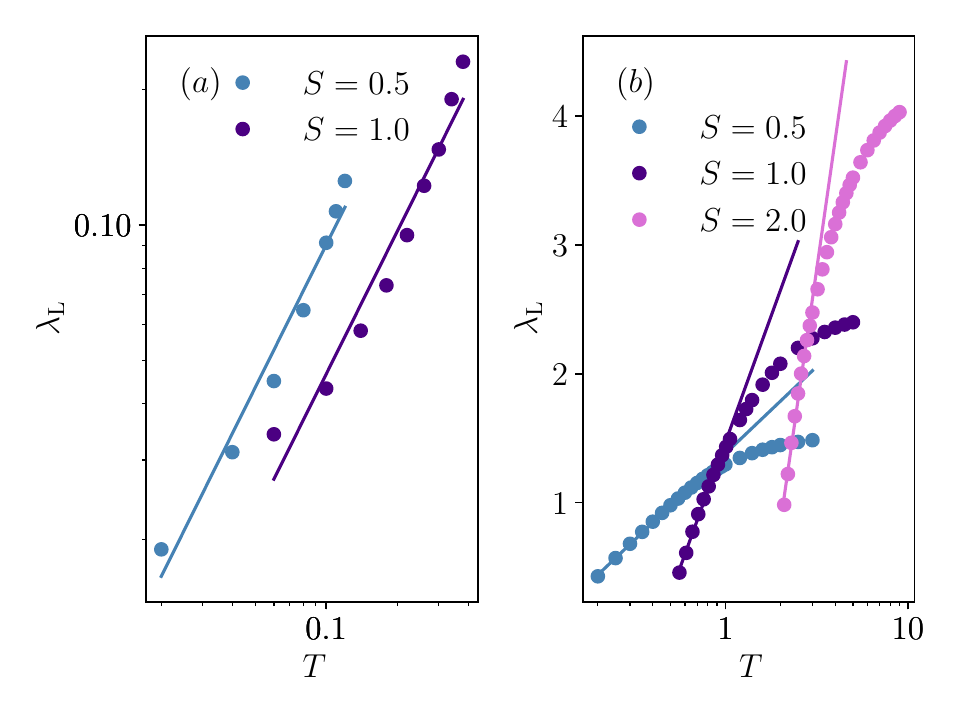}
	\caption{{\bf Temperature dependence of $\lambda_\mathrm{L}$ in PM and SG phases of SY model}. (a) $\lambda_\mathrm{L}$ increases linearly with temperature as $\lambda_\mathrm{L} \sim T$ in the SG phase for small $S$, shown for $S=0.5, 1.0$ as lines fitted to the data (circles) on log-log scale. (b) In the PM phase,  $\lambda_\mathrm{L}$ shows logarithmic dependence on temperature, $a\ln \,{T}+c$ (line) initially before it saturates to a constant value for large $T$ (Fig.\ref{fig:SY_lambdaL_T}), as shown for $S = 0.5, 1.0, 2.0$ in semi-log plot ($\lambda_\mathrm{L}$ vs. $\ln \,{T}$).}\label{fig:lambdaLfit_T}
\end{figure}

\textit{Temperature dependence of $\lambda_\mathrm{L}$.}---
We show the temperature dependence of $\lambda_\mathrm{L}$ in Fig.\ref{fig:SY_lambdaL_T} for three values of $S$, $S=0.5,~1.0,~2.0$. 
% \textcolor{magenta}{The temperature dependence of $\lambda_\mathrm{L}$ is shown in Fig.\ref{fig:SY_lambdaL_S} for three values of $S$, going from -- (1) quantum critical/spin liquid PM phase to SG at $S=0.5$, (2), (3) local moment PM phases to SG for $S=1$ (?) and $S=2$ (?), respectively}. 
In all these case, in contrast to non-monotonic dependence of $\lambda_\mathrm{L}$ on $S$, $\lambda_\mathrm{L}$ monotonically increases with temperature throughout SG and PM phases, with possibly a weak slope change at the transition [Fig.\ref{fig:SY_lambdaL_T} (inset)]. The dependence of $\lambda_\mathrm{L}$ on temperature, in SG and PM phases is summarized below. Similar to the case of $p$-spin glass \cite{Surajit22}, $\lambda_\mathrm{L}$ follows a power-law $T$ dependence in the SG phase for SY model. For large $S$, $\lambda_\mathrm{L}\sim T^{\beta_\lambda}$, where the exponent $\beta_\lambda\approx 1.5$ weakly depends on $S$, as shown in Fig.\ref{fig:lambdaLfit_T_power}. For small $S$, $\lambda_\mathrm{L}\sim T$, i.e., has a linear $T$ dependence [Fig.\ref{fig:lambdaLfit_T}(a)], albeit in a narrow range of temperature accessible for $T\lesssim T_{SG}$ for small $S$. { The deviation from linearity could be due to numerical error since $\lambda_\mathrm{L}$ is very small at low temperatures in SG phase and it is difficult to obtain an accurate estimate numerically in this regime.}
% However, $\lambda_{\mathrm{L}}$ is very small at low temperatures in SG phase and it is difficult to obtain them numerically. We think this might be the reason for  deviation from the linear behavior for smaller $T$.
% we see a surprising exponential dependence of $\lambda_\mathrm{L}$ on temperature, $\lambda_\mathrm{L}\sim \mathrm{exp}({E_g T})$ with {an effective gap $E_g=??$}, as shown in Fig.\ref{fig:lambdaLfit_T}(a), even though the marginal SG spectrum is gapless (\ref{app:SpectralFunction}). However, we note that, for small $S$,  the exponential $T$ dependence is extracted by fitting over a relatively small range of temperature $0\lesssim T\lesssim T_{SG}<\tilde{J}$ in Fig.\ref{fig:lambdaLfit_T}(a). {Nevertheless, in this range, we are not able to fit $\lambda_\mathrm{L}(T)$ with a power-law $T$ dependence}. 

In the PM phase for $T>T_{SG}$, for all values of $S$, $\lambda_\mathrm{L}$ increases near the transition logarithmically in $T$ as, $a\ln \,{T} + c$, with constants $a$ and $c$ that depend on $S$ [Fig.\ref{fig:lambdaLfit_T}(b)]. $\lambda_\mathrm{L}(T)$ eventually tends to saturate at high temperature, as shown in Fig.\ref{fig:lambdaLfit_T}. The logarithmic dependence of $\lambda_\mathrm{L}(T)$, albeit over a small temperature window near $T_{SG}$ in the PM phase, even for relatively large $S=2$, is somewhat surprising. For classical spin models~\cite{Bilitewski2018,Ruidas2021}, $\lambda_\mathrm{L}\sim \sqrt{T}$ is found at high temperature, e.g., above thermal phase transitions. The $\sqrt{T}$ behavior is expected for classical models from fairly general considerations~\cite{Kurchan2018}. In our case, the initial increase of $\lambda_\mathrm{L}(T)$ for $T\gtrsim T_{SG}$ cannot be fitted with a power law like $\sqrt{T}$. Nevertheless, qualitatively, the increase of $\lambda_\mathrm{L}$ is expected with increasing $T$ due to gradual unfreezing of spins going away from $T_{SG}$. As discussed in \ref{app:SpectralFunction}, near $T_{SG}$ in the PM phase, the spectral function has a sharp peak close to $\omega=0$ separated from high-frequency spectral weight with a pseudogap-like dip at intermediate frequencies. The sharp peak can be understood as a relic of the delta function peak $\sim \delta(\omega)$ that appears in correlation functions (see \ref{app:SpectralFunction}) in the SG phase, e.g., in the Wightmann correlation function [Eq.\eqref{eq:WightmannSG}]. The delta function peak indicates static order and spin freezing. With increasing $T$, the low- and high-frequency peaks move closer filling up the pseudogap at intermediate frequencies due to unfreezing of spins. This spectral weight transfer correlates with increase of $\lambda_\mathrm{L}$ with $T$. 
% As $T$ is increased further, the $\rho(\w)$ acquires a broad local-moment-like peak \cite{Georges2001} around, $\omega_0\sim T\ln \,[(S+1)/S]$, with almost no spectral weight at low frequency. But, since $T>\w_0$ at this elevated temperature, $\lambda_\mathrm{L}$ continues to increase.
Like in the SYK model \cite{BanerjeeAltman2016},
$\lambda_\mathrm{L}$ is expected to saturate eventually at high temperature since the model has a bounded spectrum. This is unlike the $p$-spin glass model which has an unbounded spectrum due to unbounded kinetic energy. The latter renders interaction effect negligible~\cite{Surajit22} at high temperature in the $p$-spin glass model leading to decrease of $\lambda_\mathrm{L}\sim 1/T^2$ for $T\gg \tilde{J}$.

% \begin{figure}[h!]
% 	\centering
% 	\includegraphics[width=\linewidth]{matplotlib/SY_lambdaL_T_log.pdf}
% 	\caption{ The logarithmic dependence on temperature of $\lambda_\mathrm{L}\sim \alpha \mathrm{ln} T + \beta $ is shown in the SG phase with for $S=0.5, 1.0$ }\label{fig:lambdaLfitTlog}
% \end{figure} 

\begin{figure}[h!]
	\centering
	\includegraphics[width=0.95\linewidth]{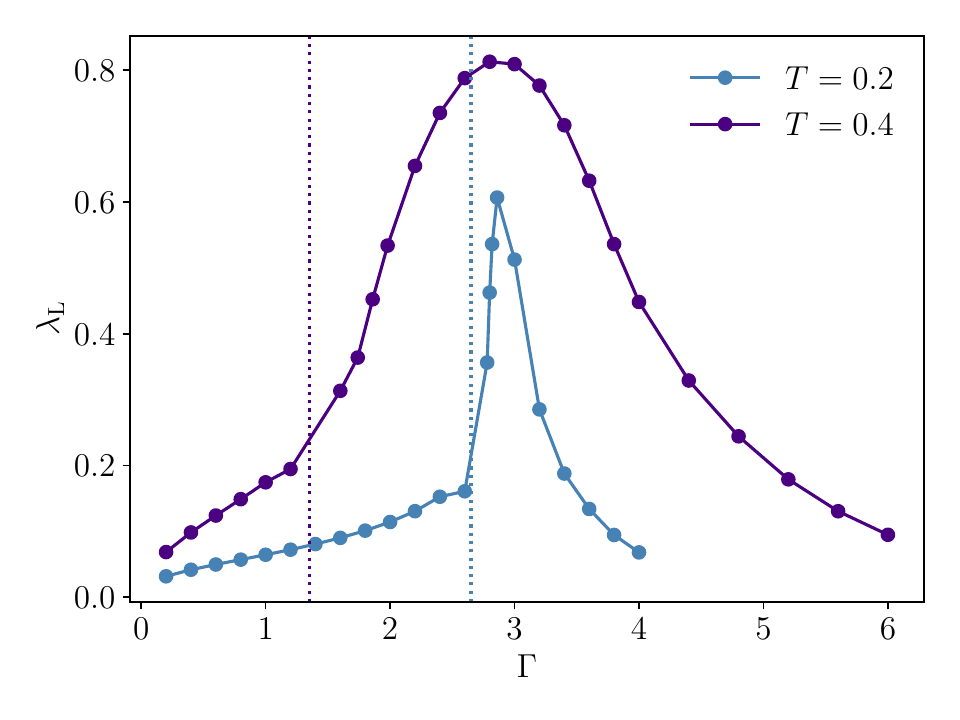}
	\caption{{\bf Variation of Lyapunov exponent with quantum parameter $\Gamma$ in the $p$-spin glass model.} $\lambda_\mathrm{L}$ as function of $\Gamma$ for $T=0.2$ and $T=0.4$ across the PM-SG phase transition marked by the vertical lines at $\Gamma_{SG}\simeq 2.65,~ 1.35$ respectively. $\lambda_\mathrm{L}$ changes non-monotonically and has a maximum above the glass transition in the PM phase~\cite{Surajit22}.}\label{fig:pspin_lambdaL_Gamma}
\end{figure}

\begin{figure}[h!]
	\centering
	\includegraphics[width=0.95\linewidth]{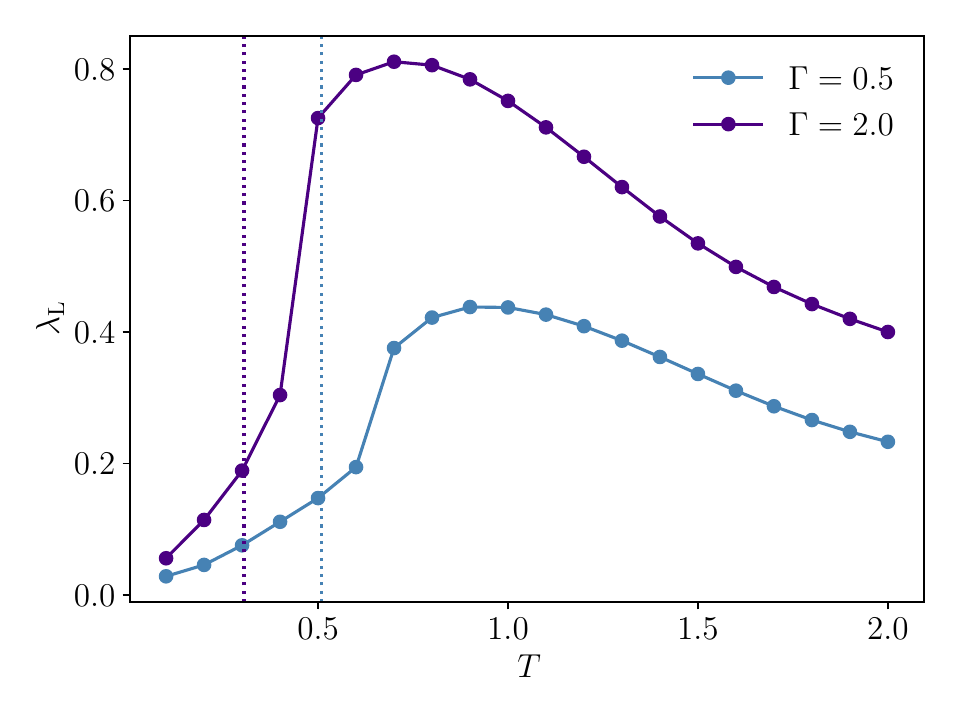}
	\caption{{\bf Temperature dependence of $\lambda_\mathrm{L}$ in the $p$-spin glass model.} $\lambda_\mathrm{L}$ as function of $T$ for $\Gamma=0.5$ and $\Gamma=2.0$ across the PM-SG phase transition marked by the vertical lines at $T_{SG}\sim 0.3, 0.51$ respectively. $\lambda_\mathrm{L}$ changes non-monotonically and has a maximum above the glass transition in the PM phase \cite{Surajit22}}\label{fig:pspin_lambdaL_T}
\end{figure}

% \begin{enumerate}
% \item {SB: We should plot similar plots for p-spin glass model, i.e. $\lambda_\mathrm{L}(\Gamma)$ for two temperatures and $\lambda_\mathrm{L}(T)$ for two gammas across the transition.}
% \item {SB: It would be good to also give results for $\lambda_\mathrm{L}(S,T)$ and $\rho(\omega)$ in the SY model for lower temperature and smaller $S$ for transition from qunatum critical region to SG, like for $T=0.05-1.0$ as a function of $S$ and for $S=0.5$ as a function of $T$. Most of the earlier works like Rozenberg et al studied this regime.}
% \end{enumerate}
 % In the SG phase, $\lambda_\mathrm{L}(S)$ decreases monotonically as $S^a$, $a\sim -1.7$ at large $S$ Fig.\ref{lambdaLfitS}. For a fixed $S$, $\lambda_\mathrm{L}$ is small due to the frozen spins in SG state.  As $T$ is increased,  $\lambda_\mathrm{L}$ increases slowly with the gradual unfreezing of spins. This leads to monotonically increasing $\lambda_\mathrm{L}$ that varies as $T^c$ as seen in Fig.\ref{lambdaLfitT} until the phase boundary with PM phase. The value of $c$ decreases for large values of $S$. In the PM phase, as $T$ is increased, the pseudogap closes and the spectrum becomes gapless, as seen in Fig. \ref{rhoGT_PM}.  As $T$ is increased further, deep in the PM phase, the $\rho(\w)$ becomes a broad peak around the same value, $w_0$ but now the $T$  $>\w_0$. Hence $\lambda_\mathrm{L}(T)$ starts to saturate and does not exhibit any maximum, as seen in Fig. \ref{lambdaL_S}. At large $T$, it saturates to a value set by the value of $J$ and $S$.

The non-monotonic behavior of {$\lambda_\mathrm{L}(S)$ [Fig.\ref{fig:SY_lambdaL_S}]} in the PM phase of the SY model is similar to the 
 non-monotonic behavior seen \cite{Surajit22} for $\lambda_\mathrm{L}(\Gamma)$ or $\lambda_\mathrm{L}(\hbar)$, i.e., as a function of quantum parameter, in the $p$-spin glass model [Eq.\eqref{eq:pSpin_1d}] as shown in Fig.\ref{fig:pspin_lambdaL_Gamma}. However, in the latter model, in contrast to the {$\lambda_\mathrm{L}(T)$ [Fig.\ref{fig:SY_lambdaL_T}]} in the SY model, $\lambda_\mathrm{L}$ also shows non-monotonic dependence on $T$ in the PM phase, as in Fig.\ref{fig:pspin_lambdaL_T}. The broad peak related to the non-monotonic $\lambda_\mathrm{L}(\Gamma,T)$ in the PM phase of the $p$-spin glass model correlates with the onset of complex glassy relaxation \cite{Surajit22}, where the dynamics start getting dominated by saddle points of the glassy energy landscape \cite{Correale2023}. In the classical limit ($\Gamma\to 0$) of the $p$-spin glass model, the origin of the broad peak was shown to be related to the interplay of the rapid increase of relaxation time approaching the glass transition, and weakening of interaction effects due to unbounded kinetic energy growth with increasing temperature. The interplay leads to a crossover from strong to weak chaos with increasing $T$, manifested as a broad peak in $\lambda_\mathrm{L}$ coinciding with the onset of two-step glassy relaxation regime \cite{Surajit22}.  {Due to its bounded spectrum}, such crossover from strong to weak chaos with temperature is absent in the SY model, and the $\lambda_\mathrm{L}(T)$ monotonically increases with $T$ in the PM phase till it saturates at { high temperature [Fig.\ref{fig:SY_lambdaL_T}]}. However, as already discussed, a non-monotonic $\lambda_\mathrm{L}(S)$ is still observed in the PM phase due to change of spectral properties as a function of $S$. 
 % For $T\simeq 1$, the spectrum become gapped or pseudogapped for both small $S$ and larger $S$, close to SG transition, with a gapless regime at intermediate $S$.
 
 % For small values of $S$ and $T\sim 1$ in units of $J$, the spectral function $\rho(\w)$ is peaked around $\w_0 \sim T ln[(S+1)/S] > T$ and has very little spectral weight at low frequencies ($\omega\sim 0$). This gapped or pseudogapped nature leads to reduced $\lambda_\mathrm{L}$ or weak chaos. As $S$ increases, $\rho(\w)$ first becomes gapless, as seen in Fig.\ref{rhoGS_PM}, and thus $\lambda_\mathrm{L}$ increases reaching a broad peak.  beyond that, close to the phase boundary, a pseudogap appears in the spectrum with a sharp peak at $\w \sim 0$ and the spectral weight spreads to higher frequencies. The appearance of peak is attributed to the formation of local moments of spins\cite{Subir2001}. Since $\lambda_\mathrm{L}$ decreases as the pseudogap deepens, it has a broad maximum in an intermediate range of $S$, as seen in Fig.\ref{lambdaL_T}. 
 
\begin{figure}[h!]
	\centering
	\includegraphics[width=\linewidth]{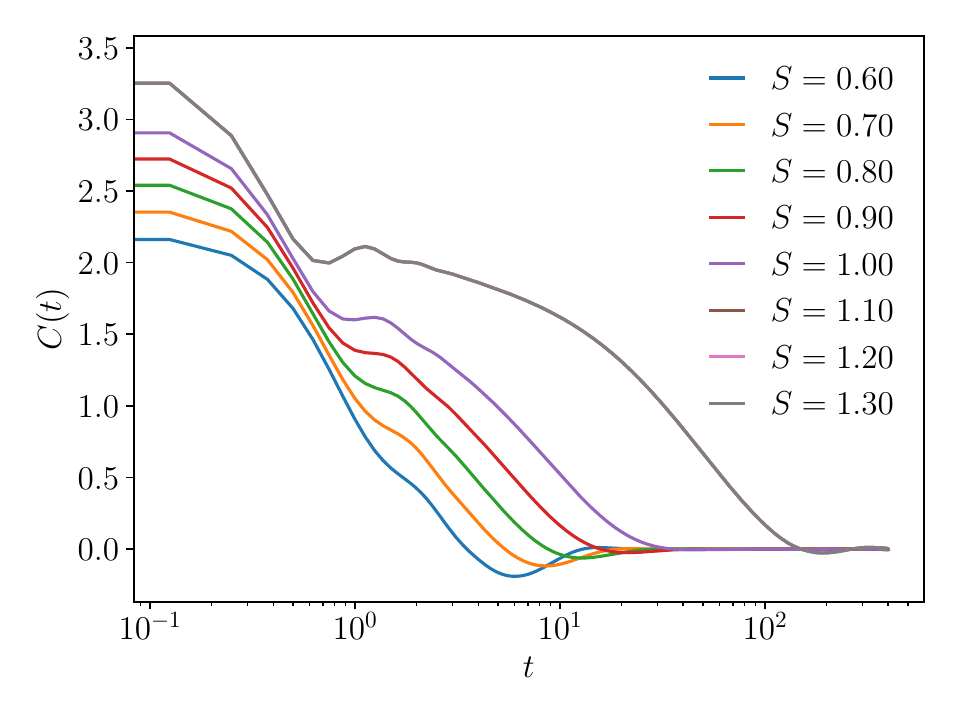}
	\caption{{\bf Onset of glassy relaxation with $S$ in SY model.} Correlation function $C(t)$ in the PM phase near the PM-SG phase transition for a fixed value $T=0.8$ and different values of $S$. As we approach the transition from the PM phase, near the transition at $S_{SG}\simeq 1.3$, the two step glassy relaxation is more pronounced.}\label{fig:CtfixedT}
\end{figure} 

%In the SG phase, $\lambda_\mathrm{L}(S)$ decreases monotonically as $S^\alpha$, $\alpha\sim 1.8$ at large S and as a function of T for a given S, $\lambda_\mathrm{L}$ increases monotonically as $T^{\beta}$, $\beta \sim 1.5$ until the phase boundary with PM phase. There is a small discontinuity in $\lambda_\mathrm{L}$ as a function of T at the phase boundary

\begin{figure}[h!]
	\centering
	\includegraphics[width=\linewidth]{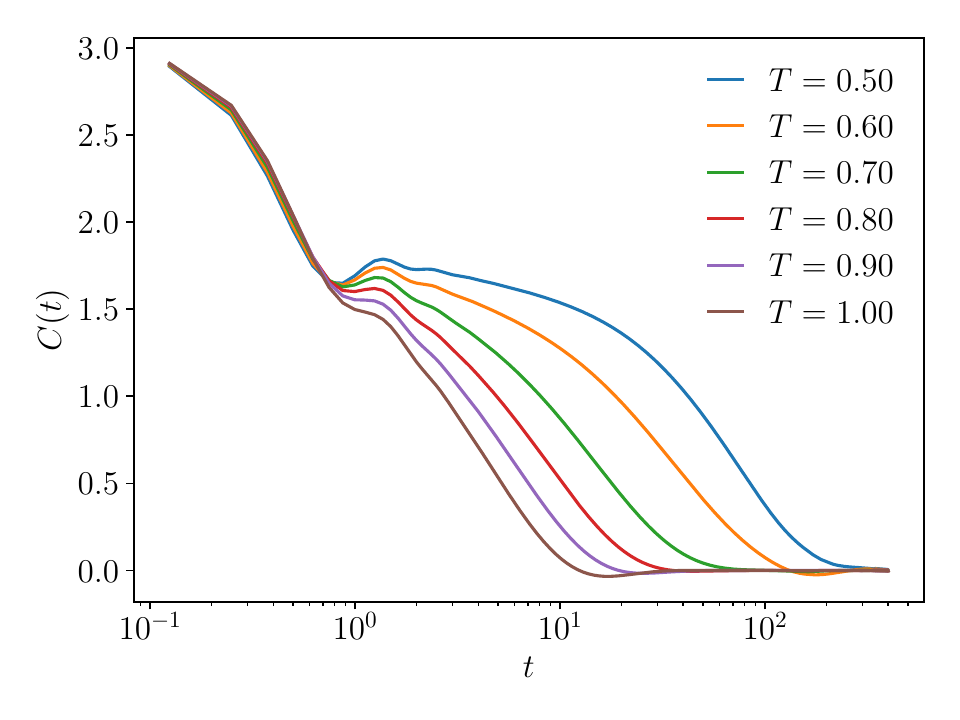}
	\caption{{\bf Onset of glassy relaxation with temperature in SY model.} Correlation function $C(t)$ in the PM phase near the PM-SG phase transition for a fixed value of $S=1.0$ and different values of $T$. The two step glassy relaxation is more pronounced as we approach $T_{SG} = 0.5$ from the PM phase.}\label{fig:CtfixedS}
\end{figure} 

The two-step glassy relaxation, seen for the $p$-spin glass model~\cite{Surajit22}, can also be observed in the PM phase of the SY model above the glass transition. To this end, we calculate the correlation function using the Fluctuation dissipation theorem {$C(t) = -(\ci/M) \sum_{\alpha}\overline{\langle\{b_{i\alpha, x}(t), b^\dagger_{i\alpha, x}(0)\} \rangle}= 2 \int (d\w/2\pi) e^{-\ci\w t}\mathrm{coth}(\beta \w/2) \mathrm{Im}\,  G_R(\w)$ }. 
% {SB: Write expression of $C(t)$ in terms of the appropriate bosoinic correlation function}. 
We show the correlation function for several $S$ in Fig .\ref{fig:CtfixedT} at $T=0.8$, and for several $T$ in Fig .\ref{fig:CtfixedS} at $S=1$, above the SG transition. The gradual onset of the two-step relaxation is evident. This suggests that the onset of two-step glassy relaxation is not a sufficient condition to realize a peak in the Lyapunov exponent.

%\begin{figure}[h!]
%\centering
%\begin{minipage}{.5\textwidth}
%  \centering
%  \includegraphics[width=1.08\linewidth]{lambdaL_Tmorediscrete}
% %\caption{$\lambda_\mathrm{L}$ as function of S for two diffrent temperatures across the PM-SG phase transition. }
%\end{minipage}%
%\begin{minipage}{.5\textwidth}
%  \centering
%  \includegraphics[width=1.08\linewidth]{lambdaL_fixedS_transition_new}
% %\caption{$\lambda_\mathrm{L}$ as function of S for two diffrent temperatures across the PM-SG phase transition.}
%\end{minipage}%
%\caption{(a)  (b)}\label{fig:lambdaL_S}
%\end{figure}

% Before we get into a spatial extended version of the above model we mention that for small values of $T$ and $S$, we notice that the numerical calculation of $\lambda_\mathrm{L}$ shows that it tends to approach the bound $2\pi T$ ($k_B = 1$ and $\hbar = 1$) as shown in the {figure of $\lambda_\mathrm{L}$ vs T\color{red}}.

\section{Butterfly velocity in quantum spin glass chains}\label{sec:ButterflyVelocity}
In this section we study the spatial spreading of chaos, first in the SY model and then in the $p$-spin glass model. 
% {\it 1D Random Heisenberg model---} We generalize the SY model from the previous section to a one dimensional model with Heisenberg interactions between nearest neighbours where at each site we have a large $N$ SY dot model with all-to-all interactions. The Hamiltonian of the 1D chain model is given by, 
% \begin{align}
% \mathcal{H}_{1D} = \frac{1}{2\sqrt{MN}}\sum_{x=0}^{L-1} \sum_{i\neq j}  J_{ij,x} S_{i\alpha \beta,x} S_{j \beta \alpha,x} \nonumber \\
% + \frac{1}{\sqrt{MN}}\sum_{x=0}^{L-1} \sum_{i,j} J^{\prime}_{ij,x} S_{i\alpha \beta,x} S_{j\beta \alpha,x+1}
% \end{align}\label{Hamiltonian1D}
% The first term represents the SY interaction at each site and the second term, the nearest neighbour Heisenberg interaction. The couplings $J_{ij,x}$ and $J^{\prime}_{ij,x}$ are drawn from Gaussian distribution, independently at each site with zero mean and variances $\overline{J_{ij}^2} = J^2$ and $\overline{J_{ij}^{\prime 2}} = J^{\prime 2}$.
% In the large $N$ limit followed by large $M$ limit, this model retains the same saddle point equations as the SY dot model, as shown in \ref{appD}, with only the coupling constant being scaled as $ J^2 + J^{\prime 2} \rightarrow J^2$. 

\begin{figure*}
	\centering
	\includegraphics[width=\linewidth]{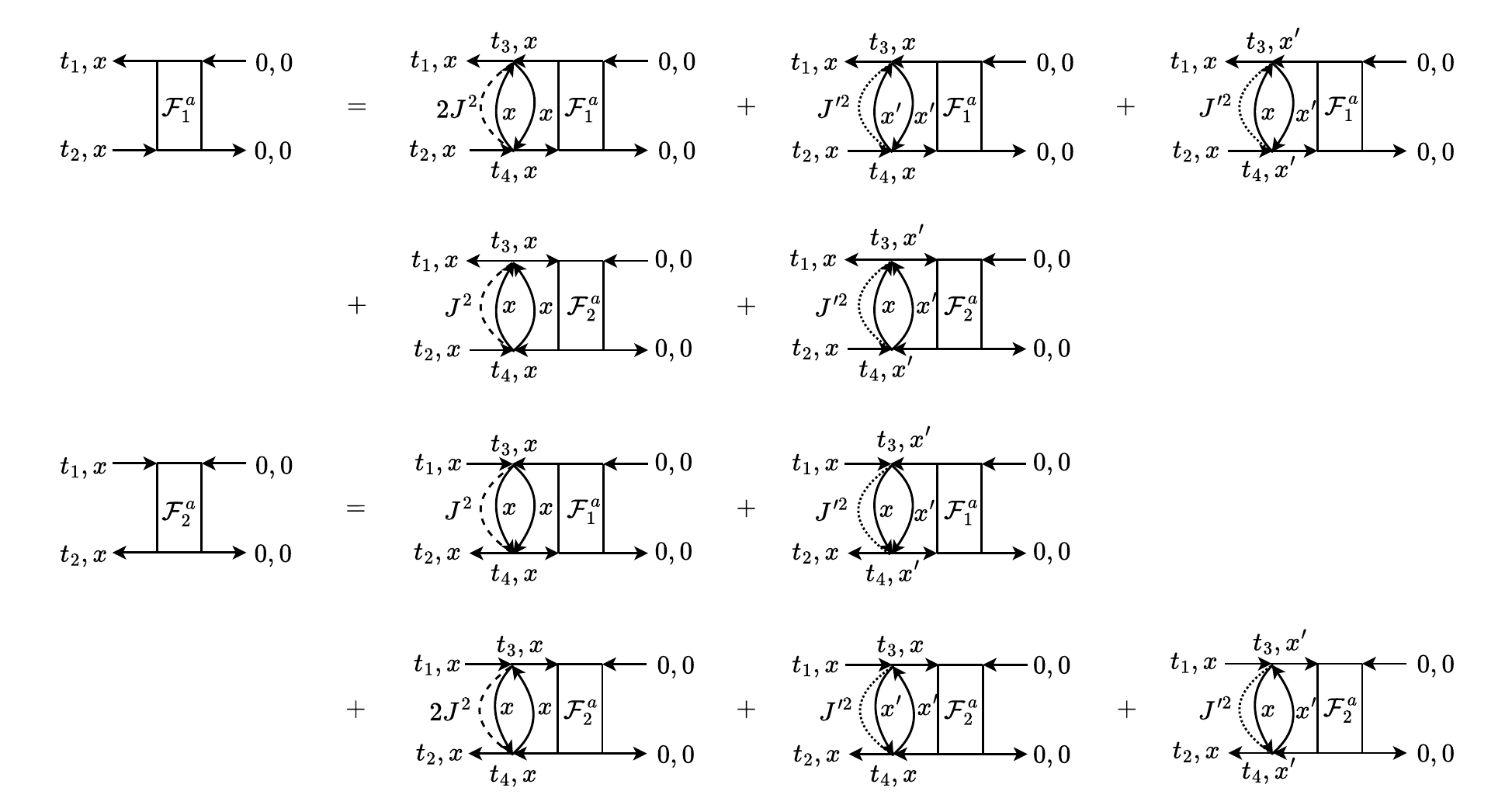}
	\caption{ Diagrammatic representation of the Kernel equation in Eq.\ref{eq:BetheSalpeter_1D} for $\mathcal{O}(1/(NM))$ term, $\mathcal{F}_{\mu,x}^a(t_1, t_2)$ in the OTOC $F_{\mu,x}^a(t_1, t_2)$ ($\mu=1,2$). The solid horizontal lines denote dressed retarded and advanced propagators $G_R$, $G_A$, and the vertical rung denotes the Wightmann correlations $G_{lr}^+$ and $G_{lr}^-$. The dashed and dotted lines denote disorder averaging over $J$ and $J'$ couplings, respectively in Eq.\eqref{eq:SY_1D}. Here $x^{\prime} = x \pm 1$. We suppress replica indices for the vertices to avoid cluttering.}\label{fig:kernel1D}
     % The third term in the first row and the fourth row contribute to the propagation of chaos between nearest neighbours.
\end{figure*} 

\subsection{Butterfly velocity across spin glass transition in the SY chain}\label{sec:SY_vB}
In the previous section, we calculated $\lambda_\mathrm{L}$ which determines the temporal growth of the four-point OTOC. Due to the spatial locality of our generalized 1D SY chain model [Eq.\eqref{eq:SY_1D}], we can study the spatio-temporal dynamics of growth and spread of chaos. The quantity which measures the propensity of spatial spreading of chaos is the butterfly velocity, $v_B$. The operators that form large commutators with an initial operator lie inside a region of space-time that forms a \emph{light cone}\cite{Gu2017,Khemani2018}, signifying a ballistic spread of chaos. The speed or velocity that {forms} the boundary of the light cone is defined as $v_B$. Like the Lyapunov exponent $\lambda_\mathrm{L}$, the butterfly speed $v_B$ provides a state-dependent measure of operator growth and spreading, and can be related~\cite{Roberts2016} to the well-known Lieb-Robinson velocity bound~\cite{Lieb1972}, that gives state-independent measure in terms of operator norm. To this end, we calculate the following disorder-averaged regularized spatio-temporal OTOCs for the SY model, 
\begin{subequations}\label{eq:OTOC1D}
\begin{align}
&F_{1, x}(t_1, t_2) \nonumber \\
&= \frac{1}{N^2 M^2}\sum_{ij\alpha\beta} \overline{\mathrm{Tr}[y b^{\dagger }_{i\alpha,x }(t_1)yb^{\dagger}_{j\beta,0}(0)y b_{i \alpha, x}(t_2) y b_{j \beta,0}(0)]} \\
&F_{2, x}(t_1, t_2) \nonumber \\
&= \frac{1}{N^2M^2}\sum_{ij\alpha\beta} \overline{\mathrm{Tr}[y b_{i\alpha,x}(t_1)y b^{\dagger}_{j\beta,0}(0)y b_{i \alpha,x}^{\dagger }(t_2)y b_{j\beta,0}(0)]}
\end{align}
\end{subequations}
As discussed in Sec.\ref{sec:LyapunovExponent} for the case of the onsite OTOCs, which is equivalent to OTOCs in zero-dimensional SY model, the above non-local OTOCs are computed via a replicated Keldysh field theory, where the OTOCs become replica diagonal up to $\mathcal{O}(1/(NM))$ diagrams. The latter are arranged as ladder series that can be written in the form of Bethe-Salpeter-like equations, as shown in Fig:\ref{fig:kernel1D},
\begin{subequations} \label{eq:BetheSalpeter_1D}
\begin{align}
&\mathcal{F}_{1,x}^{a}(t_1, t_2) = \int d x^\prime \int dt_3 dt_4 \bigg[ K_{11, xx^\prime}^{a}(t_1, t_2, t_3, t_4) \nonumber \\
&~\mathcal{F}_{1,x^\prime}^{a}(t_3, t_4)+  K_{12, x x^\prime}^{a}(t_1, t_2, t_3, t_4)\mathcal{F}_{2, x^\prime}^{a}(t_3, t_4)\bigg], \\
&\mathcal{F}_{2,x}^{a}(t_1, t_2) = \int d x^\prime \int dt_3 dt_4 \bigg[ K_{21, x x^\prime}^{a}(t_1, t_2, t_3, t_4) \nonumber \\
&~\mathcal{F}_{1,x'}^{a}(t_3, t_4)+ K_{22, x x^\prime}^{a}(t_1, t_2, t_3, t_4)\mathcal{F}_{2,x'}^{a}(t_3, t_4) \bigg].
\end{align}
\end{subequations}
In the above, the kernels, which are diagrammatically shown in Fig.\ref{fig:kernel1D}, are given in Eqs.\eqref{eq:Kernel_space}. Although we have not indicated explicitly, all the Green's functions appearing in the kernel elements are replica diagonal. Since the saddle-point Green's function (Sec.\ref{sec:SY_PhaseDiagram}) of the extended 1D SY chain model are local and independent of $x$, the kernel elements are translation invariant, as in Eqs.\eqref{eq:Kernel_space}. Thus, after Fourier transforming Eqs.\eqref{eq:BetheSalpeter_1D} to momentum space, we obtain equations for the Fourier transform of $\mathcal{F}^a_{\mu,x}(t_1,t_2)$ ($\mu=1,2$), $\mathcal{F}^a_{\mu,p}(t_1,t_2)$. The kernel elements for a given momentum $p$ are obtained by replacing $\tilde{J}^2$ in the local kernel elements of Eqs.\eqref{eq:KernelSY_Local} by momentum dependent prefactors, as shown in Eqs.\eqref{eq:Kernel_momentum}, where $\alpha = 2 J^{\prime 2}/\tilde{J}^2$. In this section, as discussed below, we calculate butterfly velocity $v_B$ using two methods --(1) a fully numerical method, which uses a straightforward generalization of the exponential growth ansatz for the local OTOC (Sec.\ref{sec:LyapunovExponent}) for each momentum mode; we refer to this as \emph{real $p$ method}, and (2) a semi-analytic method, called the \emph{single mode ansatz}, following Ref. \onlinecite{Gu2017}. Here we refer to this as \emph{imaginary $p$ method}. 
\begin{widetext}
\begin{subequations}\label{eq:Kernel_space}
\begin{align}
&K_{11, x x^\prime}^a(t_1, t_2; t_3, t_4) = G_{A}(t_{31})  G_{R}(t_{24}) G_{lr}^+(t_{43}) G_{lr}^-(t_{34}) \Big(2J^2\delta_{x, x^\prime} + J^{\prime 2}(\delta_{x, x^\prime} + \delta_{x\pm 1, x^\prime})\Big) \\
&K_{12,x x^\prime}^a(t_1, t_2; t_3, t_4) =  G_{A}(t_{31})  G_{R}(t_{24}) G_{lr}^+(t_{43}) G_{lr}^+(t_{43}) \Big(J^2\delta_{x, x^\prime} + J^{\prime 2}( \delta_{x\pm 1, x^\prime})\Big) \\
&K_{21,x x^\prime}^a(t_1, t_2; t_3, t_4) = G_{R}(t_{13})  G_{A}(t_{42}) G_{lr}^-(t_{34}) G_{lr}^-(t_{34}) \Big(J^2\delta_{x, x^\prime} + J^{\prime 2}( \delta_{x\pm 1, x^\prime})\Big) \\
&K_{22,x x^\prime}^a(t_1, t_2; t_3, t_4) = G_{R}(t_{13})  G_{A}(t_{42}) G_{lr}^-(t_{34}) G_{lr}^+(t_{43})\Big(2J^2\delta_{x, x^\prime} + J^{\prime 2}(\delta_{x, x^\prime} + \delta_{x\pm 1, x^\prime})\Big)
\end{align}
\end{subequations}
\begin{subequations}\label{eq:Kernel_momentum}
\begin{align}
K_{11, p}^a(t_1, t_2; t_3, t_4) &= \tilde{J}^2 {\Big (2 -\frac{\alpha}{2}(3-2\,\mathrm{cos }\,p)\Big)} G_{A}(t_{31})  G_{R}(t_{24}) G_{lr}^+(t_{43}) G_{lr}^-(t_{34}) \\
K_{12,p}^a(t_1, t_2; t_3, t_4) &=  \tilde{J}^2  {\Big(1 - \alpha(1-\,\mathrm{cos}\,p)\Big)} G_{A}(t_{31})  G_{R}(t_{24}) G_{lr}^+(t_{43}) G_{lr}^+(t_{43})  \\
K_{21,p}^a(t_1, t_2; t_3, t_4) &=  \tilde{J}^2  {\Big(1 - \alpha(1-\,\mathrm{cos}\,p)\Big)} G_{R}(t_{13})  G_{A}(t_{42}) G_{lr}^-(t_{34}) G_{lr}^-(t_{34}) \\
K_{22,p}^a(t_1, t_2; t_3, t_4) &= \tilde{J}^2 {\Big(2 - \frac{\alpha}{2}(3-2\,\mathrm{cos}\,p)\Big)} G_{R}(t_{13})  G_{A}(t_{42}) G_{lr}^-(t_{34}) G_{lr}^+(t_{43}) 
\end{align}
\end{subequations}
\end{widetext}

% A calculation similar to the dot model Eq.\ref{KernelSY} leads to kernels in  Eq.\ref{eq. Kernel_space} which can be diagrammatically represented as shown in Fig.\ref{fig:kernel1D}. Although we have not shown explicitly, all the Green's functions in the kernel elements are replica diagonal. Since the saddle point solution, $G_{x}(t)$ is independent of $x$, the kernel elements are modified from their dot counter parts by only the $x$ dependent coefficients. After Fourier transforming to momentum space, this gives Eq.\ref{eq. Kernel_momentum} where we have used, $J^2 + J^{\prime 2}  \rightarrow  J^2$ and $\alpha = J^{\prime 2}/J^2$.  In this section we calculate butterfly velocity $v_B$ using two methods. 

\begin{figure}[h!]
	\centering
	\includegraphics[width=0.95\linewidth]{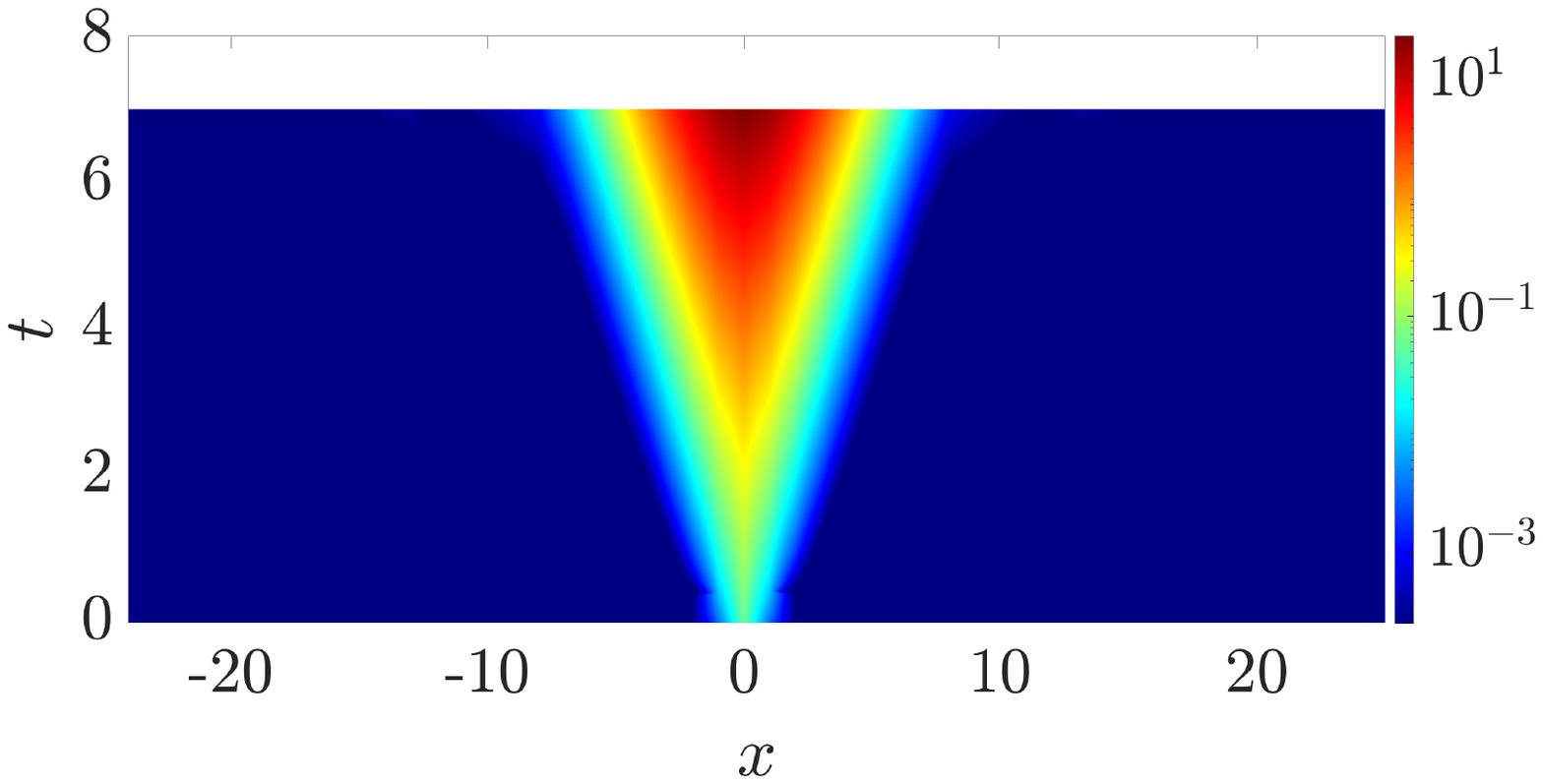}
	\caption{The growing part of OTOC, $\mathcal{F}_{1,x}(t)$, is plotted on the $x-t$ plane in logarithmic scale, as indicated in the color bar at $T=0.8$ and $S=1.0$ using the real $p$ method.}\label{fig:lightcone_color}
\end{figure}

\begin{figure}[h!]
	\centering
	\includegraphics[width=0.95\linewidth]{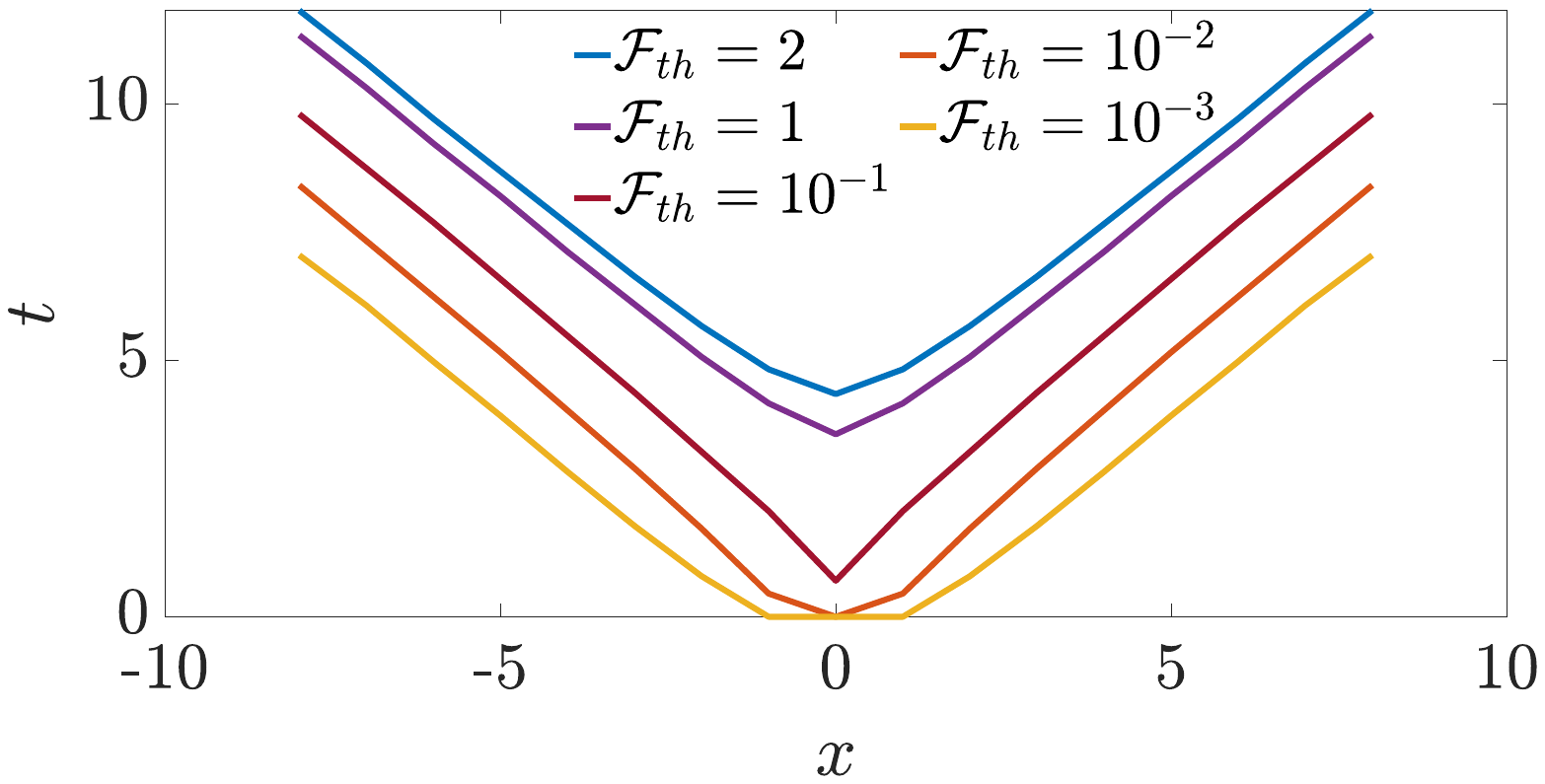}
	\caption{ Light cones at $T=0.8$ and $S=1.0$ for different values of $\mathcal{F}_{th}$ obtained using the real $p$ method. The same butterfly speed $v_B$ is obtained for a wide range of $\mathcal{F}_{th}$.}\label{fig:lightcone}
\end{figure}

{\it Real $p$ method.}--- As in Sec.\ref{sec:LyapunovExponent}, we take an exponential growth ansatz, $\mathcal{F}_{\mu,p}(t_1,t_2)= e^{\lambda_\mathrm{L}(p)(t_1+t_2)/2}f_{\mu,p}^a(t_1-t_2)$, and extract the momentum-dependent Lyapunov exponent $\lambda_\mathrm{L}(p)$ by demanding that the matrix kernel has at least one eigenvalue equal to 1 {(typically non-degenerate)}. We compute the eigenvector corresponding to the eigenvalue 1 to obtain $f_{\mu,p}^a(\omega)$. To extract $v_B$, we calculate $\mathcal{F}^a_{\mu, p}(t,t) = e^{\lambda_\mathrm{L}(p)t}f^a_{\mu, p}(0)$, where $f_{\mu, p}(0) = \int(d\omega/2\pi) f_{\mu, p}(\omega)$. We Fourier transform to the position space, $\mathcal{F}_{\mu, x}(t)\equiv \mathcal{F}^a_{\mu,x}(t,t) = (1/L)\sum_{p}e^{\ci x p} \mathcal{F}_{ \mu, p}(t,t)$ where $L$ is the number of sites or the number of discrete $p$ values taken within the Brillouin zone $[-\pi,\pi]$ for unit lattice spacing. We show in Fig.\ref{fig:lightcone_color} that $\mathcal{F}_{1,x}(t)$ exhibits quite distinctive light cone. The magnitude of $\mathcal{F}_{\mu,p}(t,t)$ cannot be fixed, i.e., the eigenvector can be multiplied by an arbitrary factor, for the kernel equations. Thus, we compute the light-cones using the locus of $(x,t)$ points corresponding to $\mathcal{F}_{1,x}(t) = \mathcal{F}_{th}$, for various threshold values $\mathcal{F}_{th}$, for which $t\propto x$. We find that this ballistic feature is violated for too small or too large values of $\mathcal{F}_{th}$. We extract $v_B(\mathcal{F}_{th})$ from the resulting light cones and find that $v_B(\mathcal{F}_{th})$ converges approximately to the same value $v_B$ over a wide range of $\mathcal{F}_{th}$, as shown in Fig.\ref{fig:lightcone}. We discuss the results for $v_B(S,T)$ obtained from the real $p$ method below along with that from the imaginary $p$ method.
% For too small or too large values of $F_{th}$ we noticed that the wave-front is not ballistic and the exponential ansatz of the OTOC doesn't apply\ref{fig:lightcone}

{\it Imaginary $p$ method.}--- To calculate butterfly velocity through this method,  we use the single-mode ansatz for OTOC and the results obtained by Gu and Kitaev \cite{Gu:2018jsv}. Using the single mode ansatz \cite{Kitaev:2017awl}  for early time OTOC, they proved a ladder identity which implies the existence of a pole and a saddle at two imaginary values of momentum that contribute to the growth of OTOC. We refer to Ref. \onlinecite{Gu:2018jsv} for the proof of the ladder identity and the details of the method. Using this ansatz we can write the solution of Eqs.\eqref{eq:BetheSalpeter_1D} for the 1D SY chain model [Eq.\eqref{eq:SY_1D}],
 \begin{align}\label{eq:SinglemodeAnsatz}
\mathcal{F}_{\mu, p}^a(t_1, t_2)\approx \frac{e^{\lambda_\mathrm{L}(p)(t_1+t_2)/2}}{C(p)} \Upsilon_{\mu, p}^R(t_{12}) \Upsilon_{\mu, p}^A(0)
\end{align}
where $\Upsilon_{\mu,p}^R(t)$ and $\Upsilon_{\mu,p}^A(t)$ are the retarded and advanced vertex functions~\cite{Gu:2018jsv} which modify the OTOC by an overall magnitude. The dominant $p$ dependence of the above comes from a pole $\lambda_\mathrm{L}=2\pi T$ ($k_\mathrm{B}=1$, $\hbar=1$) in $C(p)$, since $C(p)\propto \cos{[\lambda_\mathrm{L}(p)/4T]}$ \cite{Gu:2018jsv}.  
% is the inverse propagator of scramblon, which results in exponential growth rate of OTOC at a rate of $\lambda_\mathrm{L}(p)$. It contains the factor $\mathrm{cos}(\lambda_\mathrm{L}(p)/4T)$ and so has a pole at $\lambda_\mathrm{L}(p) = 2\pi T$($k_B = \hbar = 1$) 
This pole leads to the maximal chaos for non-Fermi liquids in SYK and related models \cite{Gu2017,Gu:2018jsv,Guo2019}. 
% We review this derivation as applicable to our extended spin glass model in \ref{appF}.  
The asymptotic form of Eq.\eqref{eq:SinglemodeAnsatz} ignores the initial correlations and any non-linear effects \cite{Gu:2018jsv,Kitaev:2017awl} and is valid only at the butterfly or chaos front, i.e., $1/N \ll \mathcal{F}_{\mu, p}(t_1, t_2)/N\ll 1$. Fourier transforming Eq.\eqref{eq:SinglemodeAnsatz} back to real space, we obtain
\begin{align}\label{eq:OTOC_vB}
    \mathcal{F}_{\mu,x}(t)\equiv \mathcal{F}^a_{\mu,x}(t,t)\approx \int_{-\infty}^\infty \frac{dp}{2\pi}\frac{e^{\lambda_\mathrm{L}(p) t + \ci p x}}{\cos(\lambda_\mathrm{L}(p)/4T)}f(p),
\end{align}
where $f(p)$, which depends on the kernel [Eqs.\eqref{eq:Kernel_momentum}] and vertex functions in Eq.\eqref{eq:SinglemodeAnsatz}, is assumed to be analytical function of $p$ and non vanishing in the momentum regime of interest~\cite{Gu:2018jsv}. 
%Dropping the index $\mu=1,2$ for convenience and transforming back to the position space we obtain, 
% \begin{align}
% F_{\mu, x}(t_1, t_2) \sim \frac{1}{N}u(x, t) \int_{k, k^\prime}  \Upsilon_{\mu, p}^{R}(t_{12})\Upsilon_{\mu, p}^{A}(0, k^\prime)\\
% \text{where } t = (t_1+t_2)/2 \text{ and } u(x, t) = \int_p \frac{e^{\lambda_\mathrm{L}(p) t + i p x}}{\mathrm{cos}(\lambda_\mathrm{L}(p)/4T)}
% \end{align}

As in the real $p$ method discussed above, we choose a fixed, albeit arbitrary, value of $\mathcal{F}_{\mu,x}(t,t)\approx 1$, which is at the lower end of the butterfly front~\cite{Gu:2018jsv}, to obtain the light cone from the resulting locus of $(x,t)$ points and extract $v_B$. For large $x$ and $t$, the integral in Eq.\eqref{eq:OTOC_vB} can be estimated by saddle-point method. The saddle-point $p_s$ for the integrand is obtained from $\lambda'_\mathrm{L}(p_s)t+\ci x=0$, where $\lambda_\mathrm{L}'(p)$ is the derivative. This together with the condition $\mathcal{F}_{\mu,x}(t,t)\approx 1$, implying $\lambda_\mathrm{L}(p)t+\ci px=0$, leads to the following equation for $p_s$
\begin{align}
    \lambda'_\mathrm{L}(p_s)=\frac{\lambda_\mathrm{L}(p_s)}{p_s},
\end{align}
and a corresponding velocity 
\begin{align}
v_s=\left(\frac{x}{t}\right)_s=\ci \lambda'_\mathrm{L}(p_s).
\end{align}
The saddle point leads to a contribution $\mathcal{F}_{\mu,x}^s(t)\sim \exp{[\lambda_\mathrm{L}(p_s)t+\ci p_s x]}$. Similarly, the pole $p_1$ of the integrand in Eq.\eqref{eq:OTOC_vB}, such that $\lambda_\mathrm{L}(p_1)=2\pi T$, gives a contribution $\mathcal{F}_{\mu,x}^p(t)\sim \exp{[2\pi T t+\ci p_1 x]}$ with an associated velocity 
\begin{align}
v_m=\ci \frac{2\pi T}{p_1}.
\end{align}
To give rise to a well defined light cone, $\mathcal{F}_{\mu,x}^s(t)$ ($\mathcal{F}_{\mu,x}^p(t)$) should be non-oscillatory in space and time, implying that $p_s$ ($p_1$) is purely imaginary, i.e., $p_s=\ci|p_s|$ ($p_1=\ci|p_1|$), and $\lambda_\mathrm{L}(\ci|p_s|)$ is purely real. In \ref{app:vB}, we show that this is indeed the case by solving for $\lambda_\mathrm{L}(p)$ for purely imaginary $p=\ci|p|$ from Eqs.\eqref{eq:BetheSalpeter_1D} using the exponential growth ansatz, as discussed earlier for the real $p$ method. $\lambda_\mathrm{L}(\ci|p|)$ is found to be a convex function of $|p|$ as expected~\cite{Gu:2018jsv}. The light cone and the associated $v_B$ are determined by either the saddle-point or the pole contribution to $\mathcal{F}_{\mu,x}(t)$ depending on whether $|p_s|<|p_1|$ or $|p_s|>|p_1|$, respectively. The conditions also translate to $v_s(\equiv v(p_s))<v_1(\equiv v(p_1))$ or $v_s>v_1$, respectively, where the velocity $v(p)=\ci \lambda'_\mathrm{L}(p)$, since $\lambda_\mathrm{L}(\ci|p|)$ is a convex (up) function of $p$ (\ref{app:vB}). 

\emph{For the condition $|p_s|<|p_1|$}, the deformed contour, that goes through the saddle point $\ci|p_s|$ on the imaginary axis in the evaluation of the integral in Eq.\eqref{eq:OTOC_vB}, does not enclose the pole and hence the latter does not contribute.  As discussed in refs.\onlinecite{Gu:2018jsv,Guo2019}, this condition is realized when $\lambda_\mathrm{L}(0)\ll 2\pi T$, i.e., when the chaos is far away from maximal. In this case $\mathcal{F}_{\mu,x}(t)\approx \mathcal{F}^s_{\mu,x}(t)$ and $v_B=v_s$. By making a small $p$ approximation, $\lambda_\mathrm{L}(p)\simeq \lambda_\mathrm{L}(0)-\tilde{D}p^2$ (with $\tilde{D}>0$), we obtain $v_B=v_s=[4\tilde{D}\lambda_\mathrm{L}(0)]^{1/2}$ and
\begin{align}
\mathcal{F}_{\mu,x}(t)\sim e^{\lambda_\mathrm{L}(0)t(1-x^2/v_B^2t^2)}
\end{align}
Here the quantity $\tilde{D}=v_B^2/4\lambda_\mathrm{L}(0)$ has the dimension of diffusion constant. For certain holographic theories \cite{Blake2016} and strongly interacting diffusive metals built from SYK-type quantum dots \cite{Gu2017,Guo2019}, $4\tilde{D}$ has been found to be exactly equal to energy diffusion constant at low temperature. $\tilde{D}$ has also been found to be same as spin diffusion constant at high temperature in a classical spin liquid~\cite{Bilitewski2018}, and even for classical XY model~\cite{Ruidas2021}. Assuming that such relation between $\tilde{D}$ and actual diffusion constant hold, at least, over a limited range of $(S,T)$ in the phased diagram of SY chain, we use $\tilde{D}$ as a proxy for diffusion constant and plot it as a function of $S$ and $T$ below.

\emph{For the other condition $|p_s|>|p_1|$}, realized for $\lambda_\mathrm{L}(0)\approx 2\pi T$, the pole contributes and dominates over the saddle-point contribution, such that 
\begin{align}
\mathcal{F}_{\mu,x}(t)\approx \mathcal{F}^p_{\mu,x}(t)\sim e^{2\pi T t(1-|x|/v_B t)} 
\end{align}
and $v_B=v_m$, as can be deduced from $\mathcal{F}_{\mu,x}(t)\approx 1$. The above corresponds to maximal chaos, realized for SYK-type non-Fermi liquids~\cite{Gu2017,Gu:2018jsv,Guo2019}.

For most part of the phase diagram [Fig.\ref{fig:phasediagramSY}] we have studied, we see that saddle point, $|p_s|$ remains well below the pole, $|p_1|$ and so the OTOC receives dominant contribution from the saddle point. See \ref{app:vB} for more details.
Therefore the chaos front travels with a speed $v_B=\ci \lambda_\mathrm{L}^\prime(\ci |p_s|)$ and the Lyapunov exponent $\lambda_\mathrm{L}$ is non-maximal, as was verified through direct calculation of $\lambda_\mathrm{L}(S,T)$ in Sec.\ref{sec:LyapunovExponent}. The SY chain can be analytically shown to be maximally chaotic in the limit $S\to0,~T\to 0$ following the same methods \cite{KitaevKITP,Maldacena2016,BanerjeeAltman2016} used for calculating $\lambda_\mathrm{L}(T)$ in the SYK model. 
% similar to approaching  that the $v_B$ reaches a value close to the  $2\pi T$ as $T\rightarrow 0$ in PM phase, but we don't explore this point in any detail in this paper.
\begin{figure}[h!]
	\centering
    \includegraphics[width=\linewidth]{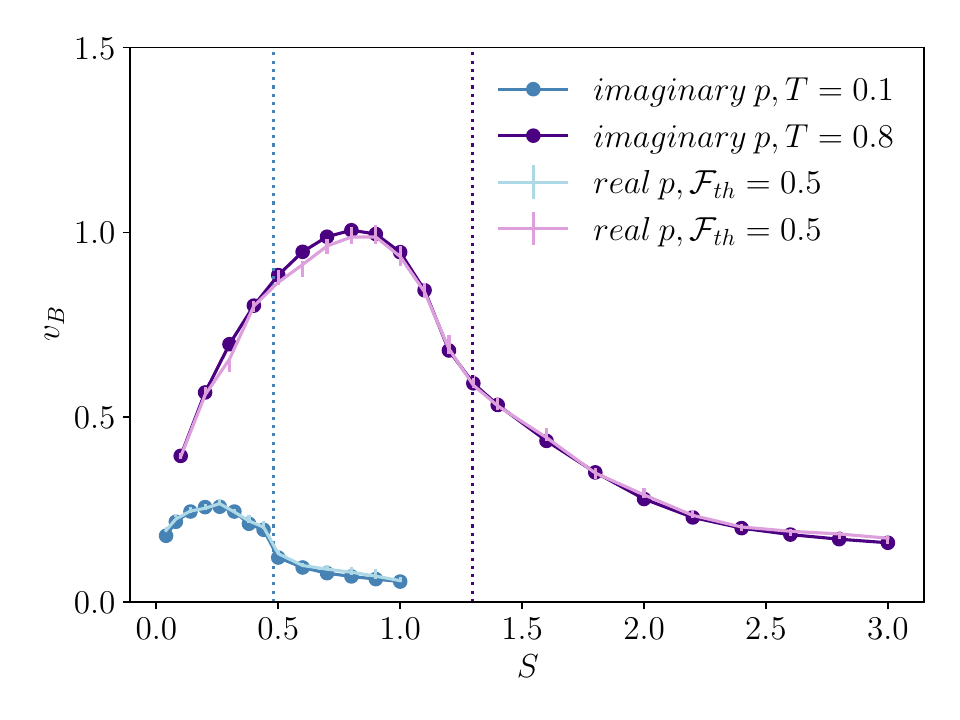}
	\caption{ {\bf Variation of butterfly velocity with $S$ in the SY model.} $v_B$ across the SG phase transition ($S_{SG}$, vertical dashed lines) as a function of $S$ at $T = 0.1, 0.8$ (in units of $\tilde{J}=1$) and $\alpha= 0.25$, calculated using both real and imaginary $p$ methods, as indicated in the legends. {For the real $p$ method, we have used $\mathcal{F}_{th} = 0.5$ with range (error bars)  $\mathcal{F}_{th}=0.1-1.0$}, for a lattice of $L=50$ sites.}\label{fig:SY_vB_S}
\end{figure}

\begin{figure}[h!]
	\centering
    \includegraphics[width=\linewidth]{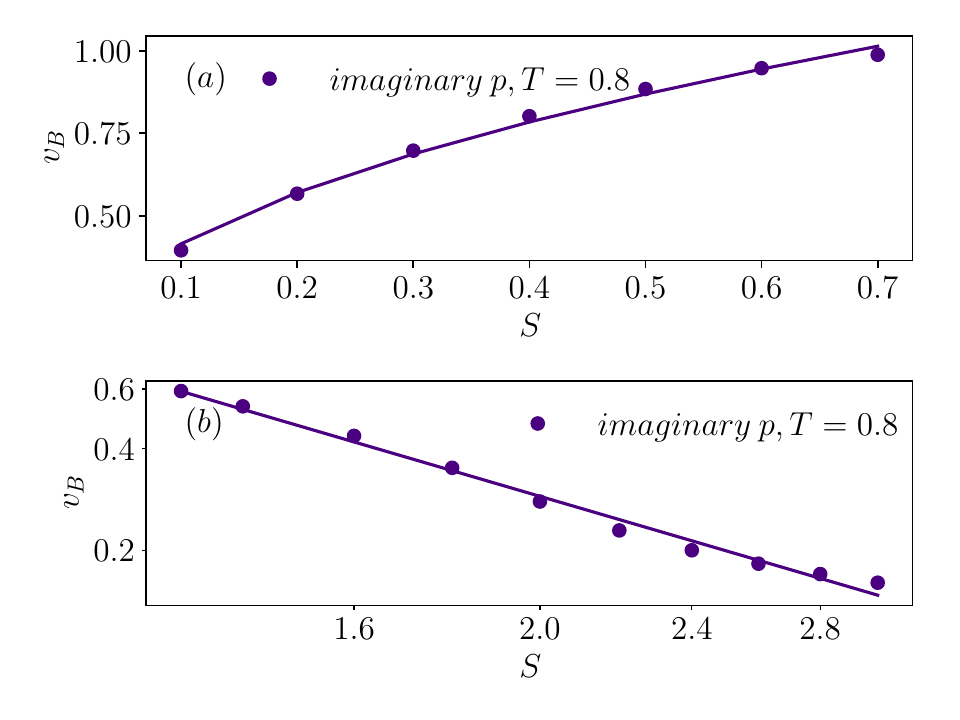}
	\caption{ {\bf $S$ dependence of $v_B$ in the PM and SG phases of SY model}. (a) The data (circles) for $v_B(S,T)$ in the PM phase, before it reaches the maximum [Fig.\ref{fig:SY_vB_S}], is fitted with $v_B(0,T)+a S^{\alpha_v}$ (line) for $T=0.8$ with {$\alpha_v\approx 0.46$}. 
 % $(\alpha_v, v_B(0,T))$ are obtained as ($0.48, 0$)
 (b) {Power law dependence of $v_B$ on $S$ in the SG phase is shown as straight line on a log-log plot. $v_B \sim 1/S^{1.7}$ for $T=0.8$.} } \label{fig:SY_vB_Sfit}
\end{figure}

We calculate $v_B$ as a function of $T$ and $S$, across the SG-PM phase transition, for $S=0.5,1.0$ and $T=0.1,0.8$ along the $a,~b$ and $d,~e$ cuts, respectively, shown in Fig.\ref{fig:phasediagramSY}, using both real and imaginary $p$ methods. 
% the methods discussed above, as shown in Fig.\ref{fig:SY_vB_T} and Fig.\ref{fig:SY_vB_S}  and 
As discussed below, we find that the results from the two methods match closely, validating the particular form of single-mode ansatz in Eq.\eqref{eq:SinglemodeAnsatz} even in the replica-symmetry broken spin glass phase. 
% To check the dependence of $v_B$ calculated from the light cone wave front on $F_{th}$, we perform the calculation for different values of $F_{th}$ and show that matching with imaginary momentum calculation holds reasonably good for a wide range of values.

%\begin{figure}[h!]
%	\centering
%	\includegraphics[width=\linewidth]{butterfly_cone_T0p1}
%	\caption{ $F_1(x,t)$ for different values of threshold at $S = 0.1$ and $T = 0.1$}\label{fig:lambdaL_S}
%\end{figure} 

\begin{figure}[h!]
	\centering
    \includegraphics[width=0.8\linewidth]{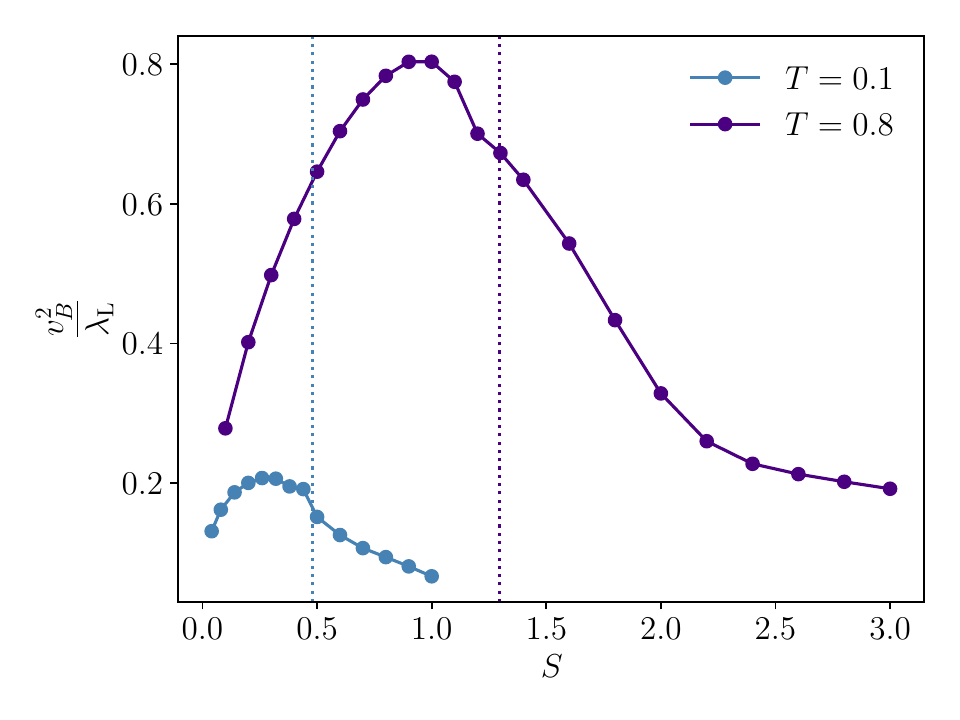}
	\caption{$\tilde{D}\sim v_B^2/\lambda_\mathrm{L}$ across the phase transition ($S_{SG}$, vertical dashed lines) as a function of $S$ calculated for $T=0.1$ and $T=0.8$.}\label{fig:SY_DC_S}
\end{figure}

\textit{$S$ dependence of butterfly velocity.}--- Fig.\ref{fig:SY_vB_S} shows $v_B$ as a function of $S$ for $T=0.1$ and $T=0.8$, going from spin liquid and spin liquid-local moment crossover region to the SG phase. As in the case of $\lambda_\mathrm{L}(S)$ [Fig.\ref{fig:SY_lambdaL_S}], $v_B(S)$ exhibits a broad maximum in the PM phase. 
% \textcolor{brown}{However, unlike $\lambda_\mathrm{L}(S)$ it has a dip at the phase transition $S_{SG}$ and then it increases monotonically with $S$, as the system becomes more classical. Such dip in $v_B$ at the phase transition has been also seen for classical spin models~\cite{Ruidas2021,Bilitewski2021}, implying that the speed of information scrambling becomes slower at the transition in all these cases.} 

\begin{figure}[h!]
	\centering
    \includegraphics[width=0.95\linewidth]{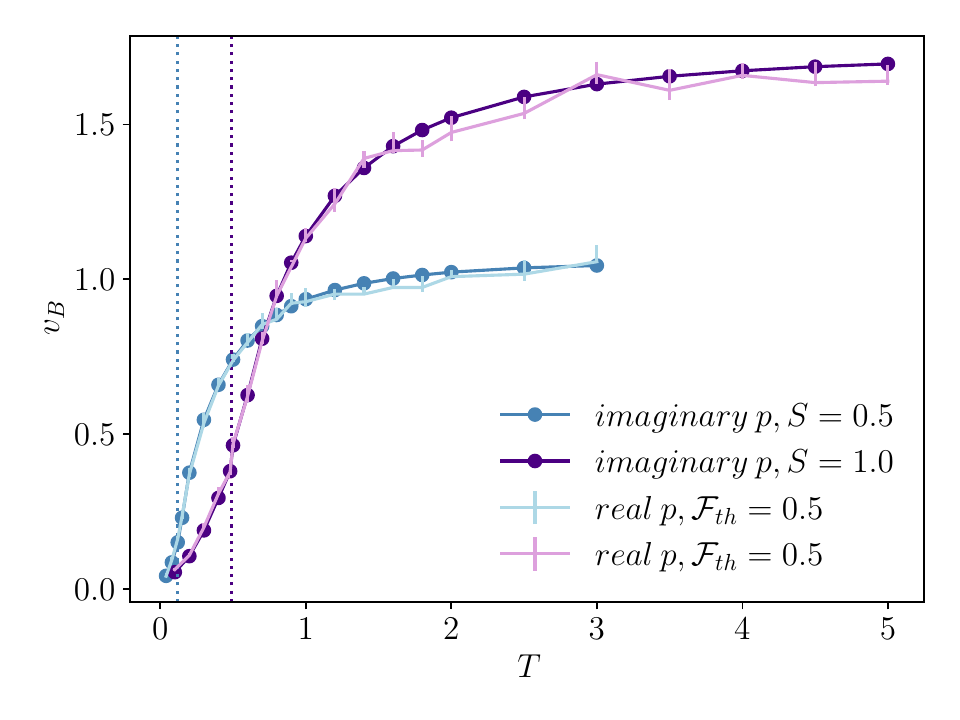}
	\caption{{\bf Temperature dependence of butterfly velocity in SY model.} $v_B$ across the phase transition ($T_{SG}$, vertical dashed lines) as a function of temperature at $S = 0.5, 1.0$, $\alpha= 0.25$, using both real and imaginary $p$ methods (see main text). The parameters and the conventions are same as those described in the caption of Fig.\ref{fig:SY_vB_S} .}\label{fig:SY_vB_T}
\end{figure}

We find $v_B$ to have power-law dependence, $v_B(S,T)-v_B(0,T) \sim \sqrt{S}$ for small $S$, similar to $\lambda_\mathrm{L}$ [Fig.\ref{fig:SY_lambdaLfit_S}(a)] with $v_B(0,T)\approx 0$ in the limit $S\to 0$, as shown in Fig.\ref{fig:SY_vB_Sfit}(a) for $T=0.8$.
% Again, this is related to the local moment like spectra at small $S\lesssim 0.2$ [Figs.\ref{fig:rho_T0.1}(a), \ref{fig:rho_T0.8}(a)]. 
{In the SG phase, $v_B$ shows a power-law decay $v_B(S) \sim 1/S^{1.7}$, as  shown in Fig.\ref{fig:SY_vB_Sfit}(b), reminiscent of power-law decay of $\lambda_\mathrm{L}(S)$ [Fig.\ref{fig:SY_lambdaLfit_S}(b)]. We also plot the proxy for the diffusion constant $\tilde{D}\sim v_B^2/\lambda_\mathrm{L}$ as a function of $S$  in Fig.\ref{fig:SY_DC_S}. Similar to $\lambda_{\mr{L}}$ and $v_B$, $\tilde{D}$ exhibits a broad maximum in within the PM phase and decays with $S$ in SG phase. From the knowledge of the behavior of  $\lambda_{\mr{L}}(S)$ and $v_B(S)$ we expect a similar power-law behavior of $\tilde{D}(S)$ in the PM and SG phase.}

\begin{figure}[h!]
	\centering
    \includegraphics[width=\linewidth]{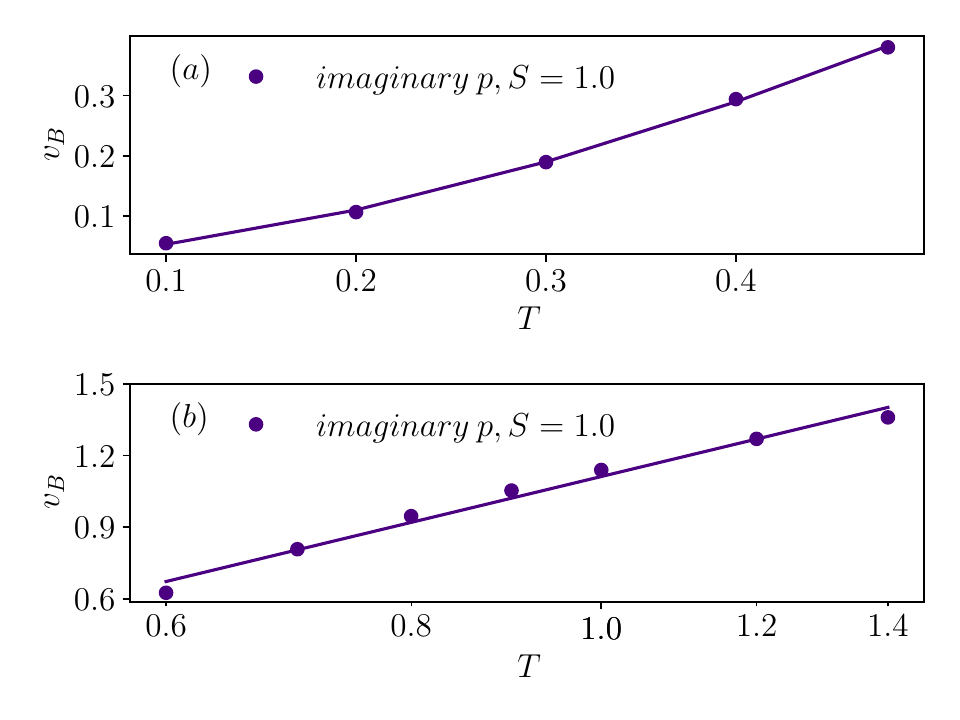}
	\caption{{\bf Temperature dependence of $v_B$ in the SG and PM phases of SY model.} (a) $v_B(S,T)$ (circles) in the SG phase is fitted with $v_B(S,T)  \simeq v_B(S,0) + a T^{\beta_v}$ (line) for $S=1.0$ where {$v_B(S,0) \simeq 0.03$ and  $\beta_v \simeq 1.65$}. (b) Logarithmic fit (line) to $v_B(T)$ in the PM phase, $v_B \sim a \ln \,{T}+c $ is shown on a semi-log plot ($v_B$ vs. $\ln \,{T}$) for $S=1.0$.}\label{fig:SY_vB_Tfit}
\end{figure}

\begin{figure}[h!]
	\centering
    \includegraphics[width=0.8\linewidth]{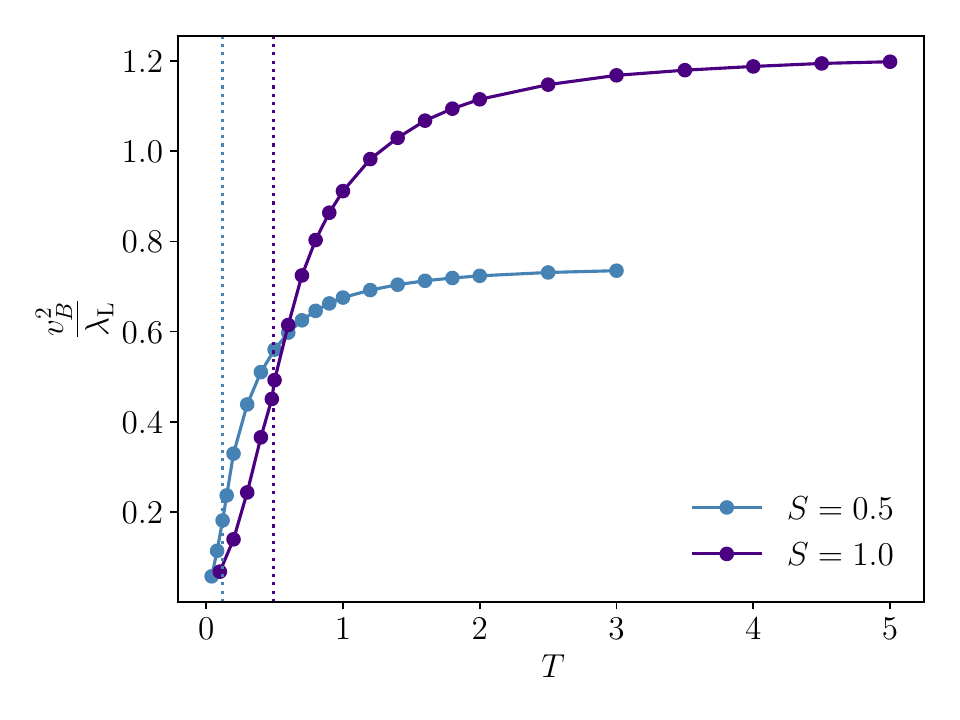}
	\caption{$\tilde{D}\sim v_B^2/\lambda_\mathrm{L}$ as a function of temperature for $S=0.5$ and $S=1.0$ in SY model. The SG transition temperatures are shown by vertical dashed lines.}\label{fig:SY_DC_T}
\end{figure}

\textit{Temperature dependence of butterfly velocity.}--- $v_B$ is shown as a function of $T$ for $S=0.5, 1.0$ across the PM-SG transition.
% \textcolor{brown}{ We show $v_B(T)$ for $S=0.5$ only in the PM phase, as we did compute $v_B$ for the narrow temperature range $T<T_{SG}$ in the SG phase for $S=0.5$.}
% {Added few points in SG as well.} 
The butterfly velocity monotonically increases with temperature through $T_{SG}$, saturating at high temperature, similar to $\lambda_\mathrm{L}(T)$ [Fig.\ref{fig:SY_lambdaL_T}]. As seen in Fig.\ref{fig:SY_vB_T}, $v_B$ approaches a value for $T\to 0$ in the SG phase. 
% {very small 0.03} 
% This implies a finite speed $v_B(S,0)$ of information scrambling at $T=0$ in the SG phase of the SY chain. 
As shown in Fig.\ref{fig:SY_vB_Tfit}(a), { $v_B(T)-v_B(S,0)\sim T^{1.65}$}, i.e., $v_B$ increases with temperature as a power law from the $T=0$ value. Like the $\lambda_\mathrm{L}(T)$ [Fig.\ref{fig:SY_vB_Tfit}(b)], the $T$ dependence of $v_B$ is logarithmic, $\sim a\ln \, {T}+c$, for $T\gtrsim T_{SG}$ in the PM phase [Fig.\ref{fig:SY_vB_Tfit}(b)]. The proxy diffusion constant $\tilde{D}\sim v_B^2/\lambda_\mathrm{L}$ is plotted in Fig.\ref{fig:SY_DC_T}. 
% \textcolor{brown}{For $S=1.0$, $\tilde{D}$ approaches a large value at low temperature $T\ll T_{SG}$ in the SG phase, since $v(S,T)\to v_B(S,0)\neq 0$ and $\lambda_\mathrm{L}\sim T$ for $T\to 0$. In the SG phase one naively expects the transport to slow down and $\tilde{D}$ to decrease with temperature. Thus, it appears that a system made out of spin glass quantum dots, like in the SY chain [Eq.\eqref{eq:SY_1D}], can lead to unusual enhanced diffusion at low temperature. However, this needs to be verified by calculating actual diffusion coefficient $D$, instead of $\tilde{D}$. The latter also has dip at the transition and saturates at high temperature in the PM phase.} 
{The behavior of $\tilde{D}(T)$ is similar to that of $\lambda_{\mr{L}}(T)$ and $v_B(T)$, where it starts from small values, as one would expect, the slower transport in the SG phase and then grows monotonically across the phase transition and saturates to an $S$ dependent value. The proxy diffusion constant $\tilde{D}(T)$  decreases with decreasing temperature as expected in a SG phase.} 
%It's hard to understand this behavior from the spectral function, since the Lyapunov exponent and butterfly velocity are calculated from the OTOC, which is a 4-point function on Keldysh contour.

\begin{figure}[h!]
	\centering
	\includegraphics[width=1.05\linewidth]{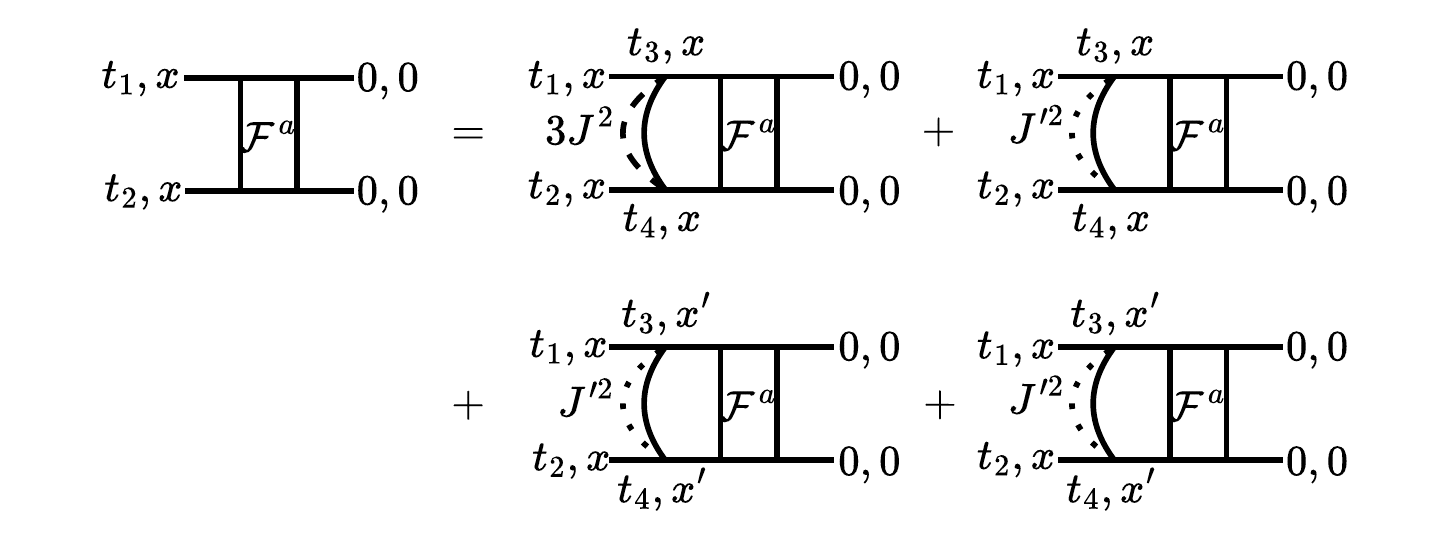}
	\caption{ Diagrammatic representation of the Kernel equation in Eq:\ref{eq:1dkernel_pspin} for $\mathcal{O}(1/N)$ term, $\mathcal{F}_{x }^a$ in the OTOC $F_{x }^a(t_1, t_2)$. The solid horizontal lines denote dressed retarded and advanced propagators $G_R$, $G_A$ and the vertical rung denotes the Wightmann correlation $G_{lr}$ (\ref{appB}). The dashed and dotted lines denote disorder averaging over $J$ and $J^{\pm}$ couplings in Eq.\eqref{eq:pSpin_1d}. Here $x^{\prime} = x \pm 1$. }\label{fig:pspin_1dkernel}
\end{figure}

\subsection{Butterfly velocity across spin glass transition in the $p$-spin glass chain}
% Following the case of SY model, we generalize the $p$-spin dot Hamiltonian in Eq:\ref{eq:Hamiltonian_pspin}, for $p=3$ to a spatially extended one dimensional model given by, 

% \begin{align} \label{Eq:1dpspin_hamiltonian}
% \mathcal{H}_{1D} =& \sum_{x=0}^{ L-1}\sum_{i}\frac{\pi_{i,x}^2}{2\mathcal{M}} + \frac{1}{3!}\sum_{x=0}^{ L-1}\sum_{ijk}J_{ijk} s_{i,x} s_{j,x} s_{k,x} \nonumber \\
% &+ \frac{1}{2!}\sum_{x=0}^{L-1}\sum_{ijk}J_{1 ijk,x}s_{i,x} s_{j,x+1} s_{k,x+1}  \nonumber \\
% &+ \frac{1}{2!}\sum_{x=0}^{L-1}\sum_{ijk} J^{\prime}_{1 ijk,x}s_{i,x+1} s_{j,x} s_{k,x}
% \end{align}
% At each site, couplings are randomly drawn from Gaussian distribution $P(J_{ijk,x}) = \mathrm{exp}\Big( -\frac{2 N^2}{3!}\frac{J_{ijk,x}^2}{2 J^2}\Big)$ and similarly nearest neighbour couplings are drawn from Gaussian distribution $P(J_{1ijk,x}) = \mathrm{exp}\Big(-\frac{2 N^2}{2!}\frac{J_{1ijk,x}^2}{2 J_1^2}\Big)$ and $P(J_{1ijk,x}^\prime ) = \mathrm{exp}\Big(- \frac{2 N^2}{2!}\frac{J_{1ijk,x}^{\prime 2}}{2 J_1^2}\Big)$. Assuming translation invariance, $Q_{ab, x}(\tau) = Q_{ab}(\tau)$ one finds that the saddle point equation remains the same as that of zero-dimensional model with the scaling of coupling, $J^2 + J_1^2 \rightarrow J^2$.  

To compute the $v_B(\Gamma,T)$ in the $p$-spin glass chain [Eq.\eqref{eq:pSpin_1d}], we use methods similar to that for the SY chain discussed above. We calculate the following regularized OTOC defined as,
\begin{align}
     F_{x}(t_1, t_2) =& \frac{1}{N^2} \sum_{i,j} \overline{\mathrm{Tr} \Big[y s_{i,x}(t_1)ys_{j,0}(0)y s_{i,x}(t_2)ys_{j,0}(0)\Big]}
 \end{align}

Following the calculations discussed earlier for the SY chain, the diagrams that contribute to the OTOC can be grouped in powers of $1/N$. The leading order $\mathcal{O}(1)$ diagrams are disconnected and do not contribute to the growth. At the next leading order $\mathcal{O}(1/N)$ we have ``ladder" diagrams which contribute to the growth of OTOC in the scrambling regime, $\lambda_\mathrm{L}^{-1} \lesssim t \lesssim \lambda_\mathrm{L}^{-1}\ln \,{N}$. As earlier, these ladder diagrams can be written in the form of a Bethe-Salpeter equation shown in Fig.\ref{fig:pspin_1dkernel}, i.e.
\begin{align}\label{eq:1dkernel_pspin}
\mathcal{F}^a_{x}(t_1, t_2) =& \int d x' \int d t_3 d t_4  K^a_{xx'}(t_1,t_2,t_3,t_4) \mathcal{F}^a_{x'}(t_3,t_4),
\end{align}
where the kernel is given by
\begin{align}
K^a_{xx'}(t_1, t_2,t_3,t_4) =& \left[(3 J^2  + {J'}^2)\delta_{x'x} + 2 {J'}^{ 2} \delta_{x',x\pm 1} \right] \nonumber \\
&G_R(t_{13}) G_R(t_{24})  G_W(t_{34}) 
\end{align}
After Fourier transforming $\mathcal{F}^a_x(t_1,t_2)$ to momentum space, we obtain $\mathcal{F}^a_p(t_1,t_2)$ and the Kernel 
\begin{align}
K^a_p(t_1,t_2,t_3,t_4)=\tilde{J}^2{[3  -\frac{\alpha}{2} (5 - 4 \cos{p})]}
\end{align}
where we have used, $J^2 + 2{J'}^2= \tilde{J}^2$ and $\alpha = {2J'}^2/\tilde{J}^2$.

To calculate the butterfly velocity we again use the real and imaginary $p$ methods discussed for the SY chain above [Sec.\ref{sec:SY_vB}]. In the imaginary $p$ method, we employ the single mode ansatz 
 \begin{align}\label{singlemode_pspin}
F_{p}^a(t_1, t_2) \approx \frac{e^{\lambda_\mathrm{L}(p)(t_1+t_2)/2}}{C(p)} \Upsilon_{p}^R(t_{12}) \Upsilon_{p}^A(0)
\end{align}
We skip the details of the implementations of the two methods here, since they are similar to those for the SY chain.
  $v_B$ is calculated across the spin glass phase transition line as a function of $T$  and $\Gamma$. 
  
 \begin{figure}[h!]
	\centering
	\includegraphics[width=\linewidth]{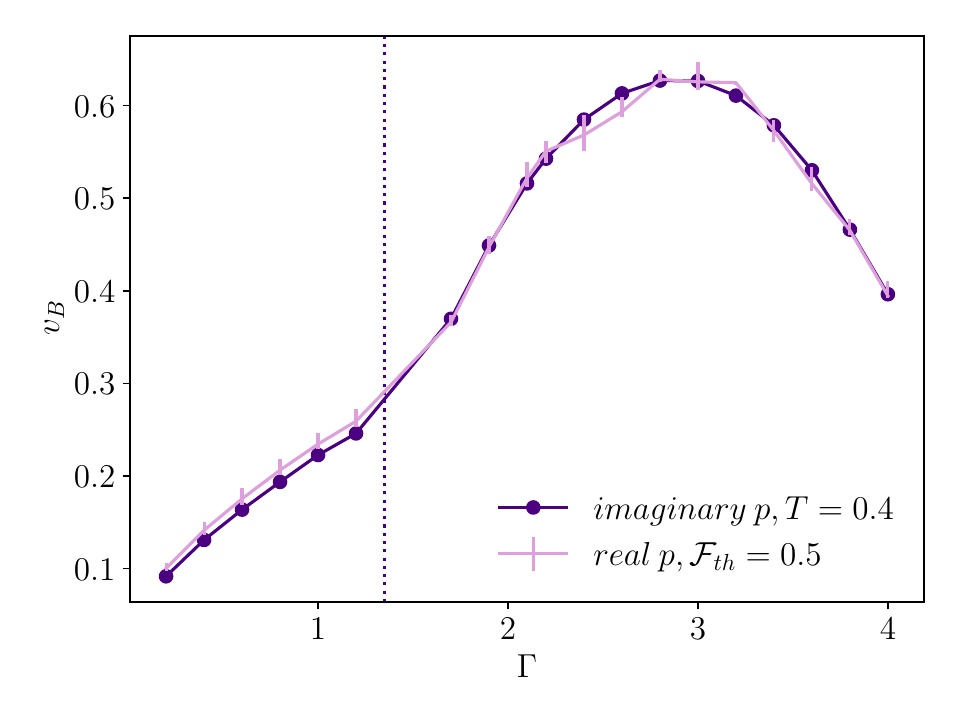}
	\caption{{\bf Variation of butterfly velocity with quantum parameter $\Gamma$ in the $p$-spin glass model.} $v_B$ across the $p$-spin glass phase transition as a function of $\Gamma$ at $T = 0.4$ , $\alpha= 0.25$. In the real $p$ calculation $v_B$ is calculated from the slope of light cone for $\mathcal{F}_{th}=0.5$, for a lattice of $L=50$ sites (see Sec.\ref{sec:SY_vB}). The error bars in the real $p$ results indicate variation of $v_B$ over a range of threshold values $\mathcal{F}_{th}=0.1-1.0$.}\label{fig:pspin_vB_Gamma}
\end{figure}

\textit{$\Gamma$ dependence of butterfly velocity.}---
Fig.\ref{fig:pspin_vB_Gamma} shows the butterfly velocity as a function of the quantum fluctuation parameter $\Gamma$ for $T=0.4$ (in units of $\tilde{J}=1$) ($y$ cut in Fig.\ref{fig:pSpin_PhaseDiagram}). 
% \textcolor{brown}{Unlike the SY model [Fig.\ref{fig:SY_vB_S}], $v_B$ does not have a dip at the transition as function of the quantum parameter}
In the PM phase, $v_B$ exhibits non-monotonic dependence on $\Gamma$ with the broad maximum above the glass transition at $\Gamma=\Gamma_{SG}$, as in the case of $\lambda_\mathrm{L}(\Gamma)$ [Fig.\ref{fig:pspin_lambdaL_Gamma}]. 
\begin{figure}[h!]
	\centering
    \includegraphics[width=\linewidth]{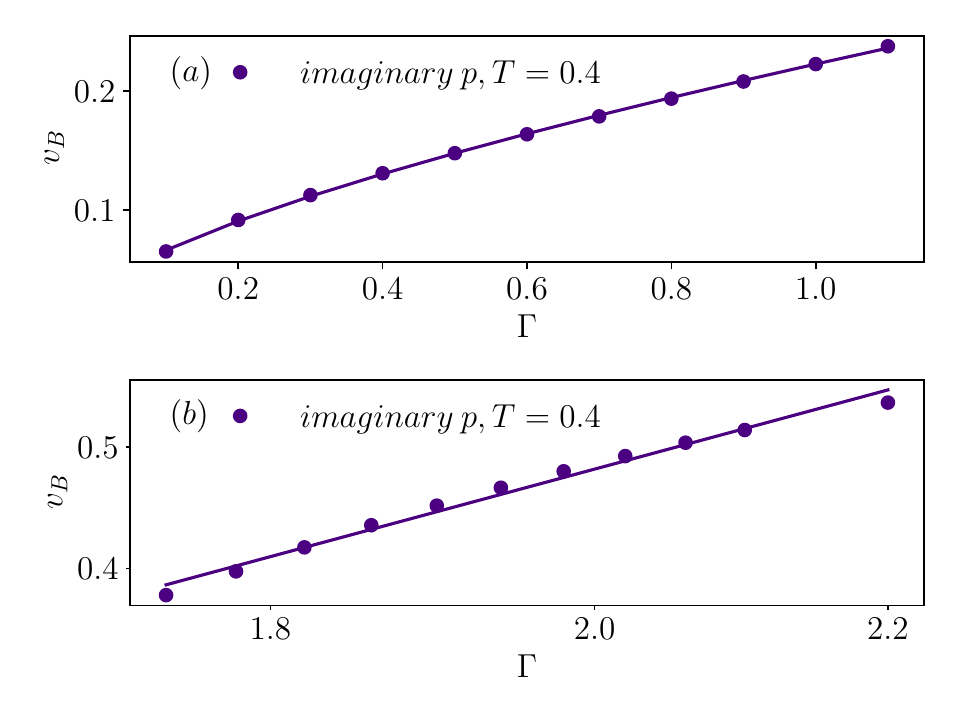}
	\caption{{\bf $\Gamma$ dependence of butterfly velocity in the SG and PM phases of $p$-spin glass model.} $v_B(\Gamma,T)$ (circles) is fitted (a) in the SG phase with $v_B  \simeq v_B(0, T) + a\Gamma^{\alpha_v}$ (line),  where {$\alpha_v \simeq 0.7$  and $ v_B(0, T) \simeq 0.027$} for $T=0.4$, and (b) in the PM phase with {$v_B \sim \Gamma^{1.5}$} (line), shown on a log-log plot for $T=0.4$.}\label{fig:pspin_vB_Gammafit}
\end{figure}
\begin{figure}[h!]
	\centering
    \includegraphics[width=0.8\linewidth]{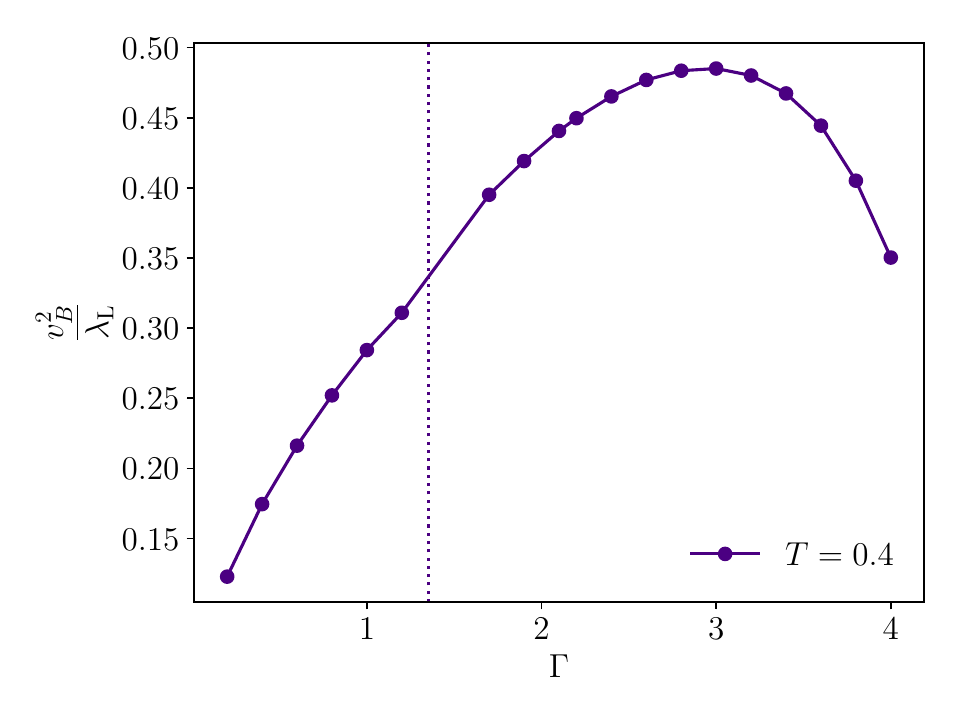}
	\caption{ $\tilde{D}\sim v_B^2/\lambda_\mathrm{L}$ across the SG phase transition as a function of $\Gamma$ calculated for $T=0.4$ }\label{fig:pspin_DC_Gamma}
\end{figure}
The butterfly velocity follows a power law, {$v_B(\Gamma,T)-v_B(0,T)\sim \Gamma^{0.7}$}, with the quantum fluctuation parameter $\Gamma$ for small $\Gamma<\Gamma_{SG}$ with extrapolated $v_B(0,T)\approx 0$ in the $\Gamma\to 0$ limit, deep in the SG phase [Fig.\ref{fig:pspin_vB_Gammafit}(a)]. 
% \textcolor{brown}{This is unlike the exponential dependence of $v_B$ on quantum/classical parameter $S$ for the SY chain [Fig.\ref{fig:SY_lambdaLfit_S}(b)].} 
$v_B$ varies as $v_B \sim \Gamma^{1.5}$ with increasing $\Gamma$ [Fig.\ref{fig:pspin_vB_Gammafit}(b)] in the PM phase close to the transition ($\Gamma\gtrsim \Gamma_{SG}$) before reaching the maximum, as shown in Fig.\ref{fig:pspin_vB_Gamma}. We also plot the proxy for diffusion constant $\tilde{D}\sim v_B^2/\lambda_\mathrm{L}$ in Fig.\ref{fig:pspin_DC_Gamma} for $T=0.4$. {In the SG phase, $\tilde{D}$ is very small for small $\Gamma$, then increases monotonically across the phase transition and exhibits a peak above the transition $\Gamma_{SG}$ in the PM phase.}.

\begin{figure}[h!]
	\centering
	\includegraphics[width=\linewidth]{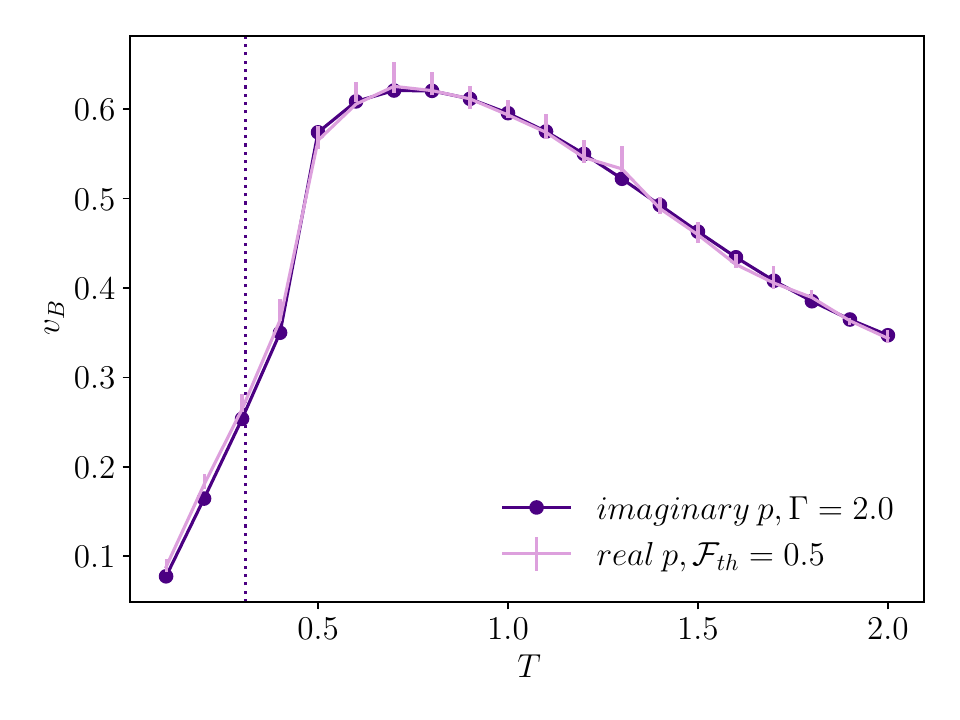}
	\caption{{\bf Temperature dependence of butterfly velocity in $p$-spin model.} $v_B$ across the phase transition as a function of temperature at $\Gamma = 2.0$, $\alpha= 0.25$. The imaginary momentum calculation uses the single mode ansatz. In the real momentum calculation $v_B$ is calculated from the edge of the light cone for values of $\mathcal{F}_{th}=0.5$, in a lattice of $L=50$ sites.  Variation in $v_B$ from nearby values of $\mathcal{F}_{th}=0.1$ and $\mathcal{F}_{th}=1.0$ is shown as error bar.  }\label{fig:pspin_vB_T}
\end{figure}

\begin{figure}[h]
	\centering
    \includegraphics[width=\linewidth]{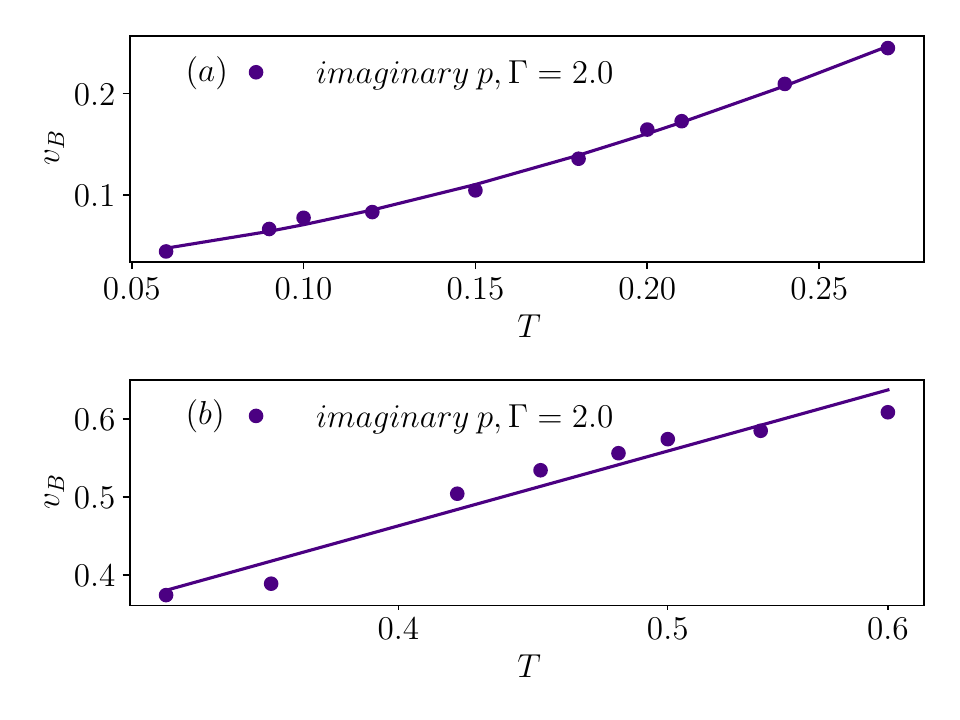}
	\caption{{\bf Temperature dependence of $v_B$ in PM and SG phases of $p$-spin glass model.}. (a) $v_B$ vs. $T$ (circles) in SG phase is fitted as $v_B \simeq v_B(\Gamma, 0 ) + aT^{\beta_{v}}$ (line) for {$\Gamma=2.0$, where $\beta_{v}\simeq 1.7$ and $v_B(\Gamma, 0 )\simeq 0.03$}. (b) Logarithmic dependence of $v_B$ on $T$ (circles) in PM phase, shown as a straight line fit on a semi-log plot for $\Gamma=2.0$.}\label{fig:pspin_vB_Tfit}
\end{figure}

\begin{figure}[h!]
	\centering
    \includegraphics[width=0.8\linewidth]{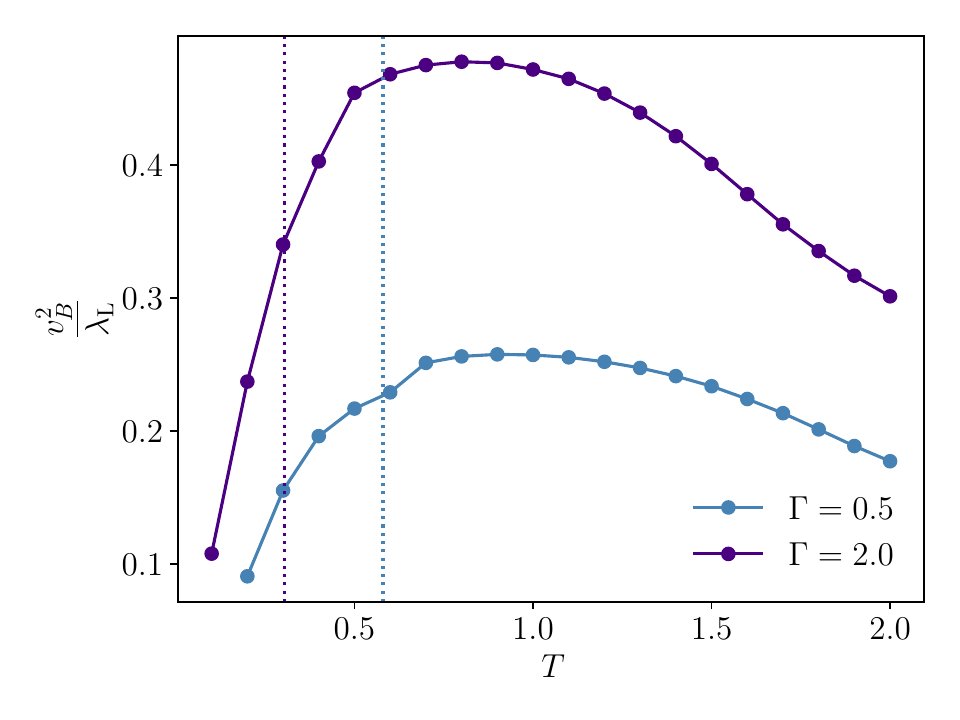}
	\caption{$\tilde{D}\sim v_B^2/\lambda_\mathrm{L}$ across the phase transition as a function of $T$ calculated for $\Gamma=2.0$. in $p$-spin glass model}\label{fig:pspin_DC_T}
\end{figure}

\textit{Temperature dependence of butterfly velocity.}--- We demonstrate the overall temperature dependence of the butterfly velocity going from the SG to the PM phase through the transition at $T_{SG}$ for $\Gamma=2.0$ in Fig.\ref{fig:pspin_vB_T}. $v_B(T)$ exhibits a broad peak in the PM phase, just like $\lambda_\mathrm{L}(T)$ in Fig.\ref{fig:pspin_lambdaL_T}. Figs.\ref{fig:pspin_vB_Tfit}(a),(b) show that the butterfly velocity follows {$v_B(\Gamma,T)-v_B(\Gamma,0)\sim T^{1.7}$, with a small extrapolated zero-temperature value $v_B(\Gamma,0)$} in the SG phase, and logarithmic $T$ dependence in the PM phase for $T\gtrsim T_{SG}$, as in the SY model [Fig.\ref{fig:SY_vB_Tfit}]. {The variation of resultant  $\tilde{D}\sim v_B^2/\lambda_\mathrm{L}$ with temperature [Fig.\ref{fig:pspin_DC_T}] is also very similar to that of $\lambda_{\mr{L}}$ and $v_B$ where it exhibits a broad peak in the PM phase.}

\section{Discussion} 

In this work, we have studied how quantum and thermal fluctuations influence propagation of quantum information, as diagnosed by butterfly velocity, along with chaotic growth rate characterized by a Lyapunov exponent, through one-dimensional generalizations of two paradigmatic zero-dimensional quantum spin glass models, Sachdev-Ye and $p$-spin glass models. These lattice models allow us to map out variations of $\lambda_\mathrm{L}$ and $v_B$ with temperature and quantum parameter deep inside the phases and near the non-trivial replica symmetry breaking spin glass transitions, as well as through crossovers between competing paramagnetic phases, e.g., spin liquid and local moment behaviors, and across the SG transition. We find distinct evolution of glassy complexity, indicated by maximum, or lack thereof, in the Lyapunov exponent and butterfly velocity in the PM phase, as a function of quantum and thermal fluctuations in the SY model. The maximum is seen only as a function of quantum parameter $S$, around the onset of two-step glassy relaxation, away from the glass transition from the PM side. The maximum is absent when the glass transition is approached by decreasing $T$. In contrast, the maximum occurs as function of both $T$ and the quantum parameter $\Gamma$ in the PM phase of $p$-spin glass. The maximum corresponds to intensified chaos due to maximal complexity arising from the sampling of exponentially large number of saddle points of the underlying glassy landscape. In the semiclassical limit, the maximal number of saddle points leads to extreme sensitivity to the initial condition~\cite{Correale2023} or chaos. 

The spin glass models discussed here and other related models can be realized in cavity quantum electrodynamics (QED) platforms \cite{Gopalakrishnan2011,Strack2011,Guo2019a,Hosseinabadi2023a,Hosseinabadi2023b}, and even scrambling dynamics can be probed in this setup \cite{Swingle2016,Marino2019}. Our work needs to be extended in future in several directions. To better understand connection between glassy complexity \cite{Correale2023,Anous_2021} and quantum many-body chaos, it will be worthwhile to compute the complexity for the SY model using both semiclassical methods~\cite{Correale2023} and quantum Thouless-Anderson-Palmer (TAP) equations~\cite{Biroli2001}, as have been done for quantum $p$-spin glass. 
% \textcolor{brown}{The asymptotic temperature dependence of $\lambda_\mathrm{L}$ and $v_B$ we obtain deep in the SG phase of SY and $p$-spin glass implies a finite $v_B$ for the $T=0$ SG ground state, and, consequently, lead to rapid increase of the quantity $\tilde{D}\sim v_B^2/\lambda_\mathrm{L}$ for $T\to 0$. As $\tilde{D}$ often acts as a proxy to diffusion constant, it will be interesting to probe the possibility of such anomalously large diffusion through actual transport calculations in a chain of spin glass quantum dots.} 
As $\tilde{D}\sim v_B^2/\lambda_\mathrm{L}$ often acts as a proxy to diffusion constant, it will be interesting to compare our predictions of $\tilde{D}$, from the asymptotic behavior of $\lambda_\mathrm{L}$ and $v_B$ from chaos calculations with actual transport calculations in a chain of spin glass quantum dots.

Our real-frequency saddle-point calculations of $\lambda_\mathrm{L}$ and $v_B$, and corresponding bosonic spectral functions, in the SY model indicate subtle competition between spin liquid and local moment solutions, as well as strong $S$ and $T$ dependent corrections to the conformal solution for the SYK-like spin liquid maximal chaotic behavior, even at small values of $S\lesssim 0.05$ and low temperature $T\sim 0.1$. This implies intriguing crossover in the small $S$ region towards the expected maximally chaotic SYK spin liquid for $S\to 0$. We have not studied this crossover and the relative stability and intricate competition between spin liquid and local moment solutions in the small $S$ region in much detail in this work. It will be important to study the chaos and spectral properties in the small-$S$ regime in future work. Finally, it will be an exciting research direction to compute entanglement properties in various crossover regions and across the replica symmetry breaking spin glass transition in these large $N$ models following methods~\cite{Haldar2020,Bera2023} similar to those applied for calculating entanglement entropy in lattice of SYK dots and in Hubbard model within dynamical mean-field theory.

\section*{Acknowldegements}
VL acknowledges support from CSIR, India and is grateful to the workstation provided by CHEP on which most of calculations in this work are done. SB acknowledges support from SERB (Grant No. CRG/2022/001062), DST, India, and QuST project of the DST, Govt. of India.

\bibliography{ChaosQuantumGlassChain_SBreview}

%----------------------------------------------------------------
%--------------------------------------------------------------
%\texttt{%\def\makeSM{1}
%%\ifdefined\makeSM
%%%%%%%%%%%%%%%%%%%%%%%%%%%%%%%%%%%%%%%%%%%%5
%%\newwrite\tempfile
%%\immediate\openout\tempfile=junkSM.\jobname
%%\immediate\write\tempfile{\noexpand{\thepage} }
%%\immediate\closeout\tempfile
%%\clearpage
%%\newpage
%%%%%%%%%%%%%%%%%%%%%%%%%%%%%%%%%%%%%%%%%%%%%%%%%%%%%%%%%%%%%%%%%%%%%%%%%%%%%%%%%%%%%%%%%%%%%%%%%%%%
%%\end{document}}
%UNCOMMENT THIS TO GENERATE FILE WITHOUT SUPPELEMENTAL MATERIAL

\appendix
\renewcommand{\appendixname}{}
\renewcommand{\thesection}{ {Appendix \Alph{section}}}
\renewcommand{\theequation}{\Alph{section}.\arabic{equation}}
\renewcommand{\thefigure}{\Alph{section}.\arabic{figure}}
\setcounter{page}{1}
\setcounter{figure}{0}
\setcounter{equation}{0}

%\widetext

%\centerline{\bf Supplemental Material}
%\centerline{\bf for}
%\centerline{\bf \titlename}
%\centerline{by \authornames}

%\email{arijit@physics.iisc.ernet.in}
%\author{Vijay B. Shenoy}
%\email{shenoy@physics.iisc.ernet.in}
\affiliation{\affiliations}
%\input{HD_SYK_supplement}
%\clearpage
%\fi %%TO MAKE SUPPLEMENTAL MATERIAL

\def  \qinv{Q^{-1}}
\def  \q0{\frac{\w_k^2}{\G}+z}

\newcommand{\lin}{linspace}
\newcommand{\m}{\Delta_0}
\newcommand{\M}{\Delta}
\newcommand{\g}{Q}
\newcommand{\ginv}{\big(Q^{-1}\big)_{ab}}
\newcommand{\qr}{q_{\text{reg}}} 
\newcommand{\sr}{\Sigma_{\text{reg}}}
\newcommand{\qt}{\tilde{q}_{{EA}}}
\newcommand{\bfig}{\begin{figure}[H]\centering}
\newcommand{\efig}{\end{figure}}
\newcommand{\s}{\hspace{0.5cm }} % s for skip
\newcommand{\sh}{\hspace{0.25cm }} % sh for half skip.

\newpage

\section{Imaginary-time path integral and saddle-point equations for SY chain}\label{appA}
We write the partition function $\mathcal{Z}(\{J_{ij,x},J'_{ij,x}\})=\mathrm{Tr}[\exp(-\beta \mathcal{H})]$ ($\beta=1/T$) of the disordered model of Eq.\eqref{eq:SY_1D} as an imaginary-time path integral using the bosonic representation $S_{i\alpha\beta, x}=b_{i\alpha, x}^\dagger b_{i\beta, x}-S\delta_{\alpha\beta}$ with the constraint $\sum_\alpha b_{i\alpha, x}^\dagger b_{i\alpha, x}=SM$ ($S\geq 0$) on the number of bosons at each $i,x$, to fix the spin of the bosonic representation to $S$. Here the (real) random couplings $\{J_{ij,x},J'_{ij,x}\}$ are drawn from Gaussian distributions with zero mean and variances $J^2$ and $J'^2$, respectively. We calculate disorder-averaged quantities by introducing replicas $a=1,\dots,n$ to obtain 
$\overline{\mathcal{Z}^n}= \int  \mathcal{D}(\bar{b},b) \mathcal{D}\lambda \mathrm{exp}[-S_{\mathrm{eff}}]$, where 
\begin{align}
&S_{\mathrm{eff}} = \int_{0}^\beta d \tau \sum_{i x} \left[  \bar{b}_{i \alpha, x}^a \partial_\tau b_{i \alpha, x}^a  -  \lambda^a_{i, x}(\tau)(\bar{b}_{i \alpha, x}^a b_{i \alpha, x}^a -S M) \right]\nonumber \\
& - \frac{1}{2MN}  \int_{0}^\beta d \tau d \tau^\prime \Bigg[\frac{J^2}{2}\sum_{i<j,x}   S_{i\alpha \beta, x}^a(\tau)  S_{i\gamma \delta, x}^b(\tau^\prime) S_{j\beta \alpha, x }^a(\tau)    \nonumber \\
&S_{j\delta \gamma, x}^b(\tau^\prime)  +J^{\prime 2} \sum_{i, j,x} S_{i\alpha \beta, x}^a(\tau)  S_{i\gamma \delta, x}^b(\tau^\prime) S_{j\beta \alpha, x+1 }^a(\tau)  \nonumber \\
&S_{j\delta \gamma, x+1}^b(\tau^\prime)  \Bigg]
\end{align}
%\end{widetext}
% \begin{widetext}
% \section{Imaginary-time path integral and saddle point equations in SY model}\label{appA}
Here the field $\lambda_{i, x}^a(\tau)$ imposes the constraint on boson number and sum over repeated $\alpha,\beta,\dots$ and $a,b$ indices are assumed. In the $N\to\infty$ limit, introducing the Hubbard-Stratonovich field $Q_{\alpha \beta \gamma \delta, x}^{ab}(\tau, \tau^\prime)$, and its conjugate $\Pi_{\alpha \beta \gamma \delta,x}^{ab}(\tau, \tau^\prime)$ to impose $Q_{\alpha \beta \gamma \delta, x}^{ab}(\tau, \tau^\prime)=(1/N) \sum_i \langle S^a_{i\alpha\beta, x}(\tau)S^b_{i\gamma\delta, x}(\tau') \rangle$, we decouple the quartic terms above. Moreover, assuming SU$(M)$ symmetry, i.e., $Q_{\alpha\beta\gamma\delta,x}^{ab}(\tau,\tau')=Q_x^{ab}(\tau,\tau')\delta_{\alpha\gamma}\delta_{\beta\delta}$, and similarly for $\Pi$, we obtain $\overline{\mathcal{Z}^n}= \int  \mathcal{D}Q\mathcal{D}\Pi\mathcal{D}(\bar{b},b) \mathcal{D}\lambda \mathrm{exp}[-S_{\mathrm{eff}}]$ and the action

\begin{align}
&S_{\mathrm{eff}} = \int_{0}^\beta d \tau \sum_{x} \Bigg[\sum_{i} \big[  \bar{b}_{i \alpha, x}^a \partial_\tau b_{i \alpha, x}^a - \lambda_{i,x}^a(\tau) (\bar{b}_{i \alpha, x}^a b_{i \alpha, x}^a \nonumber \\
& - S M) \big]  +MN \int_{0}^\beta d \tau d \tau^\prime \Pi_{ x}^{ab}(\tau^\prime, \tau)\Big[ Q_{ x}^{ab}(\tau, \tau^\prime)\nonumber\\
&- \frac{1}{N M^2}\sum_i \bar{b}_{i \alpha, x}^a(\tau) b_{i  \alpha, x}^b(\tau^\prime) b_{i \beta, x}^a(\tau) \bar{b}_{i \beta, x}^b(\tau^\prime) \Big]   \nonumber\\ 
&- MN \int_{0}^\beta d \tau d \tau^\prime Q_{x}^{ab}(\tau, \tau^\prime)\Big(\frac{J^2}{4}  Q_{x}^{ab}(\tau, \tau^\prime)  \nonumber\\ 
&+ \frac{J^{\prime 2}}{2}   Q_{x+1}^{ab}(\tau, \tau^\prime)\Big) \Bigg]
\end{align}

 Introducing $G_x^{ab}(\tau, \tau^\prime) = -(1/M)\sum_{\alpha}\langle b_{ i\alpha, x}^a(\tau) \bar{b}_{ i\alpha, x}^b(\tau^\prime )\rangle$ and the conjugate field $\Sigma_x^{ab}(\tau,\tau')$ we get $\overline{\mathcal{Z}^n}= \int  \mathcal{D}(Q,\Pi,G,\Sigma,\lambda) \mathrm{exp}[-MNS_{\mathrm{eff}}]$ with
% \begin{widetext}
% \begin{align}
% S_Q =& \sum_{x=0}^{L-1}\bigg[   \int_{0}^\beta d \tau  \sum_{i} \bar{b}_{i \alpha x}^a(\tau) \bigg(   (\partial_\tau - \lambda)\delta_{ab}\delta(t-t^\prime) + \Sigma_x^{ab}(\tau, \tau^\prime) \bigg) b_{i \alpha x}^b(\tau^\prime) \nonumber \\
% & + S MN\int_0^\beta d\tau \lambda^a_{ x}(\tau) -MN \int_{0}^\beta d \tau d \tau^\prime Q_{x}^{ab}(\tau, \tau^\prime)\Big(\frac{J^2}{4}  Q_{x}^{ab}(\tau, \tau^\prime) + \frac{J^{\prime 2}}{2}   Q_{x+1}^{ab}(\tau, \tau^\prime)\Big) \nonumber \\
% & +MN \int_{0}^\beta d \tau d \tau^\prime \Sigma_{Q, x}^{ab}(\tau^\prime, \tau)\Big( Q_{ x}^{ab}(\tau, \tau^\prime) + G^{ab}_x(\tau, \tau^\prime) G^{ba}_x(\tau^\prime, \tau) \Big) + MN \int_{0}^\beta d \tau d \tau^\prime \Sigma_x^{ab}(\tau, \tau^\prime) G_x^{ba}(\tau^\prime, \tau)\bigg]
% \end{align}
% %\end{widetext}

% where $\lambda^a_{ x}(\tau) = \lambda$. Integrating out the quadratic part of the action,
%\begin{widetext}
\begin{align}
&S_{\mathrm{eff}} =\sum_x\Bigg[\mathrm{Tr} \ln \,   \left[(\partial_\tau - \lambda_x^a(\tau))\delta_{ab}\delta(\tau-\tau^\prime) + \Sigma_x^{ab}(\tau, \tau^\prime) \right] \nonumber \\
&-\int_{0}^\beta d \tau d \tau^\prime Q_{x}^{ab}(\tau, \tau^\prime)\left(\frac{J^2}{4}  Q_{x}^{ab}(\tau, \tau^\prime) + \frac{J^{\prime 2}}{2}   Q_{x+1}^{ab}(\tau, \tau^\prime)\right) \nonumber\\
& +\int_{0}^\beta d \tau d \tau^\prime \Pi_{ x}^{ab}(\tau^\prime, \tau)\left( Q_{ x}^{ab}(\tau, \tau^\prime) - G^{ab}_x(\tau, \tau^\prime) G^{ba}_x(\tau^\prime, \tau) \right)\nonumber \\
&+ \int_{0}^\beta d \tau d \tau^\prime \Sigma_x^{ab}(\tau, \tau^\prime) G_x^{ba}(\tau^\prime, \tau)+  S\sum_a\int_0^\beta d\tau \lambda_x^a(\tau)\Bigg]
\end{align}
where we have imposed $\lambda_{ix}^a(\tau)=\lambda^a_x(\tau)$. In the large $N,M$ limit, minimization of the above action with respect to $Q,\Pi,G,\Sigma$ and $\lambda$, with the assumption of a time-translation invariant (equilibrium) saddle point, e.g., $Q_x^{ab}(\tau,\tau')=Q_x^{ab}(\tau-\tau')$ and $\lambda_x^a(\tau)=\lambda_x^a$, leads to the following self-consistency equations for the bosonic Green's function
\begin{align}
[G_x^{-1} ]^{ab}&(\ci \omega_k )= (\ci \omega_k + \lambda_x^a)\delta^{ab} - \Sigma^{ab}_x(i \omega_k)  \nonumber \\
\Sigma_x^{ab} (\tau) &= J^2 G^{ab}_x(\tau) G^{ab}_x(\tau) G^{ab}_x(-\tau) \nonumber \\
&+ J^{\prime 2} G^{ab}_x(\tau) \Big(G^{ab}_{x+1}(\tau) G^{ab}_{x+1}(-\tau) \nonumber \\
&+ {G^{ab}_{x-1}(\tau) G^{ab}_{x-1}(-\tau)} \Big)\nonumber \\
G^{aa}_x(\tau &=0^-) = -S.
\end{align}
%\end{widetext}
Here $\omega_k=2k\pi\beta$ is the bosonic Matsubara frequency with integer $k$. Assuming lattice translation invariance, $ G^{ab}_x(\tau) = G^{ab}(\tau)$ and $\lambda_x^a=\lambda^a$, the above saddle point is the same as that [Eq.\eqref{eq:SY_SaddleEq}] for the zero-dimensional model \cite{Sachdev1993,Georges2000} with the renormalized coupling ${\tilde{J}^2=J^2 + 2J^{\prime 2}}$.

{\it Saddle point equation in PM phase---}
In the PM phase the Green's function is replica diagonal and symmetric. Thus, the saddle point equations in Eq.\eqref{eq:SY_SaddleEq},  after analytic continuation to real frequency, i.e., $\ci \omega_k\to \omega+\ci 0^+$, becomes
\begin{subequations}
\begin{align}
[G_R(\w)]&^{-1}= \w - \lambda - \Sigma_R(\w) \\
\Sigma(\ci\w_k) &= \int_0^\beta d \tau e^{i \w_k t }\Sigma(\tau) = \frac{\tilde{J}^2}{\beta^2} \sum_{k_1, k_2} G(\ci \w_{k_1})G(\ci \w_{k_2}) \nonumber \\
&G(\ci \w_{k_1} + \ci \w_{k_2} - \ci \w_{k})  \label{Eq:selfE_imag}
\end{align}
\end{subequations}

Using the spectral representation $G(\ci \w_k) = \int d\w \rho(\w)/(\ci\w_k -\w)$ with $\rho(\w) = -(1/\pi) \mathrm{Im}\,  G_R(\w)$  in Eq.\ref{Eq:selfE_imag} gives,

% \begin{widetext}
\begin{align}
&\Sigma(\ci\w_k) = \tilde{J}^2 \int d\w_1d\w_2d\w_3 \frac{\rho(\w_1) \rho(\w_2) \rho(\w_3)}{\ci\w_k - \w_1-\w_2 + \w_3}\nonumber \\
&\left[n_B(\w_1) n_B(\w_2) n_B(-\w_3) + n_B(-\w_1) n_B(-\w_2) n_B(\w_3) \right],
\end{align}
where $\lambda^a=\lambda$ and $n_B(\w) = 1/(e^{\beta \w} -1)$.

Using the identity $1/(\w - \w_1 - \w_2 +\w_3 + \ci 0^+) = -\ci \int_0^\infty dt e^{i(\w -\w_1-\w_2+\w_3 + \ci 0^+)}$ and analytic continuation gives the retarded self-energy
\begin{align}
\Sigma_R(\w) =& -\ci \tilde{J}^2 \int_0^\infty dt e^{\ci\w t}\left[n_1(t)n_1(t) n_2(t) \right.\nonumber \\
&\left.+ n_3(t) n_3(t) n_4(t) \right] \label{Eq:SigmawPM_fft}
\end{align}
where $n_1(t) = \int d\w e^{-\ci \w t}\rho(\w) n_B(-\w)$, 
$n_2(t) = \int d\w e^{\ci\w t}\rho(\w) n_B(\w)$, $n_3(t) = \int d\w e^{-\ci\w t}\rho(\w) n_B(\w)$, and $n_4(t) = \int d\w e^{\ci\w t}\rho(\w) n_B(-\w)$.

{\it Saddle point equation in SG phase---}
Using the 1-RSB ansatz in Eq.\ref{Eq:1rsb}, the analytically continued saddle point equation for the regular part of the green's function is given by
\begin{align}
[G_R(\omega)]&^{-1} = \omega -\frac{\tilde{J} g }{\Theta_R} -[\Sigma_R(\omega) -\Sigma_R(\omega=0)], \nonumber \\
 m\beta & = \frac{1}{\tilde{J}g^2}\Big(\frac{1}{\Theta_R}- \Theta_R \Big)
\end{align}
Using the similar steps for analytic continuation as in the PM phase, we get the retarded self-energy from Eq.\eqref{Eq:SG_Sigmatau},
\begin{align}
\Sigma_R&(\omega) = -\ci \tilde{J}^2 \int_0^\infty dt e^{\ci\w t}\Big[\big(n_1^2(t) n_2(t) + n_3^2(t) n_4(t)\big) \nonumber \\
&-2g\big(n_1(t) n_2(t)- n_3(t) n_4(t)\big) - g\big(n_3^2(t) - n_1^2(t)\big) \Big] \nonumber \\
& + 2 g^2 \tilde{J}^2G_R(\w) + g^2 \tilde{J}^2 G_R^*(-\w)\label{Eq:SigmawSG_fft}
\end{align}
% {SB: In the above, should $\tilde{G}(\omega)$ be $G_R(\omega)$? If so, I do not understand how the last term is coming from Eq.\eqref{Eq:SG_Sigmatau}. If both are $G_R(\omega)$, should not they just add up?}{L: Fixed it now. }
where $n_i$, $i = 1,2,3,4$ are same as those defined in the PM phase.
% if we replace $\rho(\w)$ by $\tilde{\rho}(\w) = -(1/\pi) \mathrm{Im}\,  \tilde{G}_R(\w)$. 
The two equations above can be solved numerically similar to PM phase calculation, to obtain the retarded Green's function, $G_R(\w)$. 

\subsection{Numerical solution of the saddle point equations}
Here we provide the details of the numerical implementation of the self-consistent equations in imaginary as well as in real frequency.

\subsubsection{Real-frequency solution} \label{sec:RealFreqNumerics}
{\it PM phase --}
The retarded Green function in PM phase can be obtained in the frequency domain by self-consistently solving Eq.\ref{Eq:SaddlePM_real}. We write the self-energy $\Sigma_R(\w)$ in a form convenient for numerical calculation using FFT, as in Eq.\ref{Eq:SigmawPM_fft}. We start with an initial ansatz for $G_R(\w)$ on a discrete range $\w \in [-\w_{max}, \w_{max}]$ and an initial value of $\lambda$ and calculate $\Sigma_R(\w)$ from it. At each iteration, $[G_R(\w)]^{-1}$ is updated as,

\begin{align}
    G^{-1}_{R, j+1}(\w)= (1-y)\Big(\w - \lambda - \Sigma_{R,j}(\w) \Big) + yG^{-1}_{R, j}(\w)
\end{align}

where $y \in (0,1)$ is used to ensure smooth convergence, which we choose typically between $0.2$ and $0.3$. At each iteration, we monitor the error $|[G_{R, j+1}(\w)]^{-1} - [G_{R, j}(\w)]^{-1}|$ and repeat the iteration with the updated $[G_{R, j+1}(\w)]^{-1}$ until it converges. We then check the constraint, $G(\tau=0^-) = -\int_{-\infty}^\infty d \omega \rho(\omega)/(e^{\beta \omega} -1) = -S$. If it' is not met, we repeat the procedure with a different $\lambda$. The procedure is repeated until the constraint is met.

{\it SG phase --}
The retarded Green function, $G_R(\w)$ in the SG phase can be obtained by solving Eq.\ref{Eq:SaddleSG_real}. Similar to PM phase, the retarded self-energy $\Sigma_R(\w)$ is written in a form suitable for implementation using FFT, as in Eq. \eqref{Eq:SigmawSG_fft}. Starting with an initial ansatz and an initial value of $m$, we calculate $\Sigma_R(\w)$. $G_R(\w)$ is updated at each iteration as
\begin{align}
    G^{-1}_{R,j+1}(\omega) =& (1-y)\Big(\omega -\frac{\tilde{J} g }{\Theta_R} -[\Sigma_{R,j}(\omega) -\Sigma_{R,j}(\omega=0)]\Big) \\ \nonumber
    &+ y G^{-1}_{R,j}(\omega)
\end{align}
We repeat the iteration with the updated $[G_{R, j+1}(\w)]^{-1}$ until it converges and check the constraint $-\int_{-\infty}^\infty d \omega \rho(\omega)/(e^{\beta \omega} -1) -g = -S$. $m$ is varied to satisfy the constraint.
\subsubsection{Imaginary-time solution}\label{sec:ImagTimeNumerics}
 {\it PM phase --}We solve the saddle point Eq.\ref{Eq:SaddlePM} iteratively to obtain $G(\ci \w_k)$ for $k \in [0, k_{max}]$ . We start with an initial guess solution which could be the non-interacting part of the Green function and an initial guess value of the Lagrange multiplier $\lambda$ and calculate $\Sigma(\tau)$ using Eq.\ref{Eq:Sigma_tau}. We then Fourier transform $G(\tau)$ using fast-Fourier transform(FFT) and obtain  $G(\ci \w_k)$ and $\Sigma(\ci \w_k)$. After $j$th iteration, $G_i(\ci \w_k)$ is updated as, 

\begin{equation}
    G_{j+1}(\ci \w_k) = \frac{1-y}{\ci \w_k -\lambda - \Sigma_j(\ci \w_k)} + y G_j(\ci \w_k)
\end{equation}
  After each iteration we monitor the error  $e(\ci \w_k) = |G_{i+1}(\ci \w_k) - G_i(\ci \w_k)|$ for each $k$ by comparing with $\epsilon = 10^{-9}$, the tolerance. The procedure is repeated with updated $G_{i+1}(\ci \w_k)$ after Fourier transforming back to time domain until $e(\ci \w_k) < \epsilon$ for each $k$. After convergence, we check if the constraint $G(\tau = 0^-)=-S$ is met for the current value of $\lambda$. If not we repeat the procedure again after varying $\lambda$. The iterations stop only after the constraint is met within a tolerance.

{\it SG phase --} In the SG phase we solve for the regular part of the Green function, $\tilde{G}(\ci \w_k)$ in Eq.\ref{Eq:SaddleSG_imag} using the modified self-energy $\tilde{\Sigma}(\tau)$ in Eq.\ref{Eq:SG_Sigmatau}. The procedure to solve the saddle point equation in SG phase is similar to PM phase with the modified constraint, $\tilde{G}(\tau = 0^-) = g - S$.

Starting with an initial ansatz and an initial value of the breaking point, $m$ we calculate the $\tilde{\Sigma}(\tau)$ and $g$ from Eq.\ref{Eq:breakpoint}. We then Fourier transform $\tilde{G}(\tau)$ using FFT and obtain  $\tilde{G}(\ci \w_k)$ and $\tilde{\Sigma}(\ci \w_k)$. At each iteration, $\tilde{G}_i(\ci \w_k)$ is updated as

\begin{align}
\tilde{G}_{j+1}(\ci \omega_k)=& \frac{1-y}{\ci\omega_k -\tilde{J} g/\Theta -[\tilde{\Sigma}_j(\ci \omega_k) -\tilde{\Sigma}(\ci \omega_k=0)]}\nonumber \\
&+ y \tilde{G}_j(\ci \omega_k)
\end{align}
The procedure is repeated until convergence and then we check if the constraint is met with current value of $m$. If not, we vary the value of $m$, within the range $(0,1]$ and repeat the iteration procedure until the constraint is met.

\subsection{Spectral and correlation function in the SY model}\label{app:SpectralFunction}
% {Show also the spectral function evolution for different $S$ at $T=0.1$ and for different $T$ at $S=0.5$.}
% \medskip
In the SY model, we look at the nature of spectrum as a function of thermal and quantum fluctuations. As summarized in Sec.\ref{sec:SYcrossover}, following earlier works \cite{Georges2001,Camjayi2003}, there are three types of solutions of the saddle point equations of Sec.\ref{sec:SY_PhaseDiagram}, (1) spin liquid, (2) local moment, and (3) marginal spin glass. These solutions can be distinguished via the bosonic spectral function $\rho(\w) = -(1/\pi) \mathrm{Im}\, G_R(\w)$.  In the PM phase, the spin liquid solution resides in the quantum critical region $IV$, as shown in Fig.\ref{fig:phasediagramSY}, extending all the way down to $T=0$ for $S=0$. For small $S<S_{max}\approx 0.052$, the zero-temperature bosonic Green's function for complex frequency $z$ is given by $G(z)\sim e^{-\ci(\theta+\pi/4)}/\sqrt{z}$, with $\mathrm{Im}\, z>0$ and $\pi/4<\theta<3\pi/4$, where the spectral asymmetry $\theta$ is related to spin $S$ via $\theta/4+\sin{(2\theta)}/4=1/2+S$ \cite{Sachdev1993,Georges2001}. This leads to a $T=0$ spectral function, $\rho(\omega)\sim -\mathrm{sgn}(\omega)/\sqrt{|\omega|}$, with characteristic square root divergence at $\omega\to 0$, and a finite-$T$ conformal spectral function~\cite{Parcollet1998} in region $IV$ of Fig.\ref{fig:phasediagramSY},
\begin{align}
    \rho(\omega)\sim \frac{\sinh{\left(\omega/2T\right)}}{\sqrt{T}}\Gamma\left(\frac{1}{4}+\ci\frac{\omega-\alpha T}{2\pi T}\right)\Gamma\left(\frac{1}{4}-\ci\frac{\omega-\alpha T}{2\pi T}\right), \label{eq:SpinLiquidRho}
\end{align}
where $\sin(\pi/4+\theta)=1/\sqrt{1+e^{-2\alpha}}$. At finite temperature, the above spectral function, $\rho(\omega)\sim \omega$ for $|\omega|\lesssim T$, has a peak at $|\omega|\sim \mathcal{O}(T)$, and thereafter decays as $1/\sqrt{|\omega|}$ for $\mathcal{O}(T)\lesssim |\omega|\lesssim J$. 
\begin{figure}[h]
	\centering
	\includegraphics[width=0.5\textwidth]{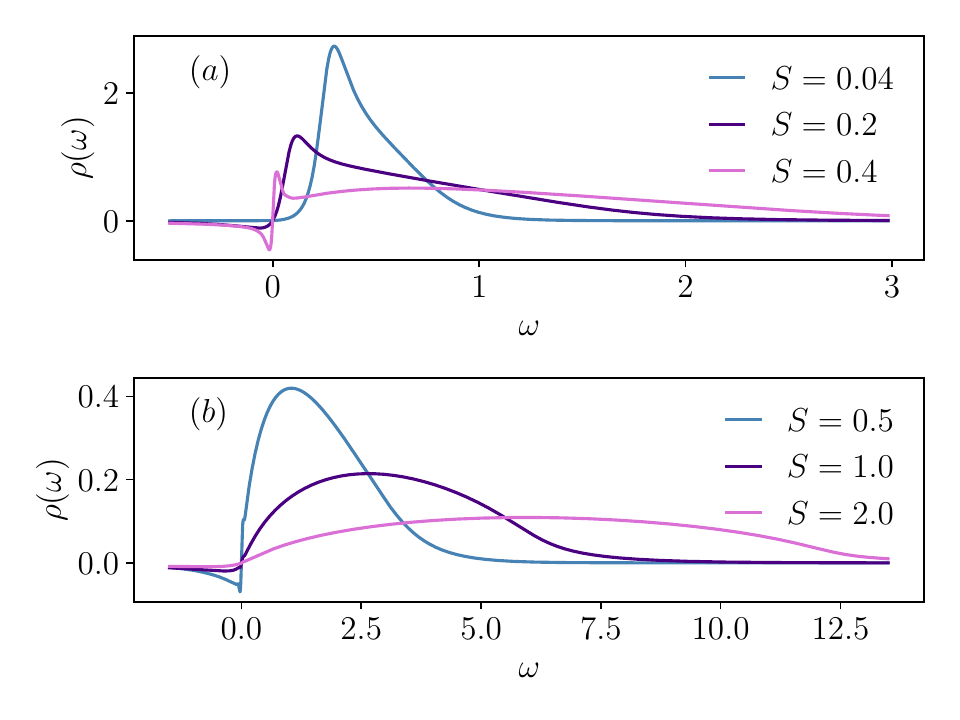}
	\caption{Spectral function as a function of $S$ at  $T = 0.1$  in (a) PM phase and (b) SG phase.}.\label{fig:rho_T0.1} 
\end{figure}

The other PM solution, the local moment solution coexist, and has intricate competition \cite{Georges2001} with the spin liquid solution in the quantum critical region $IV$ (Fig.\ref{fig:phasediagramSY}). As a result, it is not always easy to distinguish the spin liquid solution and the local moment solution for general $(S,T)$ in the quantum critical region. Ref. \onlinecite{Camjayi2003} has shown the existence of spin liquid solution for small $S<0.5$ using analytical continuation of the numerical solution of the imaginary-time saddle-point equations [Eqs.\eqref{Eq:SaddlePM}]. 
% Special care~\cite{Sachdev1993,Tikhanovskaya2021} needs to be taken to obtain the spin liquid solution [Eq.\eqref{eq:SpinLiquidRho}] for small $S$ and $T$ from the numerical solution of the real-frequency saddle-point equations [Eq.\eqref{Eq:SaddlePM_real}] due to the $1/\sqrt{|\omega|}$ divergence. This has not been done in our numerical calculations since we do not focus on the small-$S$ region in much detail. 
Our results are consistent with that of Ref. \onlinecite{Camjayi2003}.
% , however, we also get solutions which are closer to the local moment for some parameter values at small $S\lesssim 0.1$, as discussed below. 
This local moment becomes the dominant solution above the quantum critical region $T\gtrsim J$, in the PM phase. At high temperature, the local moment solution has a characteristic peak at $\omega\sim \omega_0\sim T\ln \,{[(S+1)/S]}$~\cite{Georges2001}. At very low temperature, below $T_{SG}$ (Fig.\ref{fig:phasediagramSY}), where the local moment solution is metastable, its spectral function has two peaks at $\sim \pm 1/T$ and a delta function peak with $\rho(\omega)\sim S\omega\delta(\omega)$~\cite{Georges2001}. This low-temperature local moment solution is not relevant for us. 

\begin{figure}[h]
	\centering
	\includegraphics[width=0.5\textwidth]{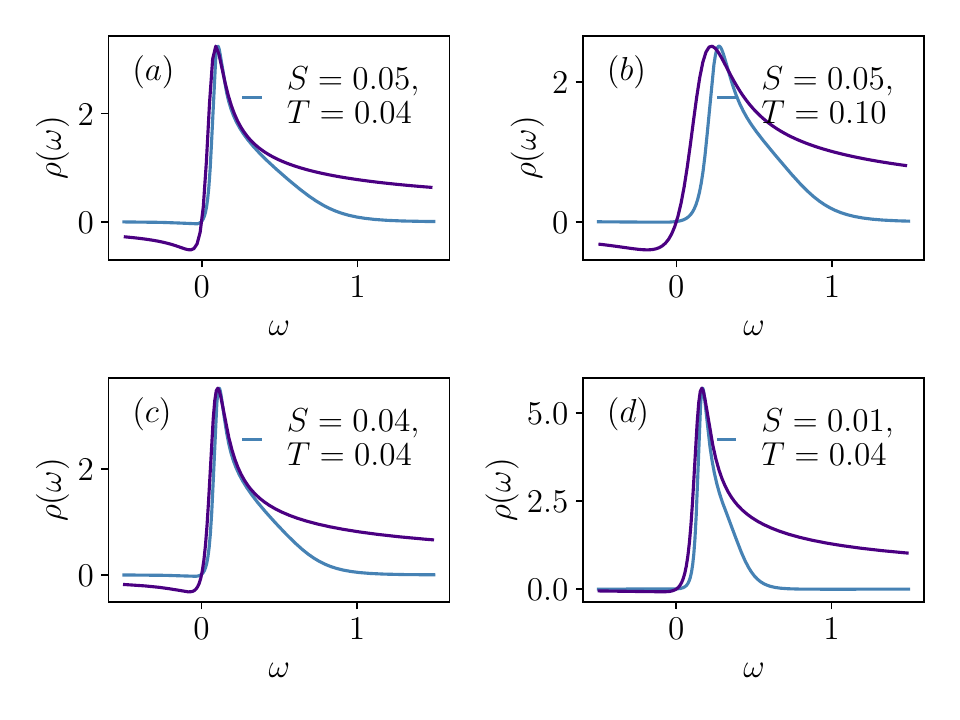}
	\caption{Comparison of numerically obtained spectral function (blue) for small $S<S_{max}$ with finite-temperature conformal spectral function (purple) of Eq.\eqref{eq:SpinLiquidRho} for the spin liquid. The conformal spectral function is scaled as necessary so that its maximum becomes equal to that of numerical results for the spectral function. 
 %We find reasonable agreement for the location of the peak and in some region around the peak between the two spectral functions, except in (b) which is at relatively high temperature, $T=0.1$.}. The scaling factors applied on conformal spectral function to match the peak value are  1.56 for (a), 1.9 for (b), 1.43 for (c) and 1.42 for (d)
 }\label{fig:rho_S0.05}
\end{figure}

For the marginal spin glass solution, on the other hand, one again obtains a linear-in-$\omega$ behavior of the spectral function, termed as pseudogap~\cite{Camjayi2003}, where $\rho(\omega)\sim \omega/S$ for low frequency and low temperature $T\ll T_{SG}$, with a high-frequency broad peak. As $T$ increases towards glass transition, a narrow low-frequency peak around $|\omega|\sim T$ appears on top of the linear-$\omega$ pseudogap background. The latter fills up with increasing temperature. The narrow peak around $\sim T$ appears from the loss of spectral weight from the $\delta$-function part in the correlation function, e.g., the Wightmann function of Eq.\eqref{eq:WightmannSG}, in the SG phase. The $\delta(\omega)$ part in the correlation functions corresponds to static order and its weight is transferred to excitations in the narrow peak of the spectral function $\rho(\omega)$ due to the gradual unfreezing of spins approaching the SG transition. The broad high-frequency peak at $\omega\sim\mathcal{O}(JS)$ originates from precession of frozen spin around their local fields~\cite{Camjayi2003}. 

\begin{figure}[h]
	\centering
	\includegraphics[width=0.5\textwidth]{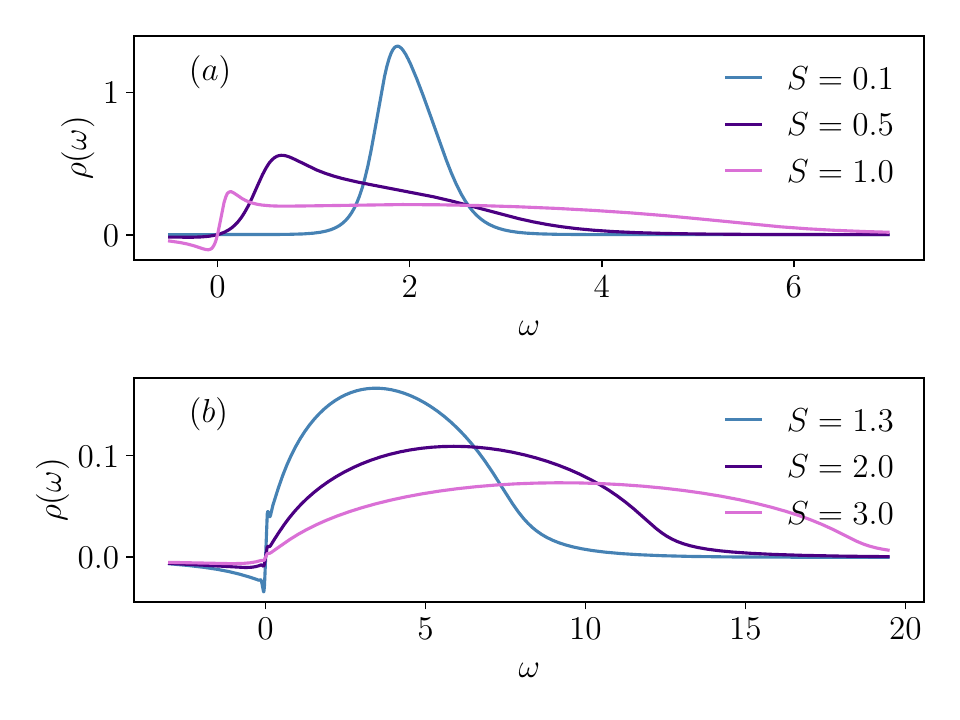}
	\caption{Spectral function as a function of $S$ at  $T = 0.8$  in (a) PM phase and (b) SG phase.}\label{fig:rho_T0.8}
\end{figure}

We have studied the evolution of the spectral function along the fixed-$T$ and fixed-$S$ cuts (Fig.\ref{fig:phasediagramSY}) employed for our chaos calculations in Secs. \ref{sec:LyapunovExponent},\ref{sec:ButterflyVelocity}. We find crossover between the PM solutions and transition from the PM to SG solution. Figs.\ref{fig:rho_T0.1}(a) and (b) show the evolution of $\rho(\omega)$ at low temperature $T=0.1$ with $S$ ($d$ cut, Fig.\ref{fig:phasediagramSY}) for the PM and SG phases, respectively. The result for $S=0.04<S_{max}$ in Fig.\ref{fig:rho_T0.1}(a) is qualitatively consistent with a spin liquid solution of Eq.\eqref{eq:SpinLiquidRho}. In Fig.\ref{fig:rho_S0.05}, we make detailed comparison of numerically obtained $\rho(\omega)$ with conformal spectral function of Eq.\eqref{eq:SpinLiquidRho} for $S<S_{max}$ at low temperature. The match is reasonable for $T\lesssim 0.05$ at low energies [Fig.\ref{fig:rho_S0.05}(a,c,d)]. However, there are large deviations from the conformal form at high energies. The comparison becomes worse for $S=0.05,~T=0.1$ [Fig.\ref{fig:rho_S0.05}(b)]. The spectral function $\rho(\omega)$ for $S=0.2>S_{max}$ [Fig.\ref{fig:rho_S0.05}(a)] also qualitatively resembles a spin liquid solution, though no analytical form is known in this case.
% However, for even smaller $S=0.04$, the $\rho(\omega)$ with a peak around $\omega\sim 0.3\sim \omega_0$ is consistent with a local moment-like solution. This solution was obtained consistently even after reaching the parameter by cooling down from high temperature. 
The spectral function acquires low-energy and high-energy peaks, with a dip in spectral weight in between, with further increase of $S$ towards $S_{SG}\simeq 0.48$, as shown in Fig.\ref{fig:rho_T0.1}(a). In the SG phase [Fig.\ref{fig:rho_T0.1}(b)], $\rho(\omega)$ follows the expected marginal SG behavior discussed above. Qualitatively similar progression of $\rho(\omega)$ with $S$ is seen in Fig.\ref{fig:rho_T0.8} for the higher $T=0.8$ ($e$ cut in Fig.\ref{fig:phasediagramSY}) with $S_{SG}\simeq 1.29$. However, at this higher temperature, $\rho(\omega)$ may be consistent with a local moment like solution with a peak at $\omega_0\sim T\ln \,{[(S+1)/S]}$.

\begin{figure}[h]
	\centering
	\includegraphics[width=0.5\textwidth]{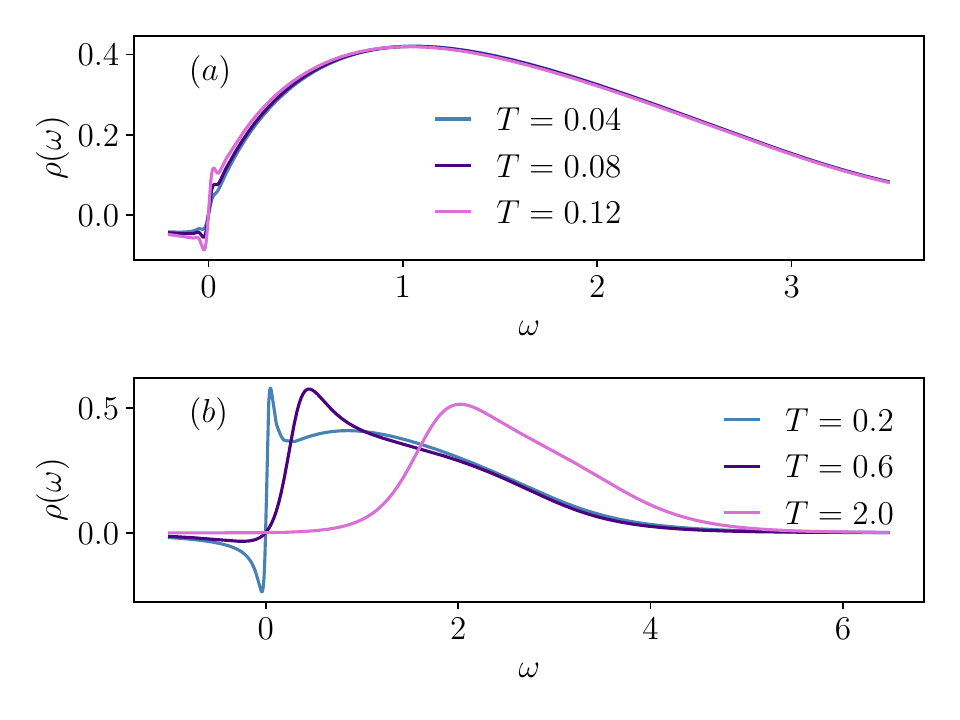}
	\caption{Spectral function as a function of $T$ at $S=0.5$ in (a) SG phase and (b) PM phase.}\label{fig:rho_S0.5}
\end{figure}

\begin{figure}[h]
	\centering
	\includegraphics[width=0.5\textwidth]{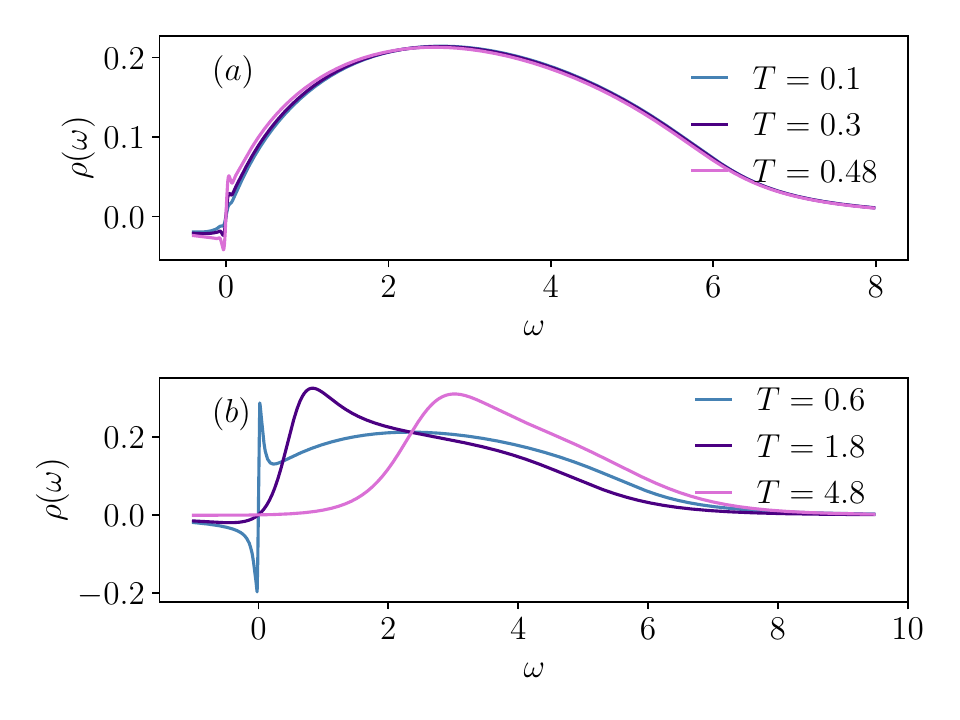}
	\caption{Spectral function as a function of $T$ at $S=1.0$ in (a) SG phase and (b) PM phase.}\label{fig:rho_S1.0}
\end{figure}

Similarly, for fixed-$S$ cuts $a$ ($S=0.5$) and $b$ ($S=1.0$) in Fig.\ref{fig:phasediagramSY}, the spectral function evolves from that of the SG solutions for $T<T_{SG}\simeq 0.12,~0.49$ in Figs.\ref{fig:rho_S0.5}(a),\ref{fig:rho_S1.0}(a) to the spin liquid or local moment spectral functions in Figs.\ref{fig:rho_S0.5}(b),\ref{fig:rho_S1.0}(b). The narrow peak close to the transition $T\gtrsim T_{SG}$ at $\omega\sim \mathcal{O}(T)$ results from the melting of static order in the SG phase, as discussed earlier. 

% within the SG phase, the specrtum $\tilde{\rho}(\w)$ has a linear in $\w$ behavior close to $\w = 0$. There is hardly any dependence of $\tilde{\rho}(\w)$  on $T$, as seen in Fig.\ref{rhoGT_SG}. \textcolor{magenta}{As $T$ is increased, in the PM phase, the pseudogap gets filled as the thermal fluctuations start melting the frozen spins in $\delta$ function part. At further high temperature, pseudogap disappears as the $\delta$ function merges with the high frequency part of the spectrum and becomes a single peak} {What is the basis of these statements? In this part on spectral function, many statements seem very ad hoc, without stating any evidence or argument or citing any reference.}.{L: This explanation is taken from \cite{Camjayi2003}. Related to Fig 2 in that paper. Although they plot susceptibility($\chi^{\prime \prime}(\w)$) instead of $\rho(\w)$.} As a function of temperature, the peak keeps shifting to higher frequencies $\sim T\mathrm{ln}[(S+1)/S]$, as seen in \ref{rhoGT_PM}. 

\begin{figure}[h]
	\centering
	\includegraphics[width=0.5\textwidth]{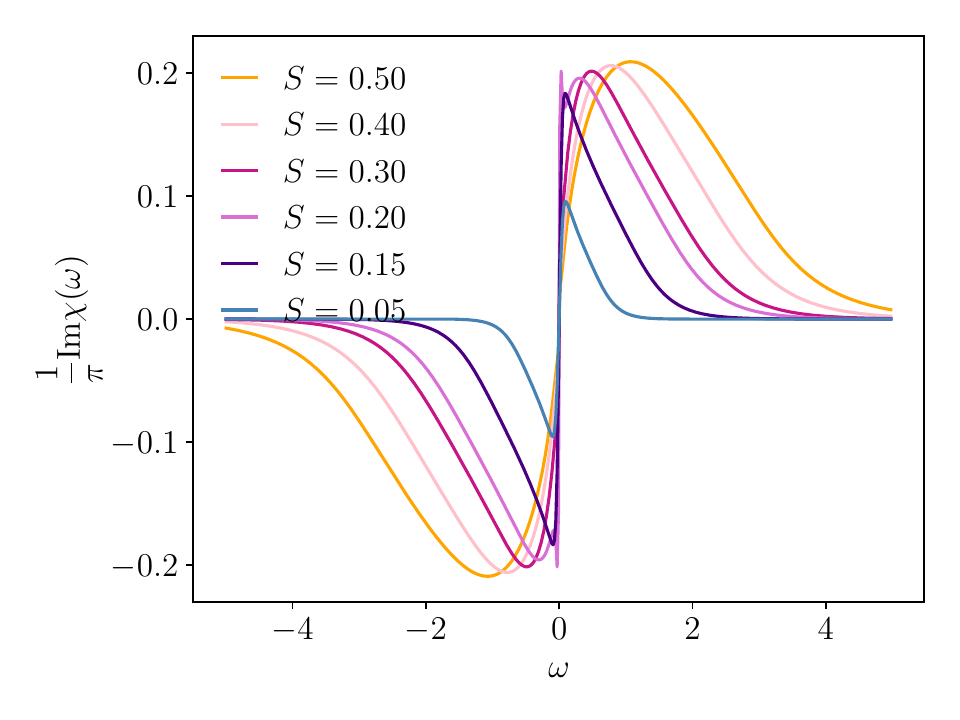}
	\caption{Susceptibility $\chi^{\prime \prime}(\w)$ across the PM-SG transition at $T=0.04$. The transition happens at critical value of $S\approx 0.28$. }\label{fig:chi_T0.04}
\end{figure}

\begin{figure}[h]
	\centering
	\includegraphics[width=0.5\textwidth]{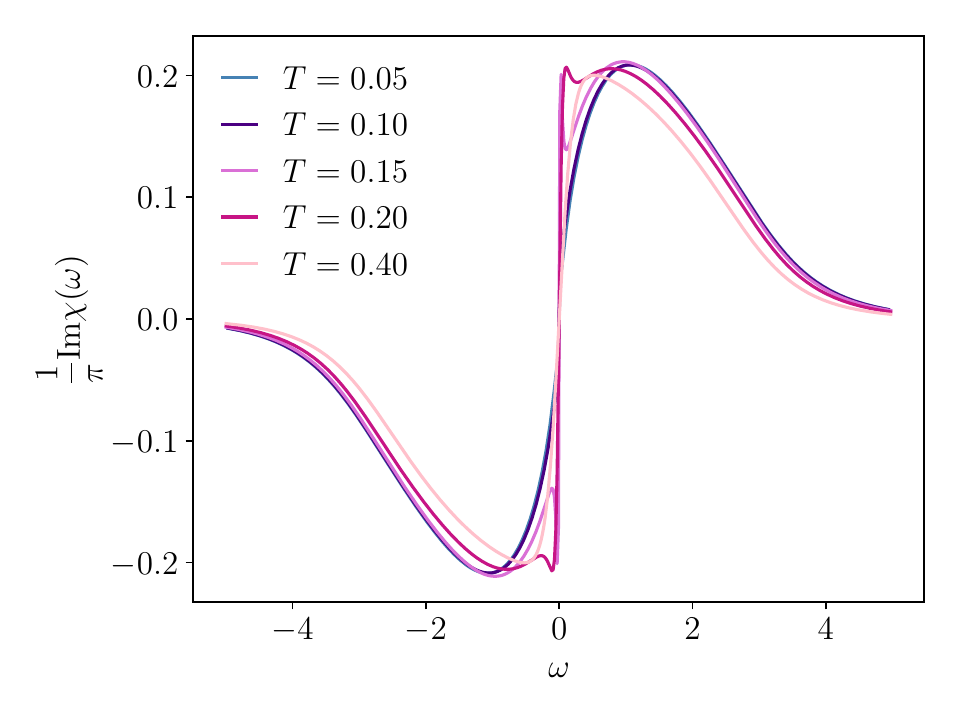}
	\caption{Susceptibility $\chi^{\prime \prime}(\w)$ across the PM-SG transition at $S=0.5$. The transition temperature is $T \approx 0.12$}\label{fig:chi_S0.5}
\end{figure}

%Similarly at a fixed $S$, starting from a lower temperature well within the SG phase, the specrtum $\tilde{\rho}(\w)$ has a linear in $\w$ behavior close to $\w = 0$ and a $\delta$ function. As $T$ is increased, the pseudogap gets filled as the thermal fluctuations start melting the frozen spins in $\delta$ function part. Also one can notice the transfer of spectral weight from higher frequencies to pseudogap region. As $T$ is increased further we enter the PM phase, where the pseudogap disappears as the $\delta$ function merges with the high frequency part of the spectrum and becomes a single feature. [Large S?][Figures?]
{\it Spin Susceptibility.}--- To probe the spin liquid regime for small $S$ further, we compute the local spin susceptibility $\chi(\tau) = \langle S(\tau) S(0)\rangle$. In the large $N$ limit, $\chi(\tau) = G^{aa}(\tau) G^{aa}(-\tau)$.  From this, we calculate imaginary part of $\w$-dependent local spin susceptibility  $\chi^{\prime \prime }(\w) = (1/\pi) \mathrm{Im}\,  \chi(\w)$, as

\begin{align}
    \chi^{\prime \prime }(\w) =&  \Big[-  \frac{\tilde{J}^2}{\pi}  \mathrm{Im}\,  \Big(\ci \int_0^\infty dt e^{\ci \w t}\big(n_1(t) n_2(t) - n_3(t) n_4(t) \big)\Big) \\ \nonumber
    &+g \rho(\w) - g\rho(-w) \Big]
\end{align}

 where $n_i(t)$ ($i=1,2,3,4$) is defined below Eq.\eqref{Eq:SigmawPM_fft} and $g$ is non-zero only in SG phase. Here $\chi^{\prime \prime }(\w)$ satisfies the constraint, $\int d \w \chi^{\prime \prime }(\w) n_B(\w) = S(S+1)$ in PM phase and $\int d \w \chi^{\prime \prime }(\w) n_B(\w) = S(S+1) - g^2$ in SG phase. In Figs.\ref{fig:chi_T0.04} and \ref{fig:chi_S0.5} we plot $\chi^{\prime \prime}(\w)$ across the PM-SG transition at a fixed $T=0.04$ and at a fixed $S=0.5$, respectively. We find that these results are in agreement with those (Figures 1 and 2) of Ref. \onlinecite{Camjayi2003} that are obtained using analytical continuation of imaginary solutions of large $N$ saddle point equations using Pad\'e approximation. 
 
 Thus, overall, our results for spectral function and local susceptibility, discussed in this section, are consistent with the spin liquid solutions for small $S$. Nevertheless, as show in Sec.\ref{sec:LyapunovExponent}, e.g., in Fig.\ref{fig:SY_lambdaL_S}, the Lyapunov exponent for these spin liquid solutions extrapolates to a value much smaller than maximal $2\pi T$ for $S\to 0$, unlike in the SYK-type models \cite{BanerjeeAltman2016}. This could be due to large non-conformal \cite{Maldacena2016} $T$ and $S$ dependent corrections to the conformal solution [Eq.\eqref{eq:SpinLiquidRho}], seen in Figs.\ref{fig:rho_S0.05}.

{\it Wightmann correlation functions---}
The correlation functions needed in the calculation of chaos can be obtained from the knowledge of the retarded Green's function or the spectral function discussed above. We calculate $G(\tau)$ using the spectral function from 
\begin{align}
G(\tau) = \begin{cases} 
						\int_{-\infty}^\infty d \omega \frac{e^{-\omega \tau} \rho(\omega)}{e^{-\beta \omega} -1} &  \quad(\tau > 0)\\
						-\int_{-\infty}^\infty d \omega \frac{e^{-\omega \tau} \rho(\omega)}{e^{\beta \omega} -1} & \quad (\tau < 0)
						\end{cases}
\end{align}

Analytically continuing $G(\tau)$ to real time gives Wightmann correlation functions, in the PM phase,

\begin{align}
G_{lr}^+(t) = - G(\ci t + \beta/2) = \int_{-\infty}^\infty \frac{d\omega}{2 \pi} e^{-\ci\omega t } \frac{ \pi\rho(\omega)}{ \mathrm{sinh}\frac{\beta \omega}{2}}\\
G_{lr}^-(t) = - G(\ci t - \beta/2)  = \int_{-\infty}^\infty \frac{d\omega}{2 \pi} e^{-\ci\omega t } \frac{ \pi\rho(\omega)}{ \mathrm{sinh}\frac{\beta \omega}{2}}
\end{align}
We notice that $G_{lr}^+(t) =  G_{lr}^-(t)$ and hence $G_{lr}^+(\w) = G_{lr}^-(\w) = \pi\rho(\omega)/ \mathrm{sinh}(\beta \omega/2)$.
Similarly, in the SG phase, 

\begin{align}
G_{lr}^{+,ab}(t) =  G_{lr}^{-,ab}(t) = g\epsilon_{ab} + \int_{-\infty}^\infty \frac{d\omega}{2 \pi} e^{-\ci\omega t } \frac{ \pi\rho(\omega)}{ \mathrm{sinh}\frac{\beta \omega}{2}}\delta_{ab},
\end{align}
that allows to define $G_{lr}^{+,ab}(\w) = G_{lr}^{-,ab}(\w) = 2\pi g\delta(\w)\epsilon_{ab}  + \pi\rho(\omega)/ (\mathrm{sinh}(\beta \omega/2))\delta_{ab}$. The delta function term proportional to the EA order parameter $g$ at $\omega=0$ in the Wightmann correlation function indicates the presence of static spin glass order. Similar delta function term appears in the Fourier transform of other correlation functions, e.g., $G^{aa}_>(t)=-(\ci/M)\sum_\alpha\langle b_{i\alpha, x}^a \bar{b}_{i\alpha, x}^a \rangle$, and similarly $G^{aa}_<(t)$. 
We note that, since retarded Green's function, $G^{aa}_R(t) = \theta(t)[G^{aa}_>(t) - G^{aa}_<(t)]$ [$\theta(t)$ is the Heaviside function] and $G^{aa}_R(\omega)$ is obtained by analytically continuing $\ci \omega_k\to \omega+\ci0^+$ in Eq.\eqref{Eq:1rsb}, $G_R^{aa}(t)$ is free of any static part even in the SG phase.

\section{Imaginary-time path integral and saddle point equations in $p$-spin glass chain}\label{appB}

As in the case of SY chain [\ref{appA}], we write the partition function $\mathcal{Z}(\{J_{ijk,x},J^\pm_{ijk,x}\})=\mathrm{Tr}[\exp(-\beta \mathcal{H})]$ ($\beta=1/T$) of the disordered model of Eq.\eqref{eq:pSpin_1d}, where $J_{ijk,x}$ and $J_{ijk,x}^\pm$ are Gaussian random number with zero mean and variances $3J^2/(2N^2)$ and $3J'^2/2N^2$, as an imaginary-time path integral. We calculate disorder-averaged quantities by introducing replicas $a=1,\dots,n$, and a Lagrangian multiplier $z(\tau)$ to impose the spherical constraint, to obtain $\overline{\mathcal{Z}^n}= \int  \mathcal{D}s \mathrm{exp}[-S_{\mathrm{eff}}]$, where 

\begin{align}
  &S_{\mathrm{eff}} =   \sum_x \bigg\{\int_\tau \left[ \frac{1}{2\Gamma} \left(\frac{\partial s_{i,x}^a}{\partial \tau}\right)^2+\frac{z^a(\tau)}{2}\left( (s_{i,x}^a)^2 -N\right)\right] \nonumber \\
  &- \frac{1}{4}\int_{\tau,\tau^\prime} \sum_{ab} \left[ \frac{J^2}{N^{2}}  \left(\sum_{i} s_{i,x}^a(\tau) s_{i,x}^b(\tau^\prime) \right)^3 \right.\nonumber \\
  &+ \frac{J^{\prime 2}}{N^{2}}\left(\sum_{i} s_{i, x}^a(\tau) s_{i,x}^b(\tau^\prime) \right)\left(\sum_{i} s_{i, x+1}^a(\tau) s_{i, x+1}^b(\tau^\prime) \right)^2 \nonumber\\
& \left. + \frac{J^{\prime 2}}{N^{2}} \left(\sum_{i} s_{i, x}^a(\tau) s_{i, x}^b(\tau^\prime) \right)^2 \left(\sum_{i} s_{i, x+1}^a(\tau) s_{i, x+1}^b(\tau^\prime) \right)\right]  \bigg\}
\end{align}
Here $\int_\tau=\int_0^\beta d\tau$. We now introduce the large $N$ field $Q$ and its conjugate $\Sigma$, where $Q_x^{ab}(\tau, \tau^\prime) = \frac{1}{N}\sum_i \langle s_{i, x}^a(\tau ) s_{i, x}^b(\tau^\prime ) \rangle$, through the identity
\begin{align}
    1 =& \int D Q_x^{ab}(\tau, \tau^\prime) \delta \Big(N Q_x^{ab}(\tau, \tau^\prime) - \sum_i s_{i, x}^a(\tau ) s_{i, x}^b(\tau^\prime ) \Big) \nonumber\\
    =& \int D Q_x^{ab}(\tau, \tau^\prime) D\Sigma_x^{ab}(\tau, \tau^\prime)\mathrm{exp} \Big\{ -\frac{1}{2}\int_{\tau, \tau^\prime} \Sigma_x^{ab}(\tau, \tau^\prime) \nonumber \\
    &\Big(N Q_x^{ab}(\tau, \tau^\prime)- \sum_i s_{i, x}^a(\tau) s_{i, x}^b(\tau^\prime) \Big) \Big\}
\end{align}
Inserting the identity and integrating  over $s_{i, x}(\tau)$ in $\overline{\mathcal{Z}^n}$ leads to the effective action
\begin{align}
    &S_{\mathrm{eff}} = \sum_x \bigg\{\frac{N}{2} \mathrm{Tr} \ln \,\Big[ \big(-\frac{1}{\Gamma}\partial_{\tau}^2+z^a(\tau)\big)\delta(\tau- \tau^\prime)\delta_{ab} \nonumber \\
    & -\Sigma_x^{ab}(\tau, \tau^\prime) \Big] -\frac{N}{2} \sum_a \int_\tau z^a(\tau) \nonumber \\
    &+\frac{N}{2} \sum_{ab}\int_{\tau,\tau^\prime}  \Big[\Sigma_x^{ab}(\tau, \tau^\prime)Q_x^{ab}(\tau, \tau^\prime) -\frac{J^2}{4}  Q_x^{ab}(\tau, \tau^\prime)^3 \nonumber\\
    &-  \frac{J^{\prime 2}}{4} \Big(Q_x^{ab}(\tau, \tau^\prime) Q_{x+1}^{ab}(\tau, \tau^\prime)^2+Q_x^{ab}(\tau, \tau^\prime)^2 Q_{x+1}^{ab}(\tau, \tau^\prime)\Big)\Big] \bigg\}
    \end{align}
 In the limit $N \rightarrow \infty$, we can take the saddle point with respect to $\Sigma_x^{ab}(\tau, \tau^\prime)$. This gives $\boldsymbol{Q}_x^{-1}=\boldsymbol{Q}_0^{-1}-\boldsymbol{\Sigma}$, where the matrix $\boldsymbol{Q}^{-1}_0$ is given by $(Q_0^{-1})^{ab}(\tau,\tau') = \big(-\frac{1}{\Gamma}\partial_{\tau}^2   + z^a(\tau)\big)\delta(\tau- \tau^\prime)\delta_{ab}$.
Using this we eliminate $\Sigma_x^{ab}(\tau, \tau^\prime)$ from the action and obtain effective action in terms of $Q$,
\begin{align}
    &S_{\mathrm{eff}} = -\sum_x \bigg\{ \frac{N}{2} \mathrm{Tr} \ln \, \boldsymbol{Q}_x - \frac{N}{2} \sum_a \int_\tau z^a(\tau) \big( Q_x^{aa}(\tau, \tau)-1 {\big)}\nonumber \\
    &+ N \sum_{a,b}\int_0^\beta  d\tau d \tau^\prime \Big( \frac{1}{2\Gamma} \partial_\tau^2 Q_x^{ab}(\tau, \tau^\prime) \delta(\tau - \tau^\prime) \delta_{ab} \nonumber\\
    &+ \frac{J^2}{4 } Q_x^{ab}(\tau, \tau^\prime)^3 +  \frac{J^{\prime 2}}{4 } Q_x^{ab}(\tau, \tau^\prime)Q_{x+1}^{ab}(\tau, \tau^\prime)^2 \nonumber \\
      &+ \frac{J^{\prime 2}}{4 } Q_x^{ab}(\tau, \tau^\prime)^2 Q_{x+1}^{ab}(\tau, \tau^\prime) \Big)\bigg\}
\end{align}
Taking the saddle point with respect to $\boldsymbol{Q}$ gives the Dyson equation,
\begin{align}
     &(Q_x^{-1})^{ab}(\tau, \tau^\prime) = \Big(-\frac{1}{\Gamma}\partial_\tau^2+z_a(\tau) \Big) \delta(\tau - \tau^\prime) \delta_{ab} \nonumber \\
     &- \frac{3 J^2}{2}Q_x^{ab}(\tau, \tau^\prime)^2 - \frac{J^{\prime 2}}{2}\Big(Q_{x+1}^{ab}(\tau, \tau^\prime)^2 + Q_{x-1}^{ab}(\tau, \tau^\prime)^2\Big)\nonumber \\
     &- J^{\prime 2}Q_x^{ab}(\tau, \tau^\prime)\Big(Q_{x+1}^{ab}(\tau, \tau^\prime)  { + Q_{x-1}^{ab}(\tau, \tau^\prime)}\Big)
 \end{align}
Considering space and time translation invariant solutions for the Dyson equation, of the form $Q_x^{ab}(\tau,\tau')=Q^{ab}(\tau-\tau')$ and $z^a(\tau)=z^a$, the above equation reduces to,
\begin{subequations}
\begin{align}
     (Q^{-1})^{ab}(\ci \omega_k) &= (\frac{1}{\Gamma}\omega_k^2+z^a) \delta_{ab}-\Sigma_{ab}(\omega_k)\\
     \Sigma^{ab}(\tau)&=\frac{3\tilde{J}^2}{2}Q^{ab}(\tau)^2
 \end{align}
 \end{subequations}
 where {$\tilde{J}^2=J^2 + 2J^{\prime 2}$}. The above large $N$ saddle-point equations are the same as that of the zero-dimensional model \cite{Cugliandolo2001}.\\
 {\it Saddle point equation in PM phase---}
 Using the replica diagonal ansatz and analytically continuing to real frequency, we obtain the saddle-point equation for the retarded Green's function, 
\begin{subequations}
     \begin{align}
     [Q_R(\w)]^{-1} =- \frac{1}{\Gamma} \w^2 + z - \Sigma_R(\w) \\
     \Sigma_R(\w) = \Sigma(\ci\w_k\to \omega+\ci 0^+)
 \end{align}
\end{subequations}
 Following a procedure similar to that in the case of SY model (\ref{appA}), the self-energy can be written as,
\begin{subequations}
     \begin{align}
     \Sigma_R(\w) &= \ci \frac{3\tilde{J}^2}{2} \int_0^\infty e^{\ci \w t}\Big(n_1^2(t) - n_2^2(t) \Big), \\
     n_1(t) &= \int d\w e^{-\ci\w t}\rho(\w) n_B(-\w), \nonumber\\
     n_2(t) &= \int d\w e^{-\ci\w t}\rho(\w) n_B(\w) \nonumber
 \end{align}
\end{subequations}
where $\rho(\omega)=-(1/\pi)\mathrm{Im}\, Q_R(\omega)$.The saddle-point equations can be solved numerically, following steps similar to that discussed in the case of SY model.

  {\it Saddle point equation in SG phase---}
  Using 1RSB ansatz, after analytical continuation to real frequency $\ci\omega_k\to \omega+\ci 0^+$ only the regular part (see Sec.\ref{sec:pSpin_PhaseDiagram}) contributes to the retarded Green's function $Q_{R}^{ab}(\omega)=Q_R(\omega)\delta_{ab}=\tilde{Q}(\ci\omega_k\to \omega+\ci 0^+)\delta_{ab}$ such that
\begin{subequations}
\begin{align}
&[Q_{R}(\omega)]^{-1} = -\frac{1}{\Gamma}\w^2 + z  -\Sigma_{R}(\omega), \\
&\Sigma_{R}(\omega)  =  \frac{3{J}^2}{2}\int_{0}^{\beta}d\tau e^{\ci \w_k\tau}[\tilde{Q}(\tau)]^2 + 3\tilde{J}^2q_\mathrm{EA}Q_{R}(\w) \nonumber \\
&= \ci \frac{3\tilde{J}^2}{2} \int_0^\infty e^{\ci \w t}\Big(n_1^2(t) - n_2^2(t) \Big) + 3\tilde{J}^2q_\mathrm{EA}Q_{R}(\w),
\end{align}
where
\begin{align}
&n_1(t) = \int d\w e^{-\ci\w t}\rho(\w) n_B(-\w), \\
&n_2(t) = \int d\w e^{-\ci \w t}\rho(\w) n_B(\w)
\end{align}
\end{subequations}
The spherical constraint now takes the form,
 \begin{align}
     -\int_{-\infty}^\infty d\w \rho(\w)n_B(\w)  = 1-q_\mathrm{EA}
 \end{align}
 {\it Wightmann correlation functions---} The Wightmann functions in the PM phase are given by, 
  \begin{align}
     Q_{lr}(t) = Q(\ci t +\beta/2) = -\int \frac{d \w}{2 \pi} e^{- \ci \w t}\frac{\pi \rho(\w)}{\mathrm{sinh}\frac{\beta \w}{2}},
 \end{align}
 and 
 \begin{align}
     Q_{lr}^{ab}(t) = \epsilon_{ab}q_\mathrm{EA}  -\delta_{ab}\int \frac{d \w}{2 \pi} e^{- \ci \w t} \frac{\pi \rho(\w)}{\mathrm{sinh}\frac{\beta \w}{2}}
 \end{align}
 in the SG phase. 

\section{Calculation of butterfly velocity from the single mode ansatz}\label{app:vB}
\begin{figure}[h]
	\centering
	\includegraphics[width=1.0\linewidth]{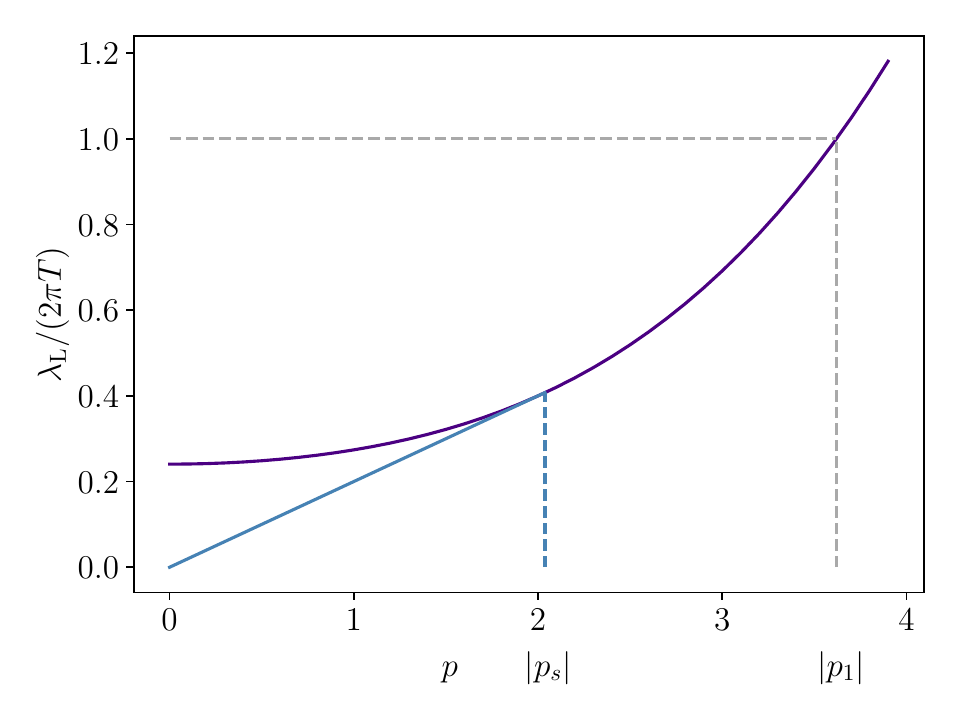}
	\caption{  $\lambda_\mathrm{L}(p=\ci|p|)$ for $T=0.8$ and $S=0.8$ where $p_1$ and $p_s$ are marked.}\label{fig:p1psS0.8}
\end{figure}

\begin{figure}[h]
	\centering
	\includegraphics[width=1.0\linewidth]{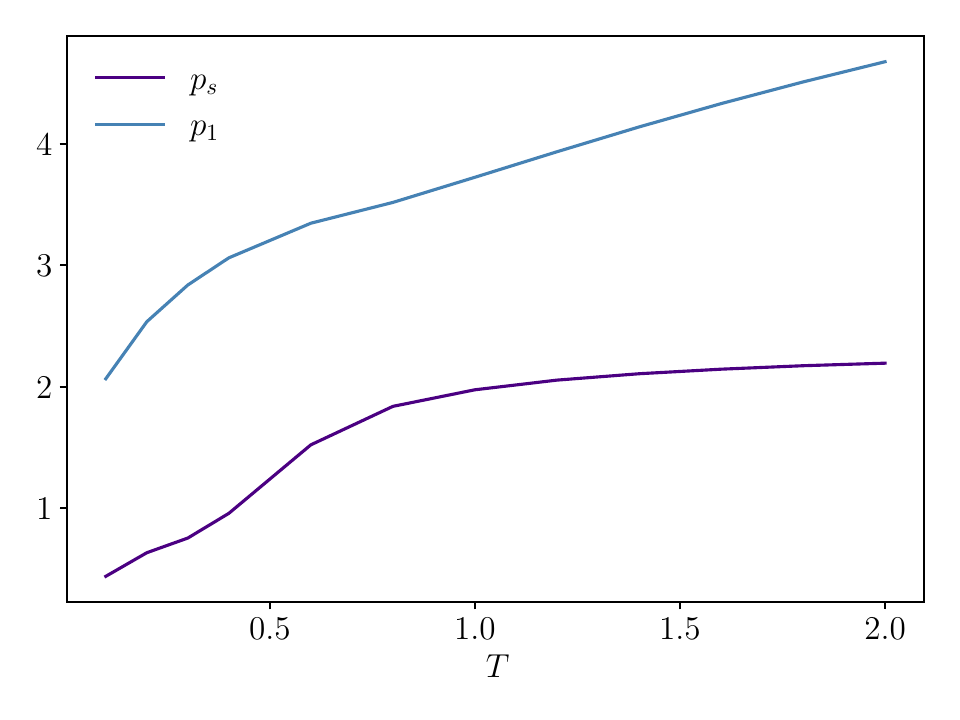}
	\caption{$p_1$, $p_s$ for a range of temperatures at $S=1.0$}\label{pspt}
\end{figure}

Here we briefly discuss the computation of $v_B$ in SY model using the imaginary $p$ method or the single mode ansatz, as described in the main text (Sec.\ref{sec:SY_vB}). By calculating $\lambda_\mathrm{L}(p)$ for imaginary values of momentum, $p=\ci|p|$, using the exponential growth ansatz $\mathcal{F}_{\mu,p}(t_1,t_2)= e^{\lambda_\mathrm{L}(p)(t_1+t_2)/2}f_{\mu,p}^a(t_1-t_2)$ in Eq.\eqref{eq:BetheSalpeter_1D}, we find $p_1$ and $p_s$, as discussed in Sec.\ref{sec:SY_vB}. At $T=0.8$ and $S=0.8$, for example, we have {$p_1 \simeq 3.62$ and $p_s\simeq 2.04$}, as shown in Fig.\ref{fig:p1psS0.8}. Also, in Fig.\ref{pspt}, we show that $p_s$ is always less than $p_1$, e.g., at $S=1.0$. 
% Further we show the light cone, the slope of which give $v_B$ at $T = 0.8$ and $S = 1.0$.

\end{document}